\documentclass[a4paper,onecolumn,11pt,accepted=2024-09-02]{quantumarticle}
\pdfoutput=1

\pdfoutput=1
\usepackage[utf8]{inputenc}
\usepackage[english]{babel}
\usepackage[T1]{fontenc}

\usepackage{amssymb,nicefrac}
\usepackage{amsfonts}
\usepackage{graphicx}
\usepackage[fleqn]{amsmath}
\usepackage{amsthm}
\usepackage[varg]{txfonts}
\usepackage{stmaryrd}

\usepackage{framed}
\usepackage{color}
\usepackage[normalem]{ulem}
\usepackage{tikzfig}
\usepackage{placeins}

\usepackage{bm}
\usepackage{braket}

\usepackage{blkarray}

\usepackage{thmtools}
\usepackage{thm-restate}

\usepackage{hyperref}

\usepackage{mathtools} 
\usepackage{bigints}

\pgfplotsset{compat=1.17}

\tikzstyle{env}=[copoint,regular polygon rotate=0,minimum width=0.2cm, fill=black]

\tikzstyle{every picture}=[baseline=-0.25em]
\tikzstyle{dotpic}=[scale=0.5]
\tikzstyle{diredges}=[every to/.style={diredge}]
\tikzstyle{dot graph}=[shorten <=-0.1mm,shorten >=-0.1mm,scale=0.6]
\tikzstyle{plot point}=[circle,fill=black,minimum width=2mm,inner sep=0]

\tikzstyle{braceedge}=[decorate,decoration={brace,amplitude=2mm,raise=-1mm}]
\tikzstyle{small braceedge}=[decorate,decoration={brace,amplitude=1mm,raise=-1mm}]
\tikzstyle{left hook arrow}=[left hook-latex]
\tikzstyle{right hook arrow}=[right hook-latex]

\tikzstyle{dtriangle}=[fill=yellow,draw=black,shape=isosceles triangle,shape border rotate=-90,isosceles triangle stretches=true,inner sep=0.8pt,minimum width=0.25cm,minimum height=2mm]
\tikzstyle{vtriang}=[fill=yellow,draw=black,shape=isosceles triangle,shape border rotate=180,isosceles triangle stretches=true,inner sep=0.8pt,minimum width=0.25cm,minimum height=2mm]
\tikzstyle{trigmc}=[fill=green,draw=black,shape=isosceles triangle,shape border rotate=90,isosceles triangle stretches=true,inner sep=0.8pt,minimum width=0.3cm,minimum height=2mm]
\tikzstyle{vrt}=[fill=yellow,draw=black,shape=isosceles triangle,shape border rotate=0,isosceles triangle stretches=true,inner sep=0.8pt,minimum width=0.25cm,minimum height=2mm]
\tikzstyle{H box}=[rectangle,fill=yellow,draw=black,xscale=1.5,yscale=1.5, inner sep=1.6pt]
\tikzstyle{gbox}=[gn_phase, rounded corners=0]
\tikzstyle{rbox}=[rectangle,fill=red,draw=black,xscale=1.0,yscale=1.0, inner sep=1.pt]
\tikzstyle{zhbx}=[rectangle,fill=white,draw=black,xscale=1.0,yscale=1.0, inner sep=1.6pt]
\tikzstyle{newh}=[rectangle,fill=yellow,draw=black,xscale=2.0,yscale=2.0, inner sep=1.6pt]
\tikzstyle{triangle}=[fill=yellow,draw=black,shape=isosceles triangle,shape border rotate=90,isosceles triangle stretches=true,inner sep=0.8pt,minimum width=0.25cm,minimum height=2mm]

\tikzstyle{bn}=[circle,fill=black,draw=black,scale=.4]
\tikzstyle{wn}=[circle,fill=white,draw=black,scale=.6]
\tikzstyle{dn}=[circle,fill=none,draw=gray]
\tikzstyle{bspider}=[fill=black,draw=black,scale=1,shape=isosceles triangle,shape border rotate=-90,isosceles triangle stretches=true,inner sep=1pt,minimum width=0.4cm,minimum height=3mm]
\tikzstyle{dbspider}=[fill=black,draw=black,scale=1,shape=isosceles triangle,shape border rotate=90,isosceles triangle stretches=true,inner sep=1pt,minimum width=0.4cm,minimum height=3mm]
\tikzstyle{L}=[rectangle,shape=rectangle,fill=green,draw=black]

\tikzstyle{Z dot}=[inner sep=0mm, minimum size=2mm, shape=circle, draw=black, fill={rgb,255: red,221; green,255; blue,221}]
\tikzstyle{Z phase dot}=[minimum size=5mm, font={\footnotesize\boldmath}, shape=rectangle, rounded corners=2mm, inner sep=0.2mm, outer sep=-2mm, scale=0.8, draw=black, fill={rgb,255: red,221; green,255; blue,221}]
\tikzstyle{X dot}=[Z dot, shape=circle, draw=black, fill={rgb,255: red,255; green,136; blue,136}]
\tikzstyle{X phase dot}=[Z phase dot, fill={rgb,255: red,255; green,136; blue,136}, font={\footnotesize\boldmath}]
\tikzstyle{hadamard edge}=[-, dashed, dash pattern=on 2pt off 0.5pt, thick, draw={rgb,255: red,68; green,136; blue,255}]

\tikzstyle{black dot}=[inner sep=0.7mm,minimum width=0pt,minimum height=0pt,fill=black,draw=black,shape=circle]

\tikzstyle{dot}=[black dot]
\tikzstyle{smalldot}=[inner sep=0.4mm,minimum width=0pt,minimum height=0pt,fill=black,draw=black,shape=circle]
\tikzstyle{white dot}=[dot,fill=white]
\tikzstyle{antipode}=[white dot,inner sep=0.3mm,font=\footnotesize]
\tikzstyle{smallwhitedot}=[smalldot,fill=white]
\tikzstyle{alt white dot}=[white dot,label={[xshift=3.07mm,yshift=-0.05mm,font=\footnotesize]left:$*$}]
\tikzstyle{gray dot}=[dot,fill=gray!40!white]
\tikzstyle{smallgraydot}=[smalldot,fill=gray!40!white]
\tikzstyle{box vertex}=[draw=black,rectangle]
\tikzstyle{small box}=[box vertex,fill=white]
\tikzstyle{whitebg}=[fill=white,inner sep=2pt]
\tikzstyle{graph state vertex}=[sg vertex,fill=black]

\tikzstyle{wide copoint}=[fill=white,draw=black,shape=isosceles triangle,shape border rotate=90,isosceles triangle stretches=true,inner sep=1pt,minimum width=1.5cm,minimum height=5mm]
\tikzstyle{wide point}=[fill=white,draw=black,shape=isosceles triangle,shape border rotate=-90,isosceles triangle stretches=true,inner sep=1pt,minimum width=1.5cm,minimum height=4mm]
\tikzstyle{very wide copoint}=[fill=white,draw=black,shape=isosceles triangle,shape border rotate=-90,isosceles triangle stretches=true,inner sep=1pt,minimum width=2.5cm,minimum height=4mm]
\tikzstyle{very wide empty copoint}=[draw=black,shape=isosceles triangle,shape border rotate=-90,isosceles triangle stretches=true,inner sep=1pt,minimum width=2.5cm,minimum height=4mm]
\tikzstyle{symm}=[ultra thick,shorten <=-1mm,shorten >=-1mm]

\tikzstyle{square box}=[rectangle,fill=white,draw=black,minimum height=5mm,minimum width=5mm,font=\small]
\tikzstyle{square gray box}=[rectangle,fill=gray!30,draw=black,minimum height=6mm,minimum width=6mm]
\tikzstyle{copoint}=[regular polygon,regular polygon sides=3,draw=black,scale=0.75,inner sep=-0.5pt,minimum width=7mm,fill=white]
\tikzstyle{point}=[regular polygon,regular polygon sides=3,draw=black,scale=0.75,inner sep=-0.5pt,minimum width=7mm,fill=white,regular polygon rotate=180]
\tikzstyle{gray point}=[point,fill=gray!40!white]
\tikzstyle{gray copoint}=[copoint,fill=gray!40!white]

\newcommand{\edgearrow}{{\arrow[black]{>}}}
\newcommand{\edgetick}{{\arrow[black,scale=0.7,very thick]{|}}}

\tikzstyle{diredge}=[->]
\tikzstyle{rdiredge}=[<-]
\tikzstyle{medium diredge}=[->]

\tikzstyle{short diredge}=[->]
\tikzstyle{halfedge}=[-)]
\tikzstyle{other halfedge}=[(-]
\tikzstyle{freeedge}=[(-)]
\tikzstyle{white edge}=[line width=5pt,white]
\tikzstyle{tick}=[postaction=decorate,decoration={markings, mark=at position 0.5 with \edgetick}]
\tikzstyle{small map edge}=[|-latex, gray!60!blue, shorten <=0.9mm, shorten >=0.5mm]
\tikzstyle{thick dashed edge}=[very thick,dashed,gray!40]
\tikzstyle{map edge}=[|-latex,very thick, gray!40, shorten <=1mm, shorten >=0.5mm]
\tikzstyle{tickedge}=[postaction=decorate,
  decoration={markings, mark=at position 0.5 with \edgetick}]
\tikzstyle{dirtickedge}=[postaction=decorate,
  decoration={markings, mark=at position 0.5 with \edgetick},
  decoration={markings, mark=at position 0.85 with \edgearrow}]
\tikzstyle{dirdoubletickedge}=[postaction=decorate,
  decoration={markings, mark=at position 0.4 with \edgetick},
  decoration={markings, mark=at position 0.6 with \edgetick},
  decoration={markings, mark=at position 0.85 with \edgearrow}]

\makeatletter
\newcommand{\boxshape}[3]{%
\pgfdeclareshape{#1}{
\inheritsavedanchors[from=rectangle] 
\inheritanchorborder[from=rectangle]
\inheritanchor[from=rectangle]{center}
\inheritanchor[from=rectangle]{north}
\inheritanchor[from=rectangle]{south}
\inheritanchor[from=rectangle]{west}
\inheritanchor[from=rectangle]{east}
\backgroundpath{
\southwest \pgf@xa=\pgf@x \pgf@ya=\pgf@y
\northeast \pgf@xb=\pgf@x \pgf@yb=\pgf@y

\@tempdima=#2
\@tempdimb=#3

\pgfpathmoveto{\pgfpoint{\pgf@xa - 5pt + \@tempdima}{\pgf@ya}}
\pgfpathlineto{\pgfpoint{\pgf@xa - 5pt - \@tempdima}{\pgf@yb}}
\pgfpathlineto{\pgfpoint{\pgf@xb + 5pt + \@tempdimb}{\pgf@yb}}
\pgfpathlineto{\pgfpoint{\pgf@xb + 5pt - \@tempdimb}{\pgf@ya}}
\pgfpathlineto{\pgfpoint{\pgf@xa - 5pt + \@tempdima}{\pgf@ya}}
\pgfpathclose
}
}}

\boxshape{NEbox}{0pt}{8pt}
\boxshape{SEbox}{0pt}{-8pt}
\boxshape{NWbox}{8pt}{0pt}
\boxshape{SWbox}{-8pt}{0pt}
\makeatother

\tikzstyle{map}=[draw,shape=NEbox,inner sep=7pt]
\tikzstyle{mapdag}=[draw,shape=SEbox,inner sep=7pt]
\tikzstyle{maptrans}=[draw,shape=SWbox,inner sep=7pt]
\tikzstyle{mapconj}=[draw,shape=NWbox,inner sep=7pt]

\tikzstyle{probs}=[shape=semicircle,fill=gray!40!white,draw=black,shape border rotate=180,minimum width=1.2cm]

\tikzstyle{arrs}=[-latex,font=\small,auto]
\tikzstyle{arrow plain}=[arrs]
\tikzstyle{arrow dashed}=[dashed,arrs]
\tikzstyle{arrow bold}=[very thick,arrs]
\tikzstyle{arrow hide}=[draw=white!0,-]
\tikzstyle{arrow reverse}=[latex-]
\tikzstyle{cdnode}=[]

\tikzstyle{rno}=[dot,fill=red,inner sep=0pt,minimum width=0.25cm]

\definecolor{zx_red}{RGB}{232, 165, 165}
\definecolor{zx_green}{RGB}{216, 248, 216}

\tikzstyle{gn}=[dot,inner sep=0pt,minimum width=2mm,fill=zx_green]
\tikzstyle{rn}=[gn,fill=zx_red]

\tikzstyle{gn_phase}=[minimum size=1.2em, font={\footnotesize\boldmath}, shape=rectangle, rounded corners=0.5em, inner sep=0.2em, outer sep=-0.2em, scale=0.8, draw=black, fill=zx_green]
\tikzstyle{rn_phase}=[gn_phase, fill=zx_red]

\tikzstyle{rc}=[dot,thick,fill=white,draw = red,minimum width=0.3cm,inner sep=0pt]
\tikzstyle{gc}=[dot,thick,fill=white,draw= green,inner sep=0pt,minimum width=0.3cm]
\tikzstyle{bc}=[dot,thick,fill=white,draw= blue,minimum width=0.3cm]

\tikzstyle{label}=[circle,fill=white,minimum width=0.3cm]

\tikzstyle{clocklabel}=[dot,fill=yellow,draw=black,font=\tiny,inner sep=0.75pt]

\tikzstyle{rsn}=[dot,fill=red,inner sep=0pt,minimum width=0.25cm] 
\tikzstyle{gsn}=[circle split,draw,fill=green,font=\tiny,inner sep=0.75pt]
\tikzstyle{bsn}=[circle split,draw,fill=blue,font=\tiny,inner sep=0.75pt]

\tikzstyle{rsc}=[circle split,thick,draw= red,draw,fill=white,font=\tiny,inner sep=0.75pt]
\tikzstyle{gsc}=[circle split,thick,draw= green,draw,fill=white,font=\tiny,inner sep=0.75pt]
\tikzstyle{bsc}=[circle split,thick,draw= blue,draw,fill=white,font=\tiny,inner sep=0.75pt]

\tikzstyle{cnot}=[fill=white,shape=circle,inner sep=-1.4pt]
\tikzstyle{wire label}=[font=\tiny, auto]

\tikzstyle{cdiag}=[matrix of math nodes, row sep=3em, column sep=3em, text height=1.5ex, text depth=0.25ex,inner sep=0.5em]
\tikzstyle{arrow above}=[transform canvas={yshift=0.5ex}]
\tikzstyle{arrow below}=[transform canvas={yshift=-0.5ex}]

\newtheorem{Th}{Theorem}
\newtheorem{theorem}[Th]{Theorem}
\newtheorem{proposition}[Th]{Proposition} 
\newtheorem{lemma}[Th]{Lemma}
\newtheorem{corollary}[Th]{Corollary}
\newtheorem{definition}[Th]{Definition} 
\newtheorem{example}[Th]{Example}
\newtheorem{remark}[Th]{Remark}

\makeatletter
\newcommand{\vast}{\bBigg@{6.5}}
\makeatother

\allowdisplaybreaks

\usepackage{parskip}
\usepackage{tabu}

\title{Differentiating and Integrating ZX Diagrams with Applications to Quantum Machine Learning}

\author{Quanlong Wang$^*$}
\email{harny.wang@quantinuum.com}
\affiliation{Quantinuum, 17 Beaumont Street, Oxford, OX1 2NA, United Kingdom}

\author{Richie Yeung$^*$}
\orcid{0000-0003-1953-8305}
\email{richie.yeung@quantinuum.com}
\affiliation{University of Oxford, United Kingdom}
\affiliation{Quantinuum, 17 Beaumont Street, Oxford, OX1 2NA, United Kingdom}

\author{Mark Koch$^*$}
\email{mark.koch@quantinuum.com}
\affiliation{Quantinuum, Terrington House, 13-15 Hills Road, CB2 1NL Cambridge, United Kingdom}

\begin{document}
\date{}\maketitle
\def\thefootnote{*}\footnotetext{Equal contribution}
\begin{abstract}
ZX-calculus has proved to be a useful tool for quantum technology with a wide range of successful applications. Most of these applications are of an algebraic nature. However, other tasks that involve differentiation and integration remain unreachable with current ZX techniques. Here we elevate ZX to an analytical perspective by realising differentiation and integration entirely within the framework of ZX-calculus. We explicitly illustrate the new analytic framework of ZX-calculus by applying it in context of quantum machine learning for the analysis of barren plateaus. 
\end{abstract}

 
\section{Introduction}

ZX-calculus is a powerful graphical rewrite system proposed by Coecke and Duncan \cite{CoeckeDuncan} for linear maps, particularly for quantum circuits. A node with $n$ edges in a ZX diagram, like in tensor network notation, represents an order $n$ tensor. Moreover, it is possible to directly evaluate the tensor by performing local rewrites (i.e., substitution of a part of a ZX diagram). Using these local rewrites, ZX-calculus has been successfully applied to circuit compilation \cite{debeaudrapbianwang, nielbianwang, duncan2020graph,sivarajah2020t}, measurement-based quantum computing \cite{kissinger2017MBQC, duncan2010rewriting}, fusion-based quantum computing \cite{bombin2021logical},  quantum error correction \cite{kissinger2022phase, horsman2017surgery}, quantum natural language processing \cite{qnlp2020, lorenz2021qnlp, kartsaklis2021lambeq}, and quantum foundations \cite{miriamali2015, bobbilledw, CDKW, Gogiosozeng19}. ZX-calculus can even be used as a concrete realisation of quantum theory \cite{COECKE20221}. These applications of ZX-calculus are algebraic in nature, and take advantage of \textit{rewriting as a form of computation}: in fact ZX-calculus is a sound, universal \cite{coeckeduncan2008} and complete \cite{amarngwanglics} proof system that serves as an alternative to traditional linear algebra, which also makes it a different formalism from tensor networks.  However, without the analytical tools of differentiation and integration, ZX-calculus fell short of tackling variational problems such as quantum machine learning or realising a comprehensive version of quantum mechanics including quantum dynamics.

In this paper we give for the first time rules for differentiating arbitrary ZX diagrams and integrating a wide class of ZX diagrams (including quantum circuits), within the framework of a slightly extended version of ZX-calculus called algebraic ZX-calculus \cite{wangalg2020} which makes it very convenient to deal with sums of ZX diagrams, thus paving the way for an analytical version of ZX-calculus. 
We apply these new techniques to develop a framework for a purely ZX-based analysis of the barren plateau phenomenon from quantum machine learning.

\subsection*{Related Work}

There have been previous attempts at providing rules for differentiating and integrating ZX diagrams \cite{yeung2020diagrammatic, toumi2021diagrammatic, Zhao2021analyzingbarren}. 
In particular, Zhao and Gao \cite{Zhao2021analyzingbarren} pioneered the use of ZX-calculus to aid in the analysis of the barren plateau phenomenon.
Similar techniques were also used in \cite{martin2023barren} to study quantum tensor network ans\"atze.
However, by explicitly using Hilbert space operations such as addition, all these previous attempts fall outside the realm of vanilla ZX-calculus as there are few techniques to further manipulate sums of diagrams.
Ultimately, all previous works studying barren plateaus in the ZX-calculus had to resort to a combination of ZX diagrams and general tensor networks in order to handle this summation problem.
The analytical ZX techniques developed in this paper on the other hand offer a unified framework to reason about differentiation and integration purely in terms of rewriting, without having to fall back to arbitrary tensor networks.

Previous attempts to formalise sums of ZX diagrams include work by Stollenwerk and Hadfield~\cite{stollenwerk2022diagrammatic} who provide notation for representing sums of ZX diagrams as a single diagram by extending the ZX calculus with a pair of sum boxes which were later formalised by Villoria, Basold, and Laarman~\cite{villoria2024enriching}.
However, this approach does not offer much additional diagrammatic reasoning power since it is merely syntactic sugar for writing linear combinations of diagrams.
Jeandel, Perdrix, and Veshchezerova \cite{jeandel2022addition} independently derived an alternate method to represent sums and derivatives with the ZX-calculus by showing how the sum of two controlled ZX diagrams can be represented as a single controlled ZX diagram.
However, their approach requires an inductive translation of diagrams to controlled diagrams such that the result will not resemble the original diagram.
Our method on the other hand preserves diagram structure and can be calculated almost instantly.
We refer to Appendix \ref{zx-compare} for a more thorough comparison of the two approaches.

Finally, our approach also offers a numerical advantage: The tensor networks considered in \cite{Zhao2021analyzingbarren, martin2023barren} have bond dimension 3 whereas all our diagrams only have dimension 2, yielding a speed-up when contracting the diagram in software.
This was exploited in \cite{koch2023speedy} to numerically detect the presence of barren plateaus by contracting diagrams representing the gradient variance of ans\"atze and observing the decay.
Crucially, that work builds on our diagrammatic barren plateau framework developed in this paper, making use of the more efficient 2-dimensional representation compared to \cite{Zhao2021analyzingbarren, martin2023barren}.
Furthermore, the techniques developed in our work have QML applications beyond barren plateaus.
For example, our results have been used in \cite{koch2022quantum} to analyse and derive novel parameter shift rules for gradient computation using ZX.

\begin{figure}
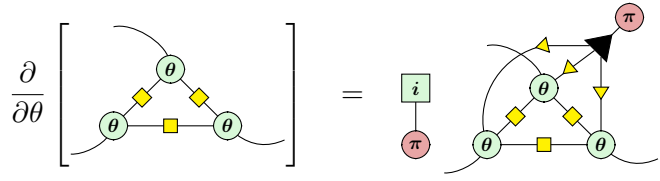

	$$\frac{\partial}{\partial\theta}\left[%
	\beginpgfgraphicnamed{graphstate}
	\InputIfFileExists{graphstate.tikz}{}{\input{./figures/graphstate.tikz}}%
	\endpgfgraphicnamed
\right] \quad=\quad %
	\beginpgfgraphicnamed{derivativegraphstate}
	\InputIfFileExists{derivativegraphstate.tikz}{}{\input{./figures/derivativegraphstate.tikz}}%
	\endpgfgraphicnamed
$$
	\caption{Example of diagrammatic differentiation.}
\end{figure}

\subsection*{Summary of results}
\begin{enumerate}
\item Differentiation of arbitrary (algebraic) ZX diagrams, with a unified diagrammatic chain and product rule. (Theorem \ref{diffthmgeneralgbox} and Theorem \ref{diffthm1thm}) 
\item Definite integration of circuit-like ZX diagrams, with up to 3 occurrences of a parameter. (Proposition \ref{1foldintglm}, Theorem \ref{2varibleintthm}, and Theorem \ref{3varibleintthm})
\item Diagrammatic formula for the expectation and variance of a quantum circuit's gradient $\frac{ \partial \braket{H}}{\partial\theta_i}$.  (Lemma \ref{1foldexpectintglm} and Theorem \ref{variancegenthm})
\item Demonstration of barren plateau analysis for an example ansatz. (Section \ref{sec:barren-example})
\end{enumerate}

From a general ZX-calculus perspective, this is the first paper to combine sums of ZX diagrams into a single ZX diagram in a methodical way. In particular, we highlight the importance of the W spider in ZX-calculus, which corresponds to the derivation structure of the product rule. These results required the combined power of the Z, X and W spiders, all 3 of which can be naturally represented within algebraic ZX-calculus.

\enlargethispage{\baselineskip}

\textbf{Note:} For presentation purposes, the proofs of some theorems and lemmas are moved to the appendix.

\section{ Algebraic ZX-calculus}\label{azxintro}

The generators of the original ZX-calculus \cite{CoeckeDuncan} are chosen with the aim to conveniently represent quantum computational models using complementary observables. On the other hand, the ZW-calculus \cite{amar1} is designed based on the GHZ and W states, two maximally entangled quantum states \cite{Coeckealeksmen}. It is known that the Z and W spiders from ZW-calculus act as the multiplication and addition monoid respectively, making it possible to perform arithmetic \cite{ghzwarith, amar1}.

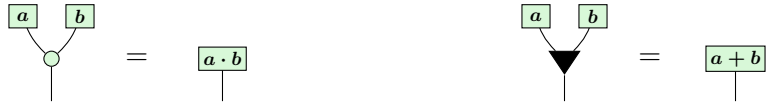
\ctikzfig{z_vs_w}

We will see that the W state is crucial for dealing with sums of diagrams, and it is in fact closely related to the product rule used in differentiation.
Conveniently, algebraic ZX-calculus \cite{wangalg2020} compactly decomposes the W spider 
and other gadgets such as the logical AND gate \cite{bobanthonywang}, into Z spiders, X spiders and triangle gates, thus giving us the benefits of ZX and ZW calculus within a single unified framework. 
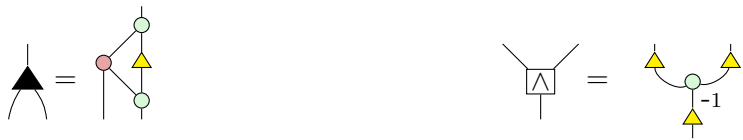
\ctikzfig{andgtplusw}

The yellow triangle of algebraic ZX-calculus is powerful as it sends the computational basis to a non-orthogonal basis, which makes diagrammatic representation and calculation of other logical gates much simpler. Conversely, the representation of the yellow triangle using other graphical calculi is more complicated. Intuitively, this is because the triangle gate is a low-level primitive in comparison to the Z, X, W and H spiders \cite{miriamaleks}.
This algebraic extension of ZX is a universal and complete language for not just complex numbers, but also commutative rings and semirings \cite{qwangrsmring}.

In this section, we give an introduction to the algebraic ZX-calculus, including its generators and rewriting rules.
In this paper ZX diagrams are either read from left to right or top to bottom.
\subsection{Generators}
The diagrams in algebraic ZX-calculus are defined by freely combining the following generating objects:
    \begin{table}[!h]
\begin{center} 
\begin{tabular}{|r@{~}r@{~}c@{~}c|r@{~}r@{~}c@{~}c|}
\hline
$R_{Z,a}^{(n,m)}$&$:$&$n\to m$ & %
	\beginpgfgraphicnamed{generalgreenspider}
	\InputIfFileExists{generalgreenspider.tikz}{}{\input{./figures/generalgreenspider.tikz}}%
	\endpgfgraphicnamed
  &  $\mathbb I$&$:$&$1\to 1$&%
	\beginpgfgraphicnamed{Id}
	\InputIfFileExists{Id.tikz}{}{\input{./figures/Id.tikz}}%
	\endpgfgraphicnamed
 \\\hline
$H$&$:$&$1\to 1$ &%
	\beginpgfgraphicnamed{newhadamard}
	\InputIfFileExists{newhadamard.tikz}{}{\input{./figures/newhadamard.tikz}}%
	\endpgfgraphicnamed
&  $\sigma$&$:$&$ 2\to 2$& %
	\beginpgfgraphicnamed{swap}
	\InputIfFileExists{swap.tikz}{}{\input{./figures/swap.tikz}}%
	\endpgfgraphicnamed
\\\hline
$C_a$&$:$&$ 0\to 2$& %
	\beginpgfgraphicnamed{cap1}
	\InputIfFileExists{cap1.tikz}{}{\input{./figures/cap1.tikz}}%
	\endpgfgraphicnamed
 &$ C_u$&$:$&$ 2\to 0$&%
	\beginpgfgraphicnamed{cup1}
	\InputIfFileExists{cup1.tikz}{}{\input{./figures/cup1.tikz}}%
	\endpgfgraphicnamed
 \\\hline
$T$&$:$&$1\to 1$&%
	\beginpgfgraphicnamed{triangle}
	\InputIfFileExists{triangle.tikz}{}{\input{./figures/triangle.tikz}}%
	\endpgfgraphicnamed
  & $T^{-1}$&$:$&$1\to 1$&%
	\beginpgfgraphicnamed{triangleinv}
	\InputIfFileExists{triangleinv.tikz}{}{\input{./figures/triangleinv.tikz}}%
	\endpgfgraphicnamed
 \\\hline
\end{tabular}\caption{Generators of algebraic ZX-calculus, where $m,n\in \mathbb N$, $ a  \in \mathbb C$.} \label{qbzxgenerator} 
\end{center}
\end{table}
\FloatBarrier

\subsection{Additional notation}
For simplicity, we introduce additional notation based on the given generators:

\begin{enumerate}
\item The green spider from the original ZX-calculus can be defined using the green box spider in algebraic ZX-calculus.
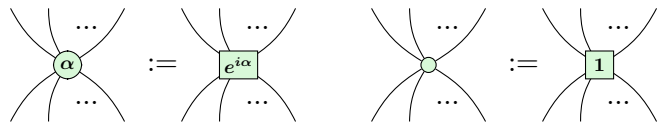
\ctikzfig{convenientdenote1}

\item The whitespace around a diagram can be interpreted as an explicit horizontal composition with the empty diagram.

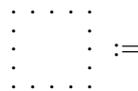
\ctikzfig{convenientdenote2}

\item The transposes of the triangle and the inverse triangle can be drawn as an inverted triangle.

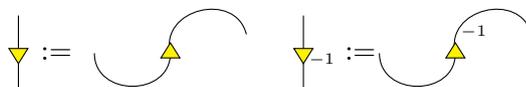
\ctikzfig{convenientdenote3}

\item The pink spider is the algebraic equivalent of the red spider from the original ZX-calculus. It is only defined for $\tau \in \{ 0, \pi \}$, and is rescaled to have integer components in its matrix representation.

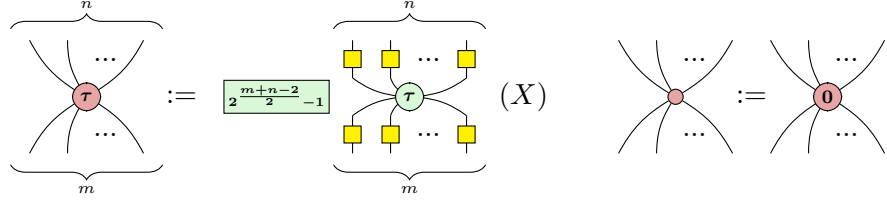
\ctikzfig{convenientdenote5}

Note that the green box represents the scalar $~%
	\beginpgfgraphicnamed{convenientdenote5-scalar}
	\begin{tikzpicture}
	\begin{pgfonlayer}{nodelayer}
		\node [style=gbox] (0) at (0, 0) {${\scriptstyle 2^{\frac{m+n-2}{2}}-1}$};
	\end{pgfonlayer}
\end{tikzpicture}
}%
	\endpgfgraphicnamed
 \,=\, 2^{\frac{m+n-2}{2}}$.

\item The W spider from ZW-calculus can be expressed as follows.

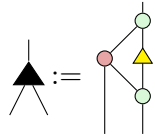
\ctikzfig{convenientdenote6}
\end{enumerate}

\subsection{Interpretation}
Although the generators in ZX-calculus are formal mathematical objects in their own right, in the context of this paper we interpret the generators as linear maps, so each ZX diagram is equivalent to a vector or matrix.
For $a  \in \mathbb C$, we have
\begin{equation} \tag{$Z$} \label{gninterp}
	\beginpgfgraphicnamed{generalgreenspider}
	\InputIfFileExists{generalgreenspider.tikz}{}{\input{./figures/generalgreenspider.tikz}}%
	\endpgfgraphicnamed
 =\ket{0}^{\otimes m}\bra{0}^{\otimes n}+a\ket{1}^{\otimes m}\bra{1}^{\otimes n},
\end{equation}

\[
	\beginpgfgraphicnamed{newhadamard}
	\begin{tikzpicture}
	\begin{pgfonlayer}{nodelayer}
		\node [style=newh] (0) at (0, 0) {};
		\node [style=none] (1) at (0, 0.5) {};
		\node [style=none] (2) at (0, -0.5) {};
	\end{pgfonlayer}
	\begin{pgfonlayer}{edgelayer}
		\draw (1.center) to (2.center);
	\end{pgfonlayer}
\end{tikzpicture}}%
	\endpgfgraphicnamed
=\frac{1}{\sqrt{2}}\begin{pmatrix}
        1 & 1 \\
        1 & -1
 \end{pmatrix}, \qquad %
	\beginpgfgraphicnamed{triangle}
	\begin{tikzpicture}
	\begin{pgfonlayer}{nodelayer}
		\node [style=none] (0) at (0, 0.5) {};
		\node [style=triangle] (1) at (0, 0) {};
		\node [style=none] (2) at (0, -0.5) {};
	\end{pgfonlayer}
	\begin{pgfonlayer}{edgelayer}
		\draw (0.center) to (2.center);
	\end{pgfonlayer}
\end{tikzpicture}}%
	\endpgfgraphicnamed
=\begin{pmatrix}
        1 & 1 \\
        0 & 1
 \end{pmatrix}, \qquad
	\beginpgfgraphicnamed{triangleinv}
	\begin{tikzpicture}
	\begin{pgfonlayer}{nodelayer}
		\node [style=none] (0) at (0.25, 0.25) {-{\scriptsize1}};
		\node [style=triangle] (1) at (0, 0) {};
		\node [style=none] (2) at (0, -0.5) {};
		\node [style=none] (3) at (0, 0.5) {};
	\end{pgfonlayer}
	\begin{pgfonlayer}{edgelayer}
		\draw (3.center) to (2.center);
	\end{pgfonlayer}
\end{tikzpicture}}%
	\endpgfgraphicnamed
=\begin{pmatrix}
        1 & -1 \\
        0 & 1
 \end{pmatrix}
\]

\[
	\beginpgfgraphicnamed{Id}
	\begin{tikzpicture}
	\begin{pgfonlayer}{nodelayer}
		\node [style=none] (1) at (0.5, 0.3) {};
		\node [style=none] (2) at (0.5, -0.3) {};
		\node [style=none] (3) at (0.5, -0.5) {};
		\node [style=none] (4) at (0.5, 0.5) {};
	\end{pgfonlayer}
	\begin{pgfonlayer}{edgelayer}
		\draw (1.center) to (2.center);
	\end{pgfonlayer}
\end{tikzpicture}}%
	\endpgfgraphicnamed
=\begin{pmatrix}
	1 & 0 \\
	0 & 1
\end{pmatrix},
\quad
	\beginpgfgraphicnamed{cap1}
	\begin{tikzpicture}
	\begin{pgfonlayer}{nodelayer}
		\node [style=none] (0) at (0, -0) {};
		\node [style=none] (1) at (1, -0) {};
	\end{pgfonlayer}
	\begin{pgfonlayer}{edgelayer}
		\draw [bend left=90, looseness=1.50] (0.center) to (1.center);
	\end{pgfonlayer}
\end{tikzpicture}}%
	\endpgfgraphicnamed
=\begin{pmatrix}
	1  \\
	0  \\
	0  \\
	1  \\
\end{pmatrix}, \quad
	\beginpgfgraphicnamed{cup1}
	\begin{tikzpicture}
	\begin{pgfonlayer}{nodelayer}
		\node [style=none] (0) at (0, 0.5) {};
		\node [style=none] (1) at (1, 0.5) {};
	\end{pgfonlayer}
	\begin{pgfonlayer}{edgelayer}
		\draw [bend right=90, looseness=1.50] (0.center) to (1.center);
	\end{pgfonlayer}
\end{tikzpicture}}%
	\endpgfgraphicnamed
=\begin{pmatrix}
	1 & 0 & 0 & 1 
\end{pmatrix}, 
\quad
	\beginpgfgraphicnamed{swap}
	\InputIfFileExists{swap.tikz}{}{\input{./figures/swap.tikz}}%
	\endpgfgraphicnamed
=\begin{pmatrix}
	1 & 0 & 0 & 0 \\
	0 & 0 & 1 & 0 \\
	0 & 1 & 0 & 0 \\
	0 & 0 & 0 & 1 
\end{pmatrix}.
\]

Furthermore, parallel and sequential composition of diagrams correspond to matrix multiplication and tensor product of the underlying linear maps, respectively.
Using the interpretation of these generators, we can compute the matrices for other frequently occurring diagrams by composition:

\[
	\beginpgfgraphicnamed{redspider01pi}
	\InputIfFileExists{redspider01pi.tikz}{}{\input{./figures/redspider01pi.tikz}}%
	\endpgfgraphicnamed
 =\sum_{\substack{0\leq i_1, \cdots, i_m,  j_1, \cdots, j_n\leq 1\\ i_1+\cdots+ i_m+k \equiv  j_1+\cdots +j_n(mod~ 2)}}\ket{i_1, \cdots, i_m}\bra{j_1, \cdots, j_n},k\in \{ 0, 1 \},
\]

\[
	\beginpgfgraphicnamed{ket0v}
	\begin{tikzpicture}
	\begin{pgfonlayer}{nodelayer}
		\node [style=none] (0) at (0, -0.25) {};
		\node [style=rn] (1) at (0, 0.25) {};
	\end{pgfonlayer}
	\begin{pgfonlayer}{edgelayer}
		\draw (1) to (0.center);
	\end{pgfonlayer}
\end{tikzpicture}}%
	\endpgfgraphicnamed
 = \ket{0} 
\qquad 
	\beginpgfgraphicnamed{bra0v}
	\begin{tikzpicture}
	\begin{pgfonlayer}{nodelayer}
		\node [style=none] (0) at (0, 0.25) {};
		\node [style=rn] (1) at (0, -0.25) {};
	\end{pgfonlayer}
	\begin{pgfonlayer}{edgelayer}
		\draw (1) to (0.center);
	\end{pgfonlayer}
\end{tikzpicture}}%
	\endpgfgraphicnamed
 = \bra{0},
\qquad
	\beginpgfgraphicnamed{ket1v}
	\begin{tikzpicture}
	\begin{pgfonlayer}{nodelayer}
		\node [style=none] (0) at (0, -0.25) {};
		\node [style={rn_phase}] (1) at (0, 0.25) {$\pi$};
	\end{pgfonlayer}
	\begin{pgfonlayer}{edgelayer}
		\draw (1) to (0.center);
	\end{pgfonlayer}
\end{tikzpicture}
}%
	\endpgfgraphicnamed
 = \ket{1},
\qquad 
	\beginpgfgraphicnamed{bra1v}
	\begin{tikzpicture}
	\begin{pgfonlayer}{nodelayer}
		\node [style=none] (0) at (0, 0.25) {};
		\node [style={rn_phase}] (1) at (0, -0.25) {$\pi$};
	\end{pgfonlayer}
	\begin{pgfonlayer}{edgelayer}
		\draw (1) to (0.center);
	\end{pgfonlayer}
\end{tikzpicture}
}%
	\endpgfgraphicnamed
 = \bra{1},
\qquad
	\beginpgfgraphicnamed{scalarpi}
	\begin{tikzpicture}
	\begin{pgfonlayer}{nodelayer}
		\node [style={rn_phase}] (0) at (0, 0) {$\pi$};
	\end{pgfonlayer}
\end{tikzpicture}
}%
	\endpgfgraphicnamed
=0,
\qquad
	\beginpgfgraphicnamed{scalartimes2}
	\begin{tikzpicture}
	\begin{pgfonlayer}{nodelayer}
		\node [style=none] (0) at (0, 0) {$=$};
		\node [style={rn_phase}] (1) at (-0.5, 0.25) {$\pi$};
		\node [style=gbox] (2) at (0.75, 0) {${ a-1}$};
		\node [style=gbox] (3) at (-0.5, -0.25) {$a$};
	\end{pgfonlayer}
	\begin{pgfonlayer}{edgelayer}
		\draw (3) to (1);
	\end{pgfonlayer}
\end{tikzpicture}
}%
	\endpgfgraphicnamed
=a.
\]

\[
	\beginpgfgraphicnamed{singleredpi}
	\begin{tikzpicture}
	\begin{pgfonlayer}{nodelayer}
		\node [style=none] (0) at (0, 0.5) {};
		\node [style={rn_phase}] (1) at (0, 0) {$\pi$};
		\node [style=none] (2) at (0, -0.5) {};
	\end{pgfonlayer}
	\begin{pgfonlayer}{edgelayer}
		\draw (0.center) to (2.center);
	\end{pgfonlayer}
\end{tikzpicture}
}%
	\endpgfgraphicnamed
=\begin{pmatrix}
	0 & 1 \\
	1 & 0
\end{pmatrix},
\qquad
	\beginpgfgraphicnamed{singlexalphanormal}
	\InputIfFileExists{singlexalphanormal.tikz}{}{\input{./figures/singlexalphanormal.tikz}}%
	\endpgfgraphicnamed
=e^{i\frac{\alpha}{2}}\begin{pmatrix}
	\cos\frac{\alpha}{2} & -i\sin\frac{\alpha}{2} \\
	-i\sin\frac{\alpha}{2} &  \cos\frac{\alpha}{2}
\end{pmatrix},
\qquad %
	\beginpgfgraphicnamed{emptysquare}
	\InputIfFileExists{emptysquare.tikz}{}{\input{./figures/emptysquare.tikz}}%
	\endpgfgraphicnamed
=1.
\]

\subsection{Rules}
Now we give the rewriting rules of algebraic ZX-calculus.
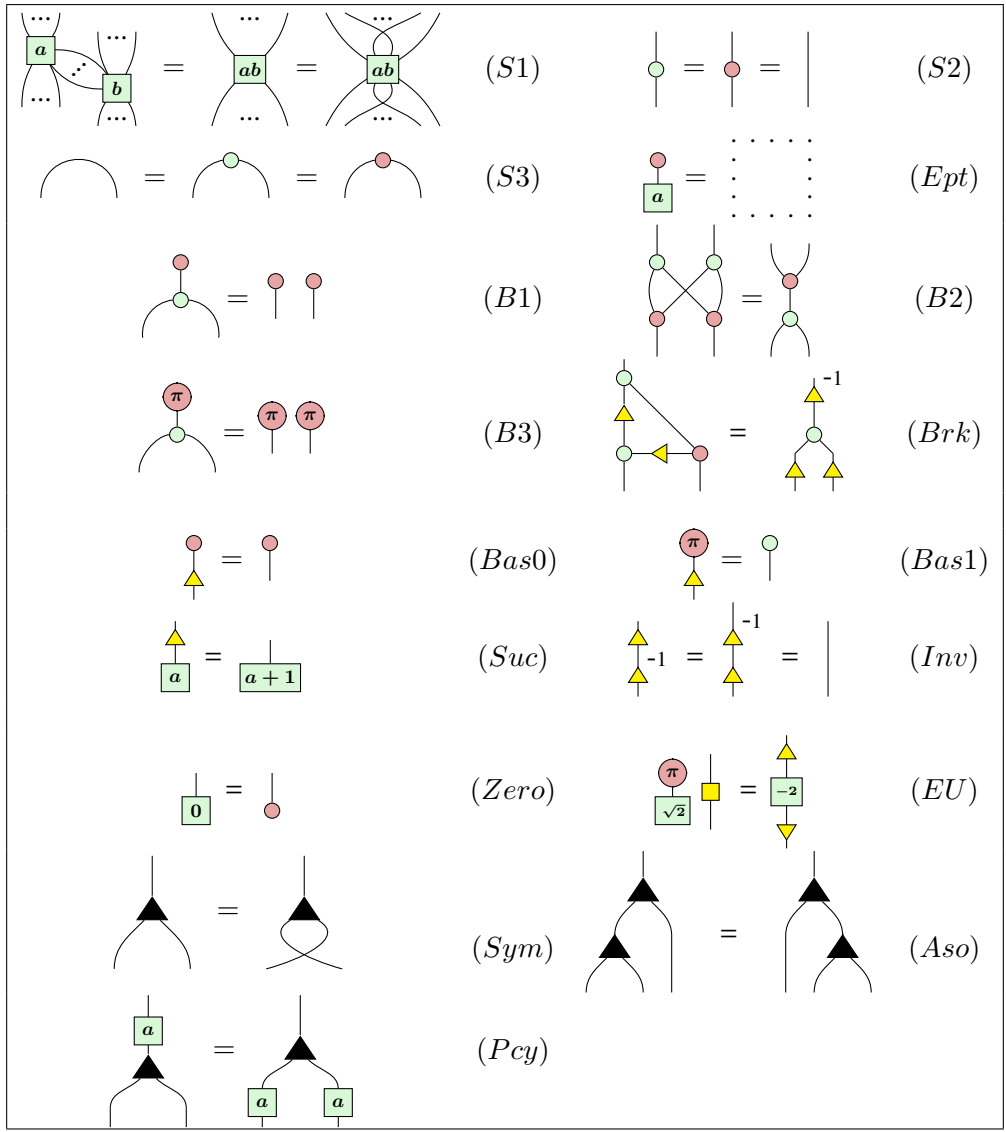
\begin{center}
$\displaystyle
\begin{array}{|cccc|}
\hline
	\beginpgfgraphicnamed{generalgreenspiderfusesym}
	\InputIfFileExists{generalgreenspiderfusesym.tikz}{}{\input{./figures/generalgreenspiderfusesym.tikz}}%
	\endpgfgraphicnamed
&(S1) &%
	\beginpgfgraphicnamed{s2new2}
	\InputIfFileExists{s2new2.tikz}{}{\input{./figures/s2new2.tikz}}%
	\endpgfgraphicnamed
 &(S2)\\
	\beginpgfgraphicnamed{induced_compact_structure}
	\InputIfFileExists{induced_compact_structure.tikz}{}{\input{./figures/induced_compact_structure.tikz}}%
	\endpgfgraphicnamed
&(S3) & %
	\beginpgfgraphicnamed{rdotaempty}
	\InputIfFileExists{rdotaempty.tikz}{}{\input{./figures/rdotaempty.tikz}}%
	\endpgfgraphicnamed
  &(Ept) \\
	\beginpgfgraphicnamed{b1ring}
	\InputIfFileExists{b1ring.tikz}{}{\input{./figures/b1ring.tikz}}%
	\endpgfgraphicnamed
&(B1)  & %
	\beginpgfgraphicnamed{b2ring}
	\InputIfFileExists{b2ring.tikz}{}{\input{./figures/b2ring.tikz}}%
	\endpgfgraphicnamed
&(B2)\\ 
	\beginpgfgraphicnamed{rpicopyns}
	\InputIfFileExists{rpicopyns.tikz}{}{\input{./figures/rpicopyns.tikz}}%
	\endpgfgraphicnamed
 &(B3)& %
	\beginpgfgraphicnamed{anddflipns}
	\InputIfFileExists{anddflipns.tikz}{}{\input{./figures/anddflipns.tikz}}%
	\endpgfgraphicnamed
&(Brk) \\
 & &&\\
	\beginpgfgraphicnamed{triangleocopy}
	\InputIfFileExists{triangleocopy.tikz}{}{\input{./figures/triangleocopy.tikz}}%
	\endpgfgraphicnamed
 &(Bas0) &%
	\beginpgfgraphicnamed{trianglepicopyns}
	\InputIfFileExists{trianglepicopyns.tikz}{}{\input{./figures/trianglepicopyns.tikz}}%
	\endpgfgraphicnamed
&(Bas1)\\
	\beginpgfgraphicnamed{plus1}
	\InputIfFileExists{plus1.tikz}{}{\input{./figures/plus1.tikz}}%
	\endpgfgraphicnamed
&(Suc)& %
	\beginpgfgraphicnamed{triangleinvers}
	\InputIfFileExists{triangleinvers.tikz}{}{\input{./figures/triangleinvers.tikz}}%
	\endpgfgraphicnamed
  & (Inv) \\
   & &&\\
	\beginpgfgraphicnamed{zerotoredns}
	\InputIfFileExists{zerotoredns.tikz}{}{\input{./figures/zerotoredns.tikz}}%
	\endpgfgraphicnamed
&(Zero)& %
	\beginpgfgraphicnamed{eunoscalar2}
	\InputIfFileExists{eunoscalar2.tikz}{}{\input{./figures/eunoscalar2.tikz}}%
	\endpgfgraphicnamed
&(EU) \\
	\beginpgfgraphicnamed{lemma4}
	\InputIfFileExists{lemma4.tikz}{}{\input{./figures/lemma4.tikz}}%
	\endpgfgraphicnamed
&(Sym) &  %
	\beginpgfgraphicnamed{associate}
	\InputIfFileExists{associate.tikz}{}{\input{./figures/associate.tikz}}%
	\endpgfgraphicnamed
 &(Aso)\\ 
	\beginpgfgraphicnamed{TR1314combine2}
	\InputIfFileExists{TR1314combine2.tikz}{}{\input{./figures/TR1314combine2.tikz}}%
	\endpgfgraphicnamed
&(Pcy) &&\\
  		  		\hline
  		\end{array}$
  	\end{center}
Where $a, b \in \mathbb C$. The vertically flipped versions of the rules are assumed to hold as well.

\subsection{Useful lemmas}

The following lemmas which will be used in the sequel can be derived from the rules.

{\tabulinesep=1.2mm
\begin{tabu}{|c|c|}
\hline
\begin{minipage}[t]{0.45\linewidth}
  \begin{lemma}\cite{qwangnormalformbit}\label{pinksf}\\
    For $\tau, \sigma \in \{0, \pi\}$, 
    pink spiders fuse.\\
$$ %
	\beginpgfgraphicnamed{redspider0pifusion}
	\InputIfFileExists{redspider0pifusion.tikz}{}{\input{./figures/redspider0pifusion.tikz}}%
	\endpgfgraphicnamed
 (S1r)  $$
    \end{lemma} 
\end{minipage}&
\begin{minipage}[t]{0.45\linewidth}
   \begin{lemma}\cite{qwangnormalformbit}\label{hadhad}\\
    Hadamard is involutive.
    \vspace{0.25cm}
    $$ %
	\beginpgfgraphicnamed{nhsquare2}
	\InputIfFileExists{nhsquare2.tikz}{}{\input{./figures/nhsquare2.tikz}}%
	\endpgfgraphicnamed
 \text{  (H2)}$$
    \end{lemma}
    \vspace{0.20cm}
\end{minipage}\\\hline
\begin{minipage}[t]{0.45\linewidth}
    \begin{lemma}\cite{qwangnormalformbit}\label{zx2elm}\\
      Pink $\pi$ transposes the triangle.\\
      \vspace{0.25cm}
      \ctikzfig{zx2e}
      \vspace{0.10cm}
    \end{lemma}
\end{minipage}&
\begin{minipage}[t]{0.45\linewidth}
  \begin{lemma}\cite{qwangnormalformbit}\label{traingleinverse}\\
    Green $\pi$ inverts the triangle.\\
\ctikzfig{definitionTriangleInverse2} 
 \end{lemma}
\end{minipage}\\\hline
\begin{minipage}[t]{0.45\linewidth}
   \begin{lemma}\cite{qwangnormalformbit}\label{trianglerpidotlm}\\
      triangle stabilises $\bra{1}$.\\
 \ctikzfig{trianglerpidot} 
    \end{lemma}
\end{minipage}&
\begin{minipage}[t]{0.45\linewidth}
  \begin{lemma}\cite{qwangnormalformbit}\label{hopfnslm}\\
      Hopf rule.\\
$$%
	\beginpgfgraphicnamed{hopfns}
	\InputIfFileExists{hopfns.tikz}{}{\input{./figures/hopfns.tikz}}%
	\endpgfgraphicnamed
 (Hopf)$$
\end{lemma}
\end{minipage}\\\hline
\begin{minipage}[t]{0.45\linewidth}
\begin{lemma}\cite{qwangnormalformbit}\label{pimultiplecplm}\\
  $\pi$ copy rule. For $m \geq 0$:\\
$$%
	\beginpgfgraphicnamed{pigrcopy}
	\InputIfFileExists{pigrcopy.tikz}{}{\input{./figures/pigrcopy.tikz}}%
	\endpgfgraphicnamed
 (Pic)$$
\end{lemma}
\end{minipage}&
\begin{minipage}[t]{0.45\linewidth}
\begin{lemma}\cite{qwangnormalformbit}\label{1iprf}\\
$\pi$ commutation rule.\\
\ctikzfig{picommutationdm} 
\vspace{0.30cm}
  \end{lemma}
\end{minipage}\\\hline
\end{tabu}}

\begin{remark}
  Due to the associative rule (Aso),  we can define the $W$ spider
  \[ %
	\beginpgfgraphicnamed{mlegsblackspider}
	\InputIfFileExists{mlegsblackspider.tikz}{}{\input{./figures/mlegsblackspider.tikz}}%
	\endpgfgraphicnamed
 \]
  and give its interpretation as follows \cite{hadzihasanovic2015diagrammatic}:
  \[ %
	\beginpgfgraphicnamed{mlegsblackspiderone}
	\InputIfFileExists{mlegsblackspiderone.tikz}{}{\input{./figures/mlegsblackspiderone.tikz}}%
	\endpgfgraphicnamed
=\underbrace{\ket{0\cdots0}}_{m}\bra{0}+\sum_{k=1}^{m}\overbrace{\ket{\underbrace{0\cdots 0}_{k-1} 1 0\cdots 0}}^{m}\bra{1}.
  \]
  As a consequence, we have
  \begin{equation}\label{mlegsblackspider0and1eq}
	\beginpgfgraphicnamed{mlegsblackspider0and1}
	\InputIfFileExists{mlegsblackspider0and1.tikz}{}{\input{./figures/mlegsblackspider0and1.tikz}}%
	\endpgfgraphicnamed

  \end{equation}
\end{remark}

For $n=2$, the state $\ket{01} + \ket{10}$ can be represented as the quantum state corresponding to the Pauli X gate according to the map-state duality:
\begin{restatable}{lemma}{piwtopicap}\label{piwtopicaplm}
\[  %
	\beginpgfgraphicnamed{piwtopicap}
	\InputIfFileExists{piwtopicap.tikz}{}{\input{./figures/piwtopicap.tikz}}%
	\endpgfgraphicnamed
\]
\end{restatable}

\section{Differentiating ZX diagrams}
In this section, we show how to differentiate any algebraic ZX diagram within algebraic ZX-calculus, and how to represent the derivative of original ZX diagrams \cite{CoeckeDuncan} and quantum circuits in algebraic ZX as a special case. 
We refer to \cite{toumi2021diagrammatic} for a formal definition of the categorical semantics of diagrammatic differentiation. 
We start by differentiating the simplest parameterised generator in original ZX-calculus: the one-legged green spider. 
\begin{restatable}{lemma}{dgngen}\label{dgngen}
Suppose $f(\theta)$ is a differentiable real function of $\theta$. Then
\[
  \frac{\partial}{\partial\theta}\left[%
	\beginpgfgraphicnamed{functiongn}
	\begin{tikzpicture}
	\begin{pgfonlayer}{nodelayer}
		\node [style=none] (0) at (0, -0.5) {};
		\node [style={gn_phase}] (1) at (0, 0.25) {${f(\theta)}$  };
	\end{pgfonlayer}
	\begin{pgfonlayer}{edgelayer}
		\draw (1) to (0.center);
	\end{pgfonlayer}
\end{tikzpicture}
}%
	\endpgfgraphicnamed
\right] = %
	\beginpgfgraphicnamed{functiongn2}
	\InputIfFileExists{functiongn2.tikz}{}{\input{./figures/functiongn2.tikz}}%
	\endpgfgraphicnamed

\]
\end{restatable}
\textbf{Note:} For presentation purposes, the proofs of some theorems and lemmas are moved to the appendix.

Using the derivative of the one-legged spider, we can differentiate any ZX diagram with only one occurrence of the parameter being differentiated against. Here is an example.

\ctikzfig{diff-example-1/d1}
\begin{center}
$$ \ket{00}\bra{0} + e^{i2\theta}\ket{11}\bra{1} \overset{\partial}{\mapsto} 2i*e^{i2\theta}\ket{11}\bra{1}$$
\end{center}
When there are multiple occurrences of the same parameter, the derivative can be expressed as a sum of ZX diagrams using the product rule. For example, the density matrix of $R_z(\theta)$ can be differentiated as follows.

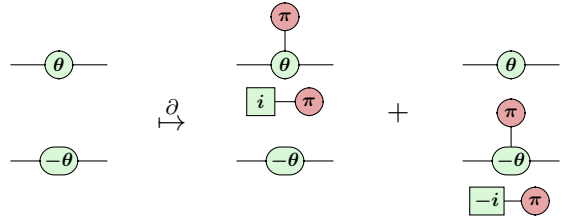
\ctikzfig{diff-example-1/d2}
Since there are no rules on how to further manipulate sums of ZX diagrams, any reasoning from this point on would need to explicitly rely on the matrix interpretation of the diagrams instead of rewriting.
In order to proceed with diagrammatic reasoning, we need to express the derivative as a single diagram.

By observing that the product rule leaves the unparameterised parts of the diagram untouched and can be ``factored out'', we only need to resynthesise the derivative of the parameterised part.

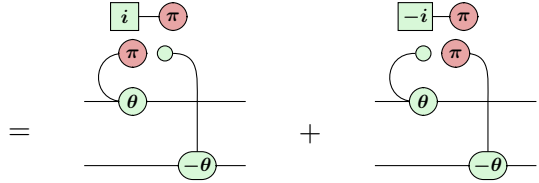
\ctikzfig{diff-example-1/d3}
After this factorisation, the diagrammatic terms in the sum (top of the diagram) can be further manipulated until we can eliminate the sum using a simple rule such as $\ket{0} + \ket{1} = \sqrt{2}\ket{+}$. (See appendix for a demonstration of this technique)
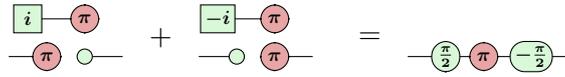
\ctikzfig{diffexampletop-res}
Therefore
$$ \frac{\partial}{\partial \theta} \left[ %
	\beginpgfgraphicnamed{diff-example-1/density-z}
	\begin{tikzpicture}[scale=0.85, baseline={([yshift=-.5ex]current bounding box.center)}]
	\begin{pgfonlayer}{nodelayer}
		\node [style=none] (0) at (2, 1.5) {};
		\node [style=none] (1) at (2, 0.5) {};
		\node [style=none] (2) at (0.5, 1.5) {};
		\node [style={gn_phase}] (3) at (1.25, 1.5) {$\theta$};
		\node [style={gn_phase}] (4) at (1.25, 0.5) {$-\theta$};
		\node [style=none] (5) at (0.5, 0.5) {};
	\end{pgfonlayer}
	\begin{pgfonlayer}{edgelayer}
		\draw (0.center) to (2.center);
		\draw (5.center) to (1.center);
	\end{pgfonlayer}
\end{tikzpicture}
}%
	\endpgfgraphicnamed
 \right] \qquad = \qquad %
	\beginpgfgraphicnamed{diff-example-1/d4}
	
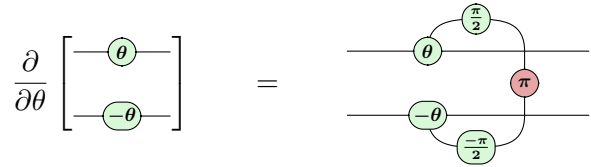
\begin{tikzpicture}[scale=0.85, baseline={([yshift=-.5ex]current bounding box.center)}]
	\begin{pgfonlayer}{nodelayer}
		\node [style=none] (84) at (17.75, 1.5) {};
		\node [style=none] (85) at (17.75, 0.5) {};
		\node [style=none] (86) at (14, 1.5) {};
		\node [style=none] (87) at (15.25, 1.5) {};
		\node [style=none] (88) at (15.25, 0.5) {};
		\node [style=none] (89) at (14, 0.5) {};
		\node [style=none] (94) at (16, 0) {};
		\node [style=none] (95) at (16, 2) {};
		\node [style={rn_phase}] (96) at (16.75, 1) {$\pi$};
		\node [style=none] (97) at (16.75, 0.5) {};
		\node [style=none] (98) at (16.75, 1.5) {};
		\node [style={gn_phase}] (99) at (15.25, 1.5) {$\theta$};
		\node [style={gn_phase}] (100) at (15.25, 0.5) {$-\theta$};
		\node [style={gn_phase}] (101) at (16, 2) {$ \frac{\pi}{2}$};
		\node [style={gn_phase}] (102) at (16, 0) {$ \frac{-\pi}{2}$};
	\end{pgfonlayer}
	\begin{pgfonlayer}{edgelayer}
		\draw (84.center) to (86.center);
		\draw (89.center) to (85.center);
		\draw [in=-90, out=0, looseness=1.25] (94.center) to (97.center);
		\draw (97.center) to (96);
		\draw (96) to (98.center);
		\draw [in=0, out=90, looseness=1.25] (98.center) to (95.center);
		\draw [in=-180, out=90, looseness=1.25] (87.center) to (95.center);
		\draw [in=180, out=-90, looseness=1.25] (88.center) to (94.center);
	\end{pgfonlayer}
\end{tikzpicture}
}%
	\endpgfgraphicnamed
 $$
This equation, first derived by Zhao et al.~\cite{Zhao2021analyzingbarren}, is essentially the parameter shift rule by Schuld et al.~\cite{schuld2019evaluating} expressed as a single ZX diagram (also see Corollary \ref{diffthm2tmcr}).

The key result of the paper allows us to express the derivative of an arbitrary ZX diagram in terms of a single diagram. It is based on the observation that the product rule and the unnormalised $\ket{W_n}$ state resemble each other: in the product rule, each term has one differentiated function, and in the W state each term has one bit set to 1 in the basis state.
$$ \partial(fgh) = (\partial f)gh + f(\partial g)h + fg(\partial h) $$
$$ \ket{W_3} = \ket{100} + \ket{010} + \ket{001}$$
We will show in Theorem \ref{diffthm1thm} that the product rule can indeed be represented using a W state supplemented with some local change of bases. The following lemma demonstrates that the difference between $f$ and $\partial f$ can be expressed as a change of basis from the computational basis $\ket{0}$, $\ket{1}$.

\begin{restatable}{lemma}{rgboxtrianglelm}\label{rgboxtrianglelm}
For any complex number $a$, we have
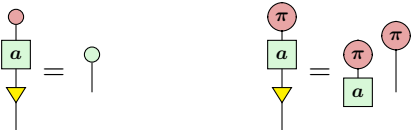
\[
	\beginpgfgraphicnamed{rgboxtriangle}
	\InputIfFileExists{rgboxtriangle.tikz}{}{\input{./figures/rgboxtriangle.tikz}}%
	\endpgfgraphicnamed

\]
\end{restatable}

To differentiate algebraic ZX diagrams, we first differentiate its parameterised generator, the one-legged green box:

 \begin{restatable}{lemma}{dgboxsingle}\label{dgboxsingle}
  Suppose  $f : \mathbb R \to \mathbb C$ is a differentiable function.Then
  \[
    \frac{\partial}{\partial \theta}\left[%
	\beginpgfgraphicnamed{functiongboxsingle}
	\begin{tikzpicture}
	\begin{pgfonlayer}{nodelayer}
		\node [style=gbox] (0) at (0, 0.25) {${f(\theta)}$  };
		\node [style=none] (1) at (0, -0.5) {};
	\end{pgfonlayer}
	\begin{pgfonlayer}{edgelayer}
		\draw (0) to (1.center);
	\end{pgfonlayer}
\end{tikzpicture}}%
	\endpgfgraphicnamed
\right] = %
	\beginpgfgraphicnamed{functiongbox1}
	\InputIfFileExists{functiongbox1.tikz}{}{\input{./figures/functiongbox1.tikz}}%
	\endpgfgraphicnamed

  \]
  \end{restatable}  
All parameterised differentiable algebraic ZX diagrams can be rewritten into the following form, where $M$ is an unparameterised ZX diagram with respect to $\theta$, and $\{f_i(\theta)\}_i$ are differentiable real functions of $\theta$. Parameterised green spiders can be written as a green box with an exponentiated phase, and parameterised red spiders can be converted to parameterised green spiders via Hadamard conjugation. We emphasise that $M$ can contain other parameterised spiders, just not with respect to $\theta$.

\ctikzfig{diffthm1/draggingv3}

\begin{theorem} \label{diffthmgeneralgbox}
Assume $f_j : \mathbb R \to \mathbb C$ are differentiable functions. Then
$$\frac{\partial}{\partial \theta}\left[%
	\beginpgfgraphicnamed{diffrentialinggbox}
	\InputIfFileExists{diffrentialinggbox.tikz}{}{\input{./figures/diffrentialinggbox.tikz}}%
	\endpgfgraphicnamed
\right]\qquad=\qquad
	\beginpgfgraphicnamed{diffrentialedgboxv2}
	\InputIfFileExists{diffrentialedgboxv2.tikz}{}{\input{./figures/diffrentialedgboxv2.tikz}}%
	\endpgfgraphicnamed
$$
\end{theorem} 
\begin{proof}
By linearity, differentiating the overall diagram amounts to differentiating the parameterised part of the diagram:

\[\frac{\partial}{\partial \theta}\left[%
	\beginpgfgraphicnamed{functiongboxes}
	\InputIfFileExists{functiongboxes.tikz}{}{\input{./figures/functiongboxes.tikz}}%
	\endpgfgraphicnamed
\right]\]
\begin{align*}
&=\quad
	\beginpgfgraphicnamed{dfunctiongboxesv2}
	\InputIfFileExists{dfunctiongboxesv2.tikz}{}{\input{./figures/dfunctiongboxesv2.tikz}}%
	\endpgfgraphicnamed
\\
&=\quad
	\beginpgfgraphicnamed{functiongboxessum1v2}
	\InputIfFileExists{functiongboxessum1v2.tikz}{}{\input{./figures/functiongboxessum1v2.tikz}}%
	\endpgfgraphicnamed
\\
\end{align*}
\begin{align*}
&=\quad
	\beginpgfgraphicnamed{functiongboxessum2v2}
	\InputIfFileExists{functiongboxessum2v2.tikz}{}{\input{./figures/functiongboxessum2v2.tikz}}%
	\endpgfgraphicnamed
\\
&=\quad%
	\beginpgfgraphicnamed{functiongboxescombinev2}
	\InputIfFileExists{functiongboxescombinev2.tikz}{}{\input{./figures/functiongboxescombinev2.tikz}}%
	\endpgfgraphicnamed

\end{align*}
The first step follows from Lemma \ref{dgboxsingle} and the product rule,  the second step  follows from Lemma \ref{wplug} and Lemma \ref{wplugrpi},
 and the third step follows from the $\pi$ commutation rule for green boxes.
The final step uses the property of W spider as given in (\ref{mlegsblackspider0and1eq}).
\end{proof}

This theorem unifies the linearity and product rules of differential calculus into a single diagram, without a blowup in diagram size: the number of nodes added to the diagram after differentiation is linearly proportional to the number of parameter occurrences. This makes the result practically useful for both calculations by hand and computer simulation.

\begin{remark} \label{diff-properties}
	Using the chain rule, we can pull out a common factor if all $f_i$ are composed with some other function $g$:
	  $$ %
	\beginpgfgraphicnamed{diffrential-chainv2}
	\InputIfFileExists{diffrential-chainv2.tikz}{}{\input{./figures/diffrential-chainv2.tikz}}%
	\endpgfgraphicnamed
 $$
	If $f_i$ is a constant function, the corresponding spider does not contribute to the derivative:
	$$ %
	\beginpgfgraphicnamed{diffrential-constantv2}
	\InputIfFileExists{diffrential-constantv2.tikz}{}{\input{./figures/diffrential-constantv2.tikz}}%
	\endpgfgraphicnamed
 $$
	Finally, the differentiation gadget nicely interacts with spider fusion:
	$$ %
	\beginpgfgraphicnamed{diffrential-fusev2}
	\InputIfFileExists{diffrential-fusev2.tikz}{}{\input{./figures/diffrential-fusev2.tikz}}%
	\endpgfgraphicnamed
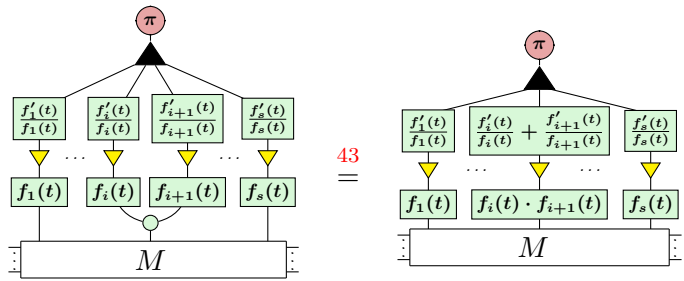
 $$
\end{remark}

Diagrammatic differentiation for regular ZX diagrams corresponds to the special case where all $f_j$ are phase functions:

\begin{theorem} \label{diffthm1thm}
The derivative of a differentiable ZX diagram can be expressed as a single ZX diagram:
$$\frac{\partial}{\partial \theta}\left[%
	\beginpgfgraphicnamed{diffthm1/draggingv2}
	\InputIfFileExists{diffthm1/draggingv2.tikz}{}{\input{./figures/diffthm1/draggingv2.tikz}}%
	\endpgfgraphicnamed
\right]\qquad=\qquad
	\beginpgfgraphicnamed{diffthm1/ddraggingv2}
	\InputIfFileExists{diffthm1/ddraggingv2.tikz}{}{\input{./figures/diffthm1/ddraggingv2.tikz}}%
	\endpgfgraphicnamed
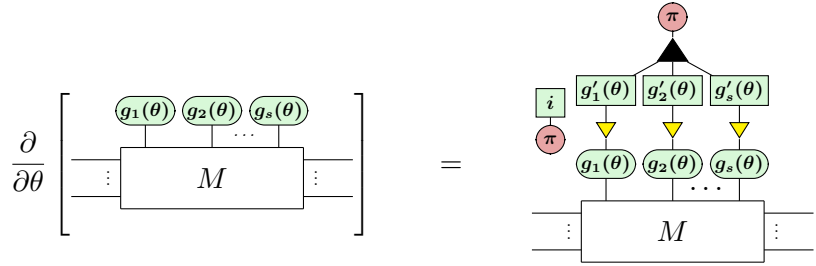
$$
\end{theorem}

\begin{proof}
This is a special case of Theorem \ref{diffthmgeneralgbox}, where $f_j(\theta) = e^{ig_j(\theta)}\neq 0$ and $\frac{f'_j(\theta)}{f_j(\theta)}=ig'_j(\theta)$, thus Lemma \ref{twtransform} applies. The ``$i$'' is common across all functions and can be factored out through the W spider using the diagrammatic chain rule from Remark \ref{diff-properties}.
\end{proof}
Since we can now differentiate arbitrary ZX diagrams, we can consider differentiating quantum circuits as a special case of Theorem \ref{diffthm1thm}. Executing a circuit $U(\boldsymbol{\theta})$ for some observable $H$ on a quantum computer estimates the expectation value
$\langle H\rangle=\left\langle 0\left|U^{\dagger}(\boldsymbol{\theta}) H U(\boldsymbol{\theta})\right| 0\right\rangle$.
Thus, the ZX diagram representing $\langle H\rangle$ has an equal number of occurrences of $\theta$ and $-\theta$.

\begin{corollary}\label{diffthm2tm}
The derivative of a parameterised quantum circuit can be expressed as a single ZX diagram:
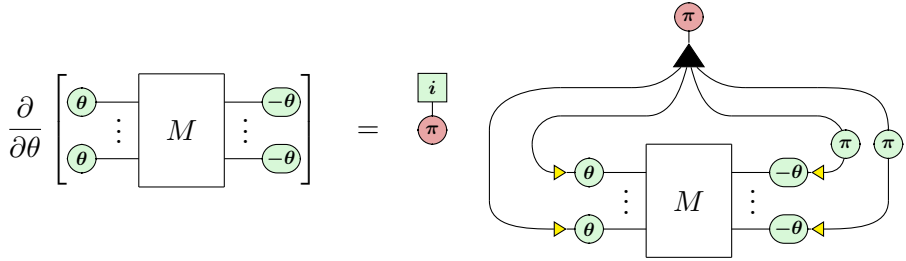
\[
  \frac{\partial}{\partial \theta}\left[%
	\beginpgfgraphicnamed{diffthm2/expect}
	\InputIfFileExists{diffthm2/expect.tikz}{}{\input{./figures/diffthm2/expect.tikz}}%
	\endpgfgraphicnamed
\right]
  \quad=\quad%
	\beginpgfgraphicnamed{diffthm2/dexpect4}
	\InputIfFileExists{diffthm2/dexpect4.tikz}{}{\input{./figures/diffthm2/dexpect4.tikz}}%
	\endpgfgraphicnamed

\]
\end{corollary}
\begin{proof}
Noting that $\frac{\partial}{\partial\theta}-\theta=-1$ and $  %
	\beginpgfgraphicnamed{minus1topi}
	\InputIfFileExists{minus1topi.tikz}{}{\input{./figures/minus1topi.tikz}}%
	\endpgfgraphicnamed
$, this follows directly from Theorem \ref{diffthm1thm}.
\end{proof}

\begin{corollary}\label{diffthm2tmcr}
As a special case of Corollary \ref{diffthm2tm}, we obtain the parameter-shift rule from \cite{schuld2019evaluating}.
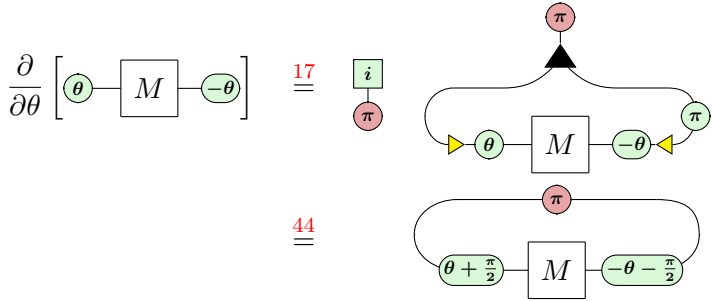
\begin{align*}
  \frac{\partial}{\partial\theta}\left[%
	\beginpgfgraphicnamed{diffthm2/expect_small}
	\InputIfFileExists{diffthm2/expect_small.tikz}{}{\input{./figures/diffthm2/expect_small.tikz}}%
	\endpgfgraphicnamed
\right]
  \quad&\overset{\ref{diffthm2tm}}{=}\quad%
	\beginpgfgraphicnamed{diffthm2/dexpect_small1}
	\InputIfFileExists{diffthm2/dexpect_small1.tikz}{}{\input{./figures/diffthm2/dexpect_small1.tikz}}%
	\endpgfgraphicnamed
\\
  \quad&\overset{\ref{w2pitritorpignslm}}{=} \quad\qquad%
	\beginpgfgraphicnamed{diffthm2/dexpect_small2}
	\InputIfFileExists{diffthm2/dexpect_small2.tikz}{}{\input{./figures/diffthm2/dexpect_small2.tikz}}%
	\endpgfgraphicnamed

\end{align*}
\end{corollary}

\begin{remark}
This result has been given as a theorem in \cite{Zhao2021analyzingbarren}, here we directly get it as a consequence of Corollary \ref{diffthm2tm} which follows from Theorem \ref{diffthm1thm}.
\end{remark}

From Corollary \ref{diffthm2tm}, we thus have a simple diagrammatic expression for the derivative of any parameterised quantum circuit. 
In general, it is not easy to obtain the derivative of a parameterised matrix in a single term bra-ket expression (with no sums,  e.g. $\bra{\phi}ABC\ket{\psi}$), thus showing the power of ZX-calculus and 2-dimensional diagrammatic reasoning.

Similar to how the decomposition of the Pauli X gate in Corollary \ref{diffthm2tmcr} gives us a 2-term parameter shift rule, decompositions of the $W$ state in Corollary \ref{diffthm2tm} are used in \cite{koch2022quantum} to obtain shift rules for gates with an arbitrary number of $\theta$-occurrences in particular generalising the existing shift rules from \cite{schuld2019evaluating} and \cite{anselmetti2021local}.
Furthermore, they use the analytical ZX technology developed here to prove the optimality of the shift rule given in \cite{anselmetti2021local}.

\section{Integrating ZX diagrams}

In this section we show how to diagrammatically integrate ZX diagrams using algebraic ZX-calculus. 
This will allow us to evaluate the expectation and variance of a quantum circuit's derivative over the uniform distribution, as demonstrated in Section~\ref{sec:qml}.

\subsection{Circuits}

As a first step, we only consider integrals of diagrams with the same number of positive and negative occurrences of a parameter $\theta$.
In particular, all diagrams arising from expectation values of circuits have this form (see Section \ref{sec:qml}).

\begin{definition}
	The Hamming weight of a bit-string $\vec x \in \{0, 1\}^n$ is defined as $w(\vec x) := \sum x_i$, i.e. the number of 1s in the bit-string.
\end{definition}

\begin{restatable}{lemma}{circsum} \label{lem:circsum}
	Let $k$ be a non-zero integer and $M$ a diagram with no occurrence of $\theta$. Then
	\[
		\frac{1}{2\pi} \int_{-\pi}^\pi %
	\beginpgfgraphicnamed{integrate/circ-sum/lhs}
	\begin{tikzpicture}
	\begin{pgfonlayer}{nodelayer}
		\node [style=none] (0) at (0.5, 1.25) {};
		\node [style=none] (1) at (-0.25, 1) {};
		\node [style=none] (2) at (0.5, -1.375) {};
		\node [style={gn_phase}, scale=1] (3) at (1.075, 1) {$-k\theta$};
		\node [style={gn_phase}, scale=1] (4) at (-0.75, 1) {$k\theta$};
		\node [style=none] (6) at (0.125, -0.05) {$M$};
		\node [style=none] (8) at (0.5, 1) {};
		\node [style=none] (9) at (-0.25, -1.375) {};
		\node [style=none] (10) at (-0.25, 1.25) {};
		\node [style=none] (12) at (-0.625, -1.125) {};
		\node [style=none] (13) at (-0.25, -0.375) {};
		\node [style=none] (14) at (-0.25, -1.125) {};
		\node [style=none] (15) at (-0.625, -0.375) {};
		\node [style=none] (16) at (-0.425, -0.625) {$\vdots$};
		\node [style=none] (18) at (0.875, -0.375) {};
		\node [style=none] (20) at (0.5, -0.375) {};
		\node [style=none] (21) at (0.5, -1.125) {};
		\node [style=none] (23) at (0.875, -1.125) {};
		\node [style=none] (24) at (0.675, -0.625) {$\vdots$};
		\node [style=none] (25) at (-0.25, 0.25) {};
		\node [style={gn_phase}, scale=1] (26) at (1.075, 0.25) {$-k\theta$};
		\node [style={gn_phase}, scale=1] (27) at (-0.75, 0.25) {$k\theta$};
		\node [style=none] (30) at (0.5, 0.25) {};
		\node [style=none] (31) at (-0.425, 0.75) {$\vdots$};
		\node [style=none] (32) at (0.675, 0.75) {$\vdots$};
		\node [style=none] (38) at (-1.2, 1.175) {};
		\node [style=none] (39) at (-1.2, 0.075) {};
		\node [style=none] (40) at (-1.45, 0.625) {\scriptsize $n$};
		\node [style=none] (41) at (-0.875, -0.325) {};
		\node [style=none] (42) at (-0.875, -1.175) {};
		\node [style=none] (43) at (-1.175, -0.775) {\scriptsize $m$};
		\node [style=none] (44) at (1.625, 1.175) {};
		\node [style=none] (45) at (1.625, 0.075) {};
		\node [style=none] (46) at (1.875, 0.625) {\scriptsize $n$};
		\node [style=none] (47) at (1.125, -0.325) {};
		\node [style=none] (48) at (1.125, -1.175) {};
		\node [style=none] (49) at (1.375, -0.75) {\scriptsize $l$};
	\end{pgfonlayer}
	\begin{pgfonlayer}{edgelayer}
		\draw (10.center) to (9.center);
		\draw (10.center) to (0.center);
		\draw (0.center) to (2.center);
		\draw (2.center) to (9.center);
		\draw (13.center) to (15.center);
		\draw (12.center) to (14.center);
		\draw (18.center) to (20.center);
		\draw (21.center) to (23.center);
		\draw (3) to (8.center);
		\draw (4) to (1.center);
		\draw (27) to (25.center);
		\draw (26) to (30.center);
		\draw [style=braceedge] (39.center) to (38.center);
		\draw [style=braceedge] (42.center) to (41.center);
		\draw [style=braceedge] (44.center) to (45.center);
		\draw [style=braceedge] (47.center) to (48.center);
	\end{pgfonlayer}
\end{tikzpicture}
}%
	\endpgfgraphicnamed
 d\theta ~= \sum_{\substack{\vec x,\vec y\in\{0, 1\}^n \\ w(\vec x) = w(\vec y)}} ~ %
	\beginpgfgraphicnamed{integrate/circ-sum/rhs}
	\begin{tikzpicture}
	\begin{pgfonlayer}{nodelayer}
		\node [style=none] (0) at (0.5, 1.25) {};
		\node [style=none] (1) at (-0.25, 1) {};
		\node [style=none] (2) at (0.5, -1.375) {};
		\node [style={rn_phase}, scale=1] (3) at (1.075, 1) {$y_1\pi$};
		\node [style={rn_phase}, scale=1] (4) at (-0.875, 1) {$x_1\pi$};
		\node [style=none] (5) at (0.125, -0.05) {$M$};
		\node [style=none] (6) at (0.5, 1) {};
		\node [style=none] (7) at (-0.25, -1.375) {};
		\node [style=none] (8) at (-0.25, 1.25) {};
		\node [style=none] (19) at (-0.25, 0.25) {};
		\node [style={rn_phase}, scale=1] (20) at (1.075, 0.25) {$y_n\pi$};
		\node [style={rn_phase}, scale=1] (21) at (-0.875, 0.25) {$x_n\pi$};
		\node [style=none] (22) at (0.5, 0.25) {};
		\node [style=none] (23) at (-0.45, 0.75) {$\vdots$};
		\node [style=none] (24) at (0.7, 0.75) {$\vdots$};
		\node [style=none] (25) at (-0.625, -1.125) {};
		\node [style=none] (26) at (-0.25, -0.375) {};
		\node [style=none] (27) at (-0.25, -1.125) {};
		\node [style=none] (28) at (-0.625, -0.375) {};
		\node [style=none] (29) at (-0.45, -0.625) {$\vdots$};
		\node [style=none] (30) at (0.875, -0.375) {};
		\node [style=none] (31) at (0.5, -0.375) {};
		\node [style=none] (32) at (0.5, -1.125) {};
		\node [style=none] (33) at (0.875, -1.125) {};
		\node [style=none] (34) at (0.7, -0.625) {$\vdots$};
		\node [style=none] (35) at (-0.875, -0.325) {};
		\node [style=none] (36) at (-0.875, -1.175) {};
		\node [style=none] (37) at (-1.175, -0.775) {\scriptsize $m$};
		\node [style=none] (38) at (1.125, -0.325) {};
		\node [style=none] (39) at (1.125, -1.175) {};
		\node [style=none] (40) at (1.375, -0.75) {\scriptsize $l$};
	\end{pgfonlayer}
	\begin{pgfonlayer}{edgelayer}
		\draw (8.center) to (7.center);
		\draw (8.center) to (0.center);
		\draw (0.center) to (2.center);
		\draw (2.center) to (7.center);
		\draw (3) to (6.center);
		\draw (4) to (1.center);
		\draw (21) to (19.center);
		\draw (20) to (22.center);
		\draw (26.center) to (28.center);
		\draw (25.center) to (27.center);
		\draw (30.center) to (31.center);
		\draw (32.center) to (33.center);
		\draw [style=braceedge] (36.center) to (35.center);
		\draw [style=braceedge] (38.center) to (39.center);
	\end{pgfonlayer}
\end{tikzpicture}
}%
	\endpgfgraphicnamed

	\]
\end{restatable}

The diagram sum above has an exponential number of terms.
We want to find a more compact representation of this integral as a single diagram.
Since the common part $M$ can be factored out of the sum, this is equivalent to finding a single diagram representation of the following projector.

\begin{definition} \label{def:P}
	The \textit{Hamming weight projector} $P_n$ is defined as
	\[
	\beginpgfgraphicnamed{integrate/projector-lhs}
	\begin{tikzpicture}
	\begin{pgfonlayer}{nodelayer}
		\node [style=none] (0) at (0.1, 0) {$P_n$};
		\node [style=none] (1) at (0.6, -0.625) {};
		\node [style=none] (2) at (-0.4, -0.625) {};
		\node [style=none] (3) at (-0.4, 0.625) {};
		\node [style=none] (4) at (0.6, 0.625) {};
		\node [style=none] (5) at (-0.575, 0.125) {$\vdots$};
		\node [style=none] (6) at (-0.75, 0.45) {};
		\node [style=none] (7) at (-0.75, -0.45) {};
		\node [style=none] (8) at (-0.4, 0.45) {};
		\node [style=none] (9) at (-0.4, -0.45) {};
		\node [style=none] (10) at (0.775, 0.125) {$\vdots$};
		\node [style=none] (11) at (0.6, 0.45) {};
		\node [style=none] (12) at (0.6, -0.45) {};
		\node [style=none] (13) at (0.95, 0.45) {};
		\node [style=none] (14) at (0.95, -0.45) {};
	\end{pgfonlayer}
	\begin{pgfonlayer}{edgelayer}
		\draw (2.center) to (1.center);
		\draw (3.center) to (2.center);
		\draw (4.center) to (1.center);
		\draw (3.center) to (4.center);
		\draw (8.center) to (6.center);
		\draw (9.center) to (7.center);
		\draw (13.center) to (11.center);
		\draw (14.center) to (12.center);
	\end{pgfonlayer}
\end{tikzpicture}
}%
	\endpgfgraphicnamed
 ~:= \sum_{\substack{\vec x, \vec y\in\{0, 1\}^n \\ w(\vec x) = w(\vec y)}} ~ %
	\beginpgfgraphicnamed{integrate/projector-rhs}
	\begin{tikzpicture}
	\begin{pgfonlayer}{nodelayer}
		\node [style={rn_phase}, scale=1] (1) at (0.375, 0.375) {$y_1\pi$};
		\node [style={rn_phase}, scale=1] (2) at (-0.5, 0.375) {$x_1\pi$};
		\node [style={rn_phase}, scale=1] (5) at (0.375, -0.375) {$y_n\pi$};
		\node [style={rn_phase}, scale=1] (6) at (-0.5, -0.375) {$x_n\pi$};
		\node [style=none] (8) at (-0.95, 0.125) {$\vdots$};
		\node [style=none] (9) at (0.8, 0.125) {$\vdots$};
		\node [style=none] (10) at (-1.175, 0.375) {};
		\node [style=none] (11) at (-1.175, -0.375) {};
		\node [style=none] (12) at (1.05, 0.375) {};
		\node [style=none] (13) at (1.05, -0.375) {};
	\end{pgfonlayer}
	\begin{pgfonlayer}{edgelayer}
		\draw (11.center) to (6);
		\draw (10.center) to (2);
		\draw (13.center) to (5);
		\draw (1) to (12.center);
	\end{pgfonlayer}
\end{tikzpicture}
}%
	\endpgfgraphicnamed

	\]
\end{definition}

We note that the Hamming weight of a bit-string of length $n$ can be encoded in $\left\lfloor \log n \right\rfloor + 1$ bits using the standard binary encoding of natural numbers.

\begin{definition}
	We write $\left[\vec x\right] := \sum_{i=1}^n 2^{i-1}x_i$ for the binary interpretation of a bit-string $\vec x \in \{0, 1\}^n$.
\end{definition}

In the following, we use techniques from classical boolean circuits for binary arithmetic to construct an algebraic ZX diagram that outputs the Hamming weight in this encoding.
We begin by giving a ZX analogue of a binary addition circuit for $n$ bits.

\begin{definition} \label{def:sum}
	We define the diagram $\Sigma_n$ recursively via
	\[ %
	\beginpgfgraphicnamed{integrate/adder/general}
	\begin{tikzpicture}
	\begin{pgfonlayer}{nodelayer}
		\node [style=none] (0) at (0.1, 0) {};
		\node [style=none] (1) at (0.1, -0.875) {};
		\node [style=none] (2) at (0.5, 0) {$\Sigma_n$};
		\node [style=none] (3) at (0.1, -1.125) {};
		\node [style=none] (4) at (0.1, 1.125) {};
		\node [style=none] (5) at (0.1, -0.25) {};
		\node [style=none] (6) at (0.875, -0.65) {};
		\node [style=none] (7) at (0.875, 0.65) {};
		\node [style=none] (10) at (0.1, 0.625) {};
		\node [style=none] (11) at (-0.2, 0) {};
		\node [style=none] (12) at (-0.2, -0.875) {};
		\node [style=none] (13) at (-0.2, -0.25) {};
		\node [style=none] (14) at (-0.2, 0.625) {};
		\node [style=none] (15) at (-0.05, 0.425) {$\vdots$};
		\node [style=none] (16) at (-0.075, -0.45) {$\vdots$};
		\node [style=none] (17) at (0.1, 0.875) {};
		\node [style=none] (18) at (-0.2, 0.875) {};
		\node [style=none] (22) at (-0.4, 0.675) {};
		\node [style=none] (23) at (-0.4, -0.05) {};
		\node [style=none] (24) at (-0.65, 0.325) {\scriptsize $n$};
		\node [style=none] (25) at (-0.4, -0.2) {};
		\node [style=none] (26) at (-0.4, -0.925) {};
		\node [style=none] (27) at (-0.65, -0.55) {\scriptsize $n$};
		\node [style=none] (28) at (0.875, -0.125) {};
		\node [style=none] (29) at (0.875, 0.475) {};
		\node [style=none] (30) at (1.025, 0.275) {$\vdots$};
		\node [style=none] (31) at (1.2, -0.125) {};
		\node [style=none] (32) at (1.2, 0.475) {};
		\node [style=none] (33) at (1.4, 0.525) {};
		\node [style=none] (34) at (1.4, -0.175) {};
		\node [style=none] (35) at (1.65, 0.125) {\scriptsize $n$};
		\node [style=none] (36) at (-0.625, 0.875) {\scriptsize carry};
		\node [style=none] (37) at (0.875, -0.425) {};
		\node [style=none] (38) at (1.225, -0.425) {};
		\node [style=none] (39) at (1.65, -0.45) {\scriptsize carry};
	\end{pgfonlayer}
	\begin{pgfonlayer}{edgelayer}
		\draw (4.center) to (3.center);
		\draw (7.center) to (6.center);
		\draw (6.center) to (3.center);
		\draw (4.center) to (7.center);
		\draw (14.center) to (10.center);
		\draw (11.center) to (0.center);
		\draw (13.center) to (5.center);
		\draw (12.center) to (1.center);
		\draw (18.center) to (17.center);
		\draw [style=braceedge] (23.center) to (22.center);
		\draw [style=braceedge] (26.center) to (25.center);
		\draw (32.center) to (29.center);
		\draw (28.center) to (31.center);
		\draw [style=braceedge] (33.center) to (34.center);
		\draw (37.center) to (38.center);
	\end{pgfonlayer}
\end{tikzpicture}
}%
	\endpgfgraphicnamed
 \qquad\quad %
	\beginpgfgraphicnamed{integrate/adder/base}
	\begin{tikzpicture}
	\begin{pgfonlayer}{nodelayer}
		\node [style=none] (2) at (0.4, 0) {$\Sigma_1$};
		\node [style=none] (3) at (0, -0.875) {};
		\node [style=none] (4) at (0, 0.875) {};
		\node [style=none] (5) at (0, 0) {};
		\node [style=none] (6) at (0.75, -0.525) {};
		\node [style=none] (7) at (0.75, 0.525) {};
		\node [style=none] (14) at (0, 0.5) {};
		\node [style=none] (15) at (0, -0.5) {};
		\node [style=none] (16) at (-0.25, -0.5) {};
		\node [style=none] (17) at (-0.25, 0) {};
		\node [style=none] (18) at (-0.25, 0.5) {};
		\node [style=none] (19) at (0.75, 0.375) {};
		\node [style=none] (20) at (0.75, -0.375) {};
		\node [style=none] (21) at (1, -0.375) {};
		\node [style=none] (22) at (1, 0.375) {};
		\node [style=none] (23) at (1.375, 0) {$:=$};
		\node [style=none] (24) at (1.875, -0.5) {};
		\node [style=none] (25) at (1.875, 0) {};
		\node [style=none] (26) at (1.875, 0.5) {};
		\node [style=rn] (27) at (2.25, 0.5) {};
		\node [style=rn] (28) at (2.25, 0) {};
		\node [style=rn] (29) at (2.25, -0.5) {};
		\node [style=bspider, rotate=90] (30) at (2.875, 0.5) {};
		\node [style=gn] (31) at (3.5, -0.5) {};
		\node [style=rn] (32) at (3.5, 0.5) {};
		\node [style=none] (33) at (3.875, 0.5) {};
		\node [style=none] (34) at (3.875, -0.5) {};
	\end{pgfonlayer}
	\begin{pgfonlayer}{edgelayer}
		\draw (4.center) to (3.center);
		\draw (7.center) to (6.center);
		\draw (6.center) to (3.center);
		\draw (4.center) to (7.center);
		\draw (16.center) to (15.center);
		\draw (17.center) to (5.center);
		\draw (18.center) to (14.center);
		\draw (22.center) to (19.center);
		\draw (20.center) to (21.center);
		\draw (24.center) to (29);
		\draw (28) to (25.center);
		\draw (26.center) to (27);
		\draw (27) to (30);
		\draw (28) to (30);
		\draw (30) to (29);
		\draw (28) to (31);
		\draw (29) to (31);
		\draw (31) to (27);
		\draw (30) to (32);
		\draw (32) to (33.center);
		\draw (32) to (31);
		\draw (31) to (34.center);
	\end{pgfonlayer}
\end{tikzpicture}
}%
	\endpgfgraphicnamed
 \qquad\quad %
	\beginpgfgraphicnamed{integrate/adder/rec}
	\begin{tikzpicture}
	\begin{pgfonlayer}{nodelayer}
		\node [style=none] (0) at (-0.25, 0) {};
		\node [style=none] (1) at (-0.25, -1.125) {};
		\node [style=none] (2) at (0.275, 0) {$\Sigma_{n+1}$};
		\node [style=none] (3) at (-0.25, -1.375) {};
		\node [style=none] (4) at (-0.25, 1.375) {};
		\node [style=none] (5) at (-0.25, -0.325) {};
		\node [style=none] (6) at (0.775, -0.775) {};
		\node [style=none] (7) at (0.775, 0.775) {};
		\node [style=none] (8) at (-0.25, 0.775) {};
		\node [style=none] (9) at (-0.55, 0) {};
		\node [style=none] (10) at (-0.55, -1.125) {};
		\node [style=none] (11) at (-0.55, -0.325) {};
		\node [style=none] (12) at (-0.55, 0.775) {};
		\node [style=none] (13) at (-0.4, 0.425) {$\vdots$};
		\node [style=none] (14) at (-0.425, -0.7) {$\vdots$};
		\node [style=none] (15) at (-0.25, 1.125) {};
		\node [style=none] (16) at (-0.55, 1.125) {};
		\node [style=none] (23) at (0.775, -0.625) {};
		\node [style=none] (24) at (0.775, 0.6) {};
		\node [style=none] (25) at (0.925, 0.1) {$\vdots$};
		\node [style=none] (26) at (1.1, -0.625) {};
		\node [style=none] (27) at (1.1, 0.6) {};
		\node [style=none] (28) at (1.5, 0) {$:=$};
		\node [style=none] (29) at (3.075, 1.25) {$\Sigma_1$};
		\node [style=none] (30) at (2.75, 0.75) {};
		\node [style=none] (31) at (2.75, 1.75) {};
		\node [style=none] (32) at (2.75, 1.25) {};
		\node [style=none] (33) at (3.375, 0.975) {};
		\node [style=none] (34) at (3.375, 1.525) {};
		\node [style=none] (35) at (2.75, 1.5) {};
		\node [style=none] (36) at (2.75, 1) {};
		\node [style=none] (40) at (3.375, 1.425) {};
		\node [style=none] (41) at (3.375, 1.075) {};
		\node [style=none] (43) at (4.7, 1.425) {};
		\node [style=none] (44) at (-0.25, -0.5) {};
		\node [style=none] (45) at (-0.55, -0.5) {};
		\node [style=none] (46) at (-0.25, 0.625) {};
		\node [style=none] (47) at (-0.55, 0.625) {};
		\node [style=none] (52) at (2, 0) {};
		\node [style=none] (53) at (2, -1.375) {};
		\node [style=none] (54) at (2, -0.575) {};
		\node [style=none] (55) at (2, 0.775) {};
		\node [style=none] (56) at (2.15, 0.425) {$\vdots$};
		\node [style=none] (57) at (2.125, -0.95) {$\vdots$};
		\node [style=none] (61) at (2, -0.75) {};
		\node [style=none] (63) at (2, 0.625) {};
		\node [style=none] (64) at (3.625, 0) {};
		\node [style=none] (65) at (3.625, -1.375) {};
		\node [style=none] (66) at (4.025, -0.25) {$\Sigma_n$};
		\node [style=none] (67) at (3.625, -1.625) {};
		\node [style=none] (68) at (3.625, 1.25) {};
		\node [style=none] (69) at (3.625, -0.75) {};
		\node [style=none] (70) at (4.4, -1.025) {};
		\node [style=none] (71) at (4.4, 0.525) {};
		\node [style=none] (72) at (3.625, 0.625) {};
		\node [style=none] (73) at (3.625, 1.075) {};
		\node [style=none] (74) at (4.4, -0.875) {};
		\node [style=none] (75) at (4.4, 0.35) {};
		\node [style=none] (76) at (2, 1.5) {};
		\node [style=none] (79) at (4.575, -0.15) {$\vdots$};
		\node [style=none] (80) at (4.7, -0.875) {};
		\node [style=none] (81) at (4.7, 0.35) {};
	\end{pgfonlayer}
	\begin{pgfonlayer}{edgelayer}
		\draw (4.center) to (3.center);
		\draw (7.center) to (6.center);
		\draw (6.center) to (3.center);
		\draw (4.center) to (7.center);
		\draw (12.center) to (8.center);
		\draw (9.center) to (0.center);
		\draw (11.center) to (5.center);
		\draw (10.center) to (1.center);
		\draw (16.center) to (15.center);
		\draw (27.center) to (24.center);
		\draw (23.center) to (26.center);
		\draw (31.center) to (30.center);
		\draw (34.center) to (33.center);
		\draw (33.center) to (30.center);
		\draw (31.center) to (34.center);
		\draw (43.center) to (40.center);
		\draw (45.center) to (44.center);
		\draw (47.center) to (46.center);
		\draw [in=-135, out=0, looseness=1.25] (55.center) to (32.center);
		\draw [in=-120, out=0, looseness=0.75] (54.center) to (36.center);
		\draw (68.center) to (67.center);
		\draw (71.center) to (70.center);
		\draw (70.center) to (67.center);
		\draw (68.center) to (71.center);
		\draw (41.center) to (73.center);
		\draw (63.center) to (72.center);
		\draw (64.center) to (52.center);
		\draw (61.center) to (69.center);
		\draw (53.center) to (65.center);
		\draw (76.center) to (35.center);
		\draw (74.center) to (80.center);
		\draw (81.center) to (75.center);
	\end{pgfonlayer}
\end{tikzpicture}
}%
	\endpgfgraphicnamed
 \]
\end{definition}

The construction of $\Sigma_n$ is analogous to a classical ripple-carry adder, where $\Sigma_1$ acts as a full-adder~\cite{mano1972digital}.
Its action on computational basis states thus corresponds to binary addition.

\begin{restatable}{lemma}{sigmacorrect} \label{lem:sigmacorrect}
	$\Sigma_n$ performs binary addition, i.e. for all $\vec x, \vec y \in \{0, 1\}^n$ and $c \in  \{0, 1\}$, we have $\Sigma_n \ket{c,\vec x,\vec y} = \ket{\vec z}$ where $\left[\vec z\right] = \left[\vec x\right] + \left[\vec y\right] + c$.
\end{restatable}

Using this adder, we define a diagram that computes the binary Hamming weight of its input using a divide-and-conquer strategy.

\begin{definition} \label{def:w}
	We define the diagram $%
	\beginpgfgraphicnamed{integrate/W/general}
	\InputIfFileExists{integrate/W/general.tikz}{}{\input{./figures/integrate/W/general.tikz}}%
	\endpgfgraphicnamed
$ recursively via
	\[ %
	\beginpgfgraphicnamed{integrate/W/base}
	\InputIfFileExists{integrate/W/base.tikz}{}{\input{./figures/integrate/W/base.tikz}}%
	\endpgfgraphicnamed
 \qquad %
	\beginpgfgraphicnamed{integrate/W/even}
	\InputIfFileExists{integrate/W/even.tikz}{}{\input{./figures/integrate/W/even.tikz}}%
	\endpgfgraphicnamed
 \qquad %
	\beginpgfgraphicnamed{integrate/W/odd}
	\InputIfFileExists{integrate/W/odd.tikz}{}{\input{./figures/integrate/W/odd.tikz}}%
	\endpgfgraphicnamed
 \]
	where $k = \left\lfloor \log n \right\rfloor + 1$.
\end{definition}

\begin{restatable}{proposition}{wsize} \label{prop:wsize}
	The diagram size of $W_n$ only grows linearly with increasing $n$.
\end{restatable}

\begin{restatable}{lemma}{wcorrect} \label{lem:wcorrect}
	$W_n$ computes the binary Hamming weight, i.e. for all $\vec x \in \{0, 1\}^n$, we have $W_n \ket{\vec x} = \ket{\vec z}$ where $\left[\vec z\right] = w(\vec x)$.
\end{restatable}

This immediately yields a single diagram representation of the Hamming weight projector $P_n$.

\begin{corollary} \label{cor:psplit}
	We can represent $P_n$ as a single diagram in terms of $W_n$:
	\[ %
	\beginpgfgraphicnamed{integrate/split/lhs}
	\InputIfFileExists{integrate/split/lhs.tikz}{}{\input{./figures/integrate/split/lhs.tikz}}%
	\endpgfgraphicnamed
 ~=~ %
	\beginpgfgraphicnamed{integrate/split/rhs}
	\InputIfFileExists{integrate/split/rhs.tikz}{}{\input{./figures/integrate/split/rhs.tikz}}%
	\endpgfgraphicnamed
 \]
\end{corollary}
\begin{proof}
	Follows by comparing the action on all computational basis states. For all $\vec x, \vec y\in\{0,1\}^n$, we have
	\begin{gather*}
	\beginpgfgraphicnamed{integrate/split/proof-left1}
	\InputIfFileExists{integrate/split/proof-left1.tikz}{}{\input{./figures/integrate/split/proof-left1.tikz}}%
	\endpgfgraphicnamed

		~\overset{\ref{def:P}}{=}~
		\begin{cases}
			1 &\text{if } w(\vec x) = w(\vec y) \\
			0 &\text{otherwise}
		\end{cases}
		\\
	\beginpgfgraphicnamed{integrate/split/proof-right1}
	\InputIfFileExists{integrate/split/proof-right1.tikz}{}{\input{./figures/integrate/split/proof-right1.tikz}}%
	\endpgfgraphicnamed

		~\overset{\ref{lem:wcorrect}}{=}~ %
	\beginpgfgraphicnamed{integrate/split/proof-right2}
	\InputIfFileExists{integrate/split/proof-right2.tikz}{}{\input{./figures/integrate/split/proof-right2.tikz}}%
	\endpgfgraphicnamed

		~=~
		\begin{cases}
			1 &\text{if } \vec a = \vec b \\
			0 &\text{otherwise}
		\end{cases}
		~=~
		\begin{cases}
			1 &\text{if } w(\vec x) = w(\vec y) \\
			0 &\text{otherwise}
		\end{cases}
	\end{gather*}
	where $k = \lfloor \log(n) \rfloor + 1$, $[\vec a] = w(\vec x)$, and $[\vec b] = w(\vec y)$.
	The last step follows since the binary interpretation $[\cdot]$ is a bijective mapping, so $\vec a = \vec b \iff [\vec a] = [\vec b] \iff w(\vec x) = w(\vec y)$.
\end{proof}

\begin{theorem} \label{thm:circ-int}
	Let $k$ be a non-zero integer and $M$ a diagram with no occurrence of $\theta$. Then
	\[
	\frac{1}{2\pi} \int_{-\pi}^\pi %
	\beginpgfgraphicnamed{integrate/circ-sum/lhs}
	}%
	\endpgfgraphicnamed
 d\theta ~= %
	\beginpgfgraphicnamed{integrate/circ-int/rhs}
	\begin{tikzpicture}
	\begin{pgfonlayer}{nodelayer}
		\node [style=none] (0) at (0.5, 0.5) {};
		\node [style=none] (1) at (-0.25, 0.25) {};
		\node [style=none] (2) at (0.5, -1.875) {};
		\node [style=none] (5) at (0.125, -0.675) {$M$};
		\node [style=none] (6) at (0.5, 0.25) {};
		\node [style=none] (7) at (-0.25, -1.875) {};
		\node [style=none] (8) at (-0.25, 0.5) {};
		\node [style=none] (19) at (-0.25, -0.5) {};
		\node [style=none] (22) at (0.5, -0.5) {};
		\node [style=none] (24) at (0.7, 0) {$\vdots$};
		\node [style=none] (25) at (-0.625, -1.625) {};
		\node [style=none] (26) at (-0.25, -0.875) {};
		\node [style=none] (27) at (-0.25, -1.625) {};
		\node [style=none] (28) at (-0.625, -0.875) {};
		\node [style=none] (29) at (-0.45, -1.125) {$\vdots$};
		\node [style=none] (30) at (0.875, -0.875) {};
		\node [style=none] (31) at (0.5, -0.875) {};
		\node [style=none] (32) at (0.5, -1.625) {};
		\node [style=none] (33) at (0.875, -1.625) {};
		\node [style=none] (34) at (0.7, -1.125) {$\vdots$};
		\node [style=none] (35) at (-0.875, -0.825) {};
		\node [style=none] (36) at (-0.875, -1.675) {};
		\node [style=none] (37) at (-1.175, -1.275) {\scriptsize $m$};
		\node [style=none] (38) at (1.125, -0.825) {};
		\node [style=none] (39) at (1.125, -1.675) {};
		\node [style=none] (40) at (1.5, -1.25) {\scriptsize $l$};
		\node [style=none] (57) at (-0.825, 0.25) {};
		\node [style=none] (58) at (-0.825, -0.5) {};
		\node [style=none] (64) at (-0.45, 0) {$\vdots$};
		\node [style=none] (70) at (-0.425, 1.375) {$W_n$};
		\node [style=none] (71) at (-0.825, 0.625) {};
		\node [style=none] (72) at (-0.825, 2.125) {};
		\node [style=none] (73) at (-1, 1.5) {$\vdots$};
		\node [style=none] (76) at (-0.825, 1.825) {};
		\node [style=none] (77) at (-0.825, 0.975) {};
		\node [style=none] (78) at (-0.05, 0.875) {};
		\node [style=none] (79) at (-0.05, 1.875) {};
		\node [style=none] (80) at (0.125, 1.475) {$\vdots$};
		\node [style=none] (81) at (-0.05, 1.7) {};
		\node [style=none] (82) at (-0.05, 1.05) {};
		\node [style=none] (83) at (0.675, 1.375) {$W_n^\dagger$};
		\node [style=none] (84) at (1.075, 0.625) {};
		\node [style=none] (85) at (1.075, 2.125) {};
		\node [style=none] (86) at (1.25, 1.5) {$\vdots$};
		\node [style=none] (89) at (1.075, 1.825) {};
		\node [style=none] (90) at (1.075, 0.975) {};
		\node [style=none] (91) at (0.3, 0.875) {};
		\node [style=none] (92) at (0.3, 1.875) {};
		\node [style=none] (93) at (0.3, 1.7) {};
		\node [style=none] (94) at (0.3, 1.05) {};
		\node [style=none] (95) at (1.075, 0.25) {};
		\node [style=none] (96) at (1.075, -0.5) {};
	\end{pgfonlayer}
	\begin{pgfonlayer}{edgelayer}
		\draw (8.center) to (7.center);
		\draw (8.center) to (0.center);
		\draw (0.center) to (2.center);
		\draw (2.center) to (7.center);
		\draw (26.center) to (28.center);
		\draw (25.center) to (27.center);
		\draw (30.center) to (31.center);
		\draw (32.center) to (33.center);
		\draw [style=braceedge] (36.center) to (35.center);
		\draw [style=braceedge] (38.center) to (39.center);
		\draw (72.center) to (71.center);
		\draw (79.center) to (78.center);
		\draw (78.center) to (71.center);
		\draw (72.center) to (79.center);
		\draw (85.center) to (84.center);
		\draw (92.center) to (91.center);
		\draw (91.center) to (84.center);
		\draw (85.center) to (92.center);
		\draw (81.center) to (93.center);
		\draw (94.center) to (82.center);
		\draw (22.center) to (96.center);
		\draw (95.center) to (6.center);
		\draw [bend right=90, looseness=2.00] (95.center) to (90.center);
		\draw [bend left=90, looseness=1.75] (89.center) to (96.center);
		\draw (57.center) to (1.center);
		\draw (58.center) to (19.center);
		\draw [bend right=270, looseness=1.75] (57.center) to (77.center);
		\draw [bend left=90, looseness=1.50] (58.center) to (76.center);
	\end{pgfonlayer}
\end{tikzpicture}
}%
	\endpgfgraphicnamed

	\]
\end{theorem}
\begin{proof}
	Immediately follows from Lemma \ref{lem:circsum} and Corollary \ref{cor:psplit}. 
\end{proof}

Similar to the differentiation diagrams from the previous section, the size of this integration diagram is linear in $n$ (follows from Proposition \ref{prop:wsize}).

\subsection{General Integration}

Now, we can easily extend the integration result from circuits to arbitrary ZX diagrams with a different number of positive and negative occurrences of $\theta$.

\begin{theorem}
	Let $k$ be a non-zero integer, $M$ a diagram with no occurrence of $\theta$, and w.l.o.g. $m \leq n$. Then
	\[
		\frac{1}{2\pi} \int_{-\pi}^\pi %
	\beginpgfgraphicnamed{integrate/general-int/lhs}
	\begin{tikzpicture}
	\begin{pgfonlayer}{nodelayer}
		\node [style=none] (0) at (0.5, 1.25) {};
		\node [style=none] (1) at (-0.25, 1) {};
		\node [style=none] (2) at (0.5, -1.375) {};
		\node [style={gn_phase}, scale=1] (3) at (1.075, 1) {$-k\theta$};
		\node [style={gn_phase}, scale=1] (4) at (-0.75, 1) {$k\theta$};
		\node [style=none] (6) at (0.125, -0.05) {$M$};
		\node [style=none] (8) at (0.5, 1) {};
		\node [style=none] (9) at (-0.25, -1.375) {};
		\node [style=none] (10) at (-0.25, 1.25) {};
		\node [style=none] (12) at (-0.625, -1.125) {};
		\node [style=none] (13) at (-0.25, -0.375) {};
		\node [style=none] (14) at (-0.25, -1.125) {};
		\node [style=none] (15) at (-0.625, -0.375) {};
		\node [style=none] (16) at (-0.425, -0.625) {$\vdots$};
		\node [style=none] (18) at (0.875, -0.375) {};
		\node [style=none] (20) at (0.5, -0.375) {};
		\node [style=none] (21) at (0.5, -1.125) {};
		\node [style=none] (23) at (0.875, -1.125) {};
		\node [style=none] (24) at (0.675, -0.625) {$\vdots$};
		\node [style=none] (25) at (-0.25, 0.25) {};
		\node [style={gn_phase}, scale=1] (26) at (1.075, 0.25) {$-k\theta$};
		\node [style={gn_phase}, scale=1] (27) at (-0.75, 0.25) {$k\theta$};
		\node [style=none] (30) at (0.5, 0.25) {};
		\node [style=none] (31) at (-0.425, 0.75) {$\vdots$};
		\node [style=none] (32) at (0.675, 0.75) {$\vdots$};
		\node [style=none] (38) at (-1.2, 1.175) {};
		\node [style=none] (39) at (-1.2, 0.075) {};
		\node [style=none] (40) at (-1.45, 0.625) {\scriptsize $n$};
		\node [style=none] (41) at (-0.875, -0.325) {};
		\node [style=none] (42) at (-0.875, -1.175) {};
		\node [style=none] (43) at (-1.175, -0.775) {\scriptsize $l$};
		\node [style=none] (44) at (1.625, 1.175) {};
		\node [style=none] (45) at (1.625, 0.075) {};
		\node [style=none] (46) at (2.25, 0.625) {\scriptsize $m \leq n$};
		\node [style=none] (47) at (1.125, -0.325) {};
		\node [style=none] (48) at (1.125, -1.175) {};
		\node [style=none] (49) at (1.375, -0.75) {\scriptsize $p$};
	\end{pgfonlayer}
	\begin{pgfonlayer}{edgelayer}
		\draw (10.center) to (9.center);
		\draw (10.center) to (0.center);
		\draw (0.center) to (2.center);
		\draw (2.center) to (9.center);
		\draw (13.center) to (15.center);
		\draw (12.center) to (14.center);
		\draw (18.center) to (20.center);
		\draw (21.center) to (23.center);
		\draw (3) to (8.center);
		\draw (4) to (1.center);
		\draw (27) to (25.center);
		\draw (26) to (30.center);
		\draw [style=braceedge] (39.center) to (38.center);
		\draw [style=braceedge] (42.center) to (41.center);
		\draw [style=braceedge] (44.center) to (45.center);
		\draw [style=braceedge] (47.center) to (48.center);
	\end{pgfonlayer}
\end{tikzpicture}
}%
	\endpgfgraphicnamed
 d\theta 
		~= %
	\beginpgfgraphicnamed{integrate/general-int/rhs}
	\begin{tikzpicture}
	\begin{pgfonlayer}{nodelayer}
		\node [style=none] (0) at (0.475, 0.5) {};
		\node [style=none] (1) at (-0.25, 0.25) {};
		\node [style=none] (2) at (0.5, -1.875) {};
		\node [style=none] (5) at (0.125, -0.675) {$M$};
		\node [style=none] (6) at (0.5, 0.25) {};
		\node [style=none] (7) at (-0.25, -1.875) {};
		\node [style=none] (8) at (-0.25, 0.5) {};
		\node [style=none] (19) at (-0.25, -0.5) {};
		\node [style=none] (22) at (0.5, -0.5) {};
		\node [style=none] (23) at (-1.075, 1.625) {$\vdots$};
		\node [style=none] (24) at (0.7, 0) {$\vdots$};
		\node [style=none] (25) at (-0.625, -1.625) {};
		\node [style=none] (26) at (-0.25, -0.875) {};
		\node [style=none] (27) at (-0.25, -1.625) {};
		\node [style=none] (28) at (-0.625, -0.875) {};
		\node [style=none] (29) at (-0.45, -1.125) {$\vdots$};
		\node [style=none] (30) at (0.875, -0.875) {};
		\node [style=none] (31) at (0.5, -0.875) {};
		\node [style=none] (32) at (0.5, -1.625) {};
		\node [style=none] (33) at (0.875, -1.625) {};
		\node [style=none] (34) at (0.7, -1.125) {$\vdots$};
		\node [style=none] (35) at (-0.875, -0.825) {};
		\node [style=none] (36) at (-0.875, -1.675) {};
		\node [style=none] (37) at (-1.15, -1.25) {\scriptsize $l$};
		\node [style=none] (38) at (1.125, -0.825) {};
		\node [style=none] (39) at (1.125, -1.675) {};
		\node [style=none] (40) at (1.45, -1.275) {\scriptsize $p$};
		\node [style=none] (41) at (-0.475, 1.5) {$W_n$};
		\node [style=none] (42) at (-0.875, 0.625) {};
		\node [style=none] (43) at (-0.875, 2.375) {};
		\node [style=none] (44) at (1.075, 0.8) {};
		\node [style=none] (45) at (1.075, 1.35) {};
		\node [style=none] (46) at (-0.1, 1.05) {};
		\node [style=none] (47) at (-0.1, 1.95) {};
		\node [style=none] (50) at (0.675, 1.5) {$W_n^\dagger$};
		\node [style=none] (51) at (1.075, 0.625) {};
		\node [style=none] (52) at (1.075, 2.375) {};
		\node [style=none] (53) at (-0.875, 2.125) {};
		\node [style=none] (54) at (-0.875, 0.875) {};
		\node [style=none] (55) at (0.3, 1.05) {};
		\node [style=none] (56) at (0.3, 1.95) {};
		\node [style=none] (63) at (0.1, 1.625) {$\vdots$};
		\node [style=none] (66) at (-0.1, 1.825) {};
		\node [style=none] (67) at (-0.1, 1.175) {};
		\node [style=none] (68) at (0.3, 1.825) {};
		\node [style=none] (69) at (0.3, 1.175) {};
		\node [style=rn] (70) at (1.4, 1.6) {};
		\node [style=none] (71) at (1.075, 1.6) {};
		\node [style=rn] (72) at (1.4, 2.15) {};
		\node [style=none] (73) at (1.075, 2.15) {};
		\node [style=none] (74) at (1.225, 1.975) {$\vdots$};
		\node [style=none] (75) at (1.675, 1.475) {};
		\node [style=none] (76) at (1.675, 2.275) {};
		\node [style=none] (77) at (2.275, 1.875) {\scriptsize $n-m$};
		\node [style=none] (78) at (-0.875, 0.25) {};
		\node [style=none] (79) at (-0.875, -0.5) {};
		\node [style=none] (80) at (-0.45, 0) {$\vdots$};
		\node [style=none] (81) at (1.075, 0.25) {};
		\node [style=none] (82) at (1.075, -0.5) {};
		\node [style=none] (83) at (1.225, 1.15) {$\vdots$};
	\end{pgfonlayer}
	\begin{pgfonlayer}{edgelayer}
		\draw (8.center) to (7.center);
		\draw (8.center) to (0.center);
		\draw (0.center) to (2.center);
		\draw (2.center) to (7.center);
		\draw (26.center) to (28.center);
		\draw (25.center) to (27.center);
		\draw (30.center) to (31.center);
		\draw (32.center) to (33.center);
		\draw [style=braceedge] (36.center) to (35.center);
		\draw [style=braceedge] (38.center) to (39.center);
		\draw (43.center) to (42.center);
		\draw (47.center) to (46.center);
		\draw (46.center) to (42.center);
		\draw (43.center) to (47.center);
		\draw (52.center) to (51.center);
		\draw (56.center) to (55.center);
		\draw (55.center) to (51.center);
		\draw (52.center) to (56.center);
		\draw (71.center) to (70);
		\draw (73.center) to (72);
		\draw [style=braceedge] (76.center) to (75.center);
		\draw (66.center) to (68.center);
		\draw (69.center) to (67.center);
		\draw (78.center) to (1.center);
		\draw (19.center) to (79.center);
		\draw [bend right=270, looseness=1.75] (78.center) to (54.center);
		\draw [bend left=90, looseness=1.25] (79.center) to (53.center);
		\draw [bend right=90, looseness=1.50] (81.center) to (44.center);
		\draw [bend left=90, looseness=1.75] (45.center) to (82.center);
		\draw (81.center) to (6.center);
		\draw (22.center) to (82.center);
	\end{pgfonlayer}
\end{tikzpicture}
}%
	\endpgfgraphicnamed

	\]
\end{theorem}
\begin{proof}
	We can rewrite the diagram into one with the same number of positive and negative occurrences of $\theta$ by exploiting the fact that $	%
	\beginpgfgraphicnamed{integrate/general-int/id1}
	\begin{tikzpicture}
	\begin{pgfonlayer}{nodelayer}
		\node [style={gn_phase}] (0) at (0, 0) {$k\theta$};
		\node [style=rn] (1) at (0.625, 0) {};
	\end{pgfonlayer}
	\begin{pgfonlayer}{edgelayer}
		\draw (0) to (1);
	\end{pgfonlayer}
\end{tikzpicture}
}%
	\endpgfgraphicnamed
 = %
	\beginpgfgraphicnamed{integrate/general-int/id2}
	\begin{tikzpicture}
	\begin{pgfonlayer}{nodelayer}
		\node [style={gn_phase}] (0) at (0, 0) {$-k\theta$};
		\node [style=rn] (1) at (0.75, 0) {};
	\end{pgfonlayer}
	\begin{pgfonlayer}{edgelayer}
		\draw (0) to (1);
	\end{pgfonlayer}
\end{tikzpicture}
}%
	\endpgfgraphicnamed
 = 1$.
	\[ %
	\beginpgfgraphicnamed{integrate/general-int/lhs}
	}%
	\endpgfgraphicnamed
 ~=~ %
	\beginpgfgraphicnamed{integrate/general-int/middle}
	\begin{tikzpicture}
	\begin{pgfonlayer}{nodelayer}
		\node [style=none] (0) at (0.5, 1.75) {};
		\node [style=none] (1) at (-0.25, 1.5) {};
		\node [style=none] (2) at (0.5, -2.125) {};
		\node [style={gn_phase}, scale=1] (3) at (1.45, 1.5) {$-k\theta$};
		\node [style={gn_phase}, scale=1] (4) at (-0.75, 1.5) {$k\theta$};
		\node [style=none] (6) at (0.125, -0.15) {$M$};
		\node [style=none] (8) at (0.5, 1.5) {};
		\node [style=none] (9) at (-0.25, -2.125) {};
		\node [style=none] (10) at (-0.25, 1.75) {};
		\node [style=none] (12) at (-0.625, -1.875) {};
		\node [style=none] (13) at (-0.25, -1.125) {};
		\node [style=none] (14) at (-0.25, -1.875) {};
		\node [style=none] (15) at (-0.625, -1.125) {};
		\node [style=none] (16) at (-0.425, -1.375) {$\vdots$};
		\node [style=none] (18) at (0.875, -1.125) {};
		\node [style=none] (20) at (0.5, -1.125) {};
		\node [style=none] (21) at (0.5, -1.875) {};
		\node [style=none] (23) at (0.875, -1.875) {};
		\node [style=none] (24) at (0.675, -1.375) {$\vdots$};
		\node [style=none] (25) at (-0.25, -0.5) {};
		\node [style={gn_phase}, scale=1] (26) at (1.45, 0.75) {$-k\theta$};
		\node [style={gn_phase}, scale=1] (27) at (-0.75, -0.5) {$k\theta$};
		\node [style=none] (30) at (0.5, 0.75) {};
		\node [style=none] (31) at (-0.55, 0.625) {$\vdots$};
		\node [style=none] (32) at (0.975, -0.05) {$\vdots$};
		\node [style=none] (38) at (-1.2, 1.675) {};
		\node [style=none] (39) at (-1.2, -0.675) {};
		\node [style=none] (40) at (-1.45, 0.5) {\scriptsize $n$};
		\node [style=none] (41) at (-0.875, -1.075) {};
		\node [style=none] (42) at (-0.875, -1.925) {};
		\node [style=none] (43) at (-1.125, -1.5) {\scriptsize $l$};
		\node [style=none] (44) at (2, 1.675) {};
		\node [style=none] (45) at (2, 0.575) {};
		\node [style=none] (46) at (2.3, 1.125) {\scriptsize $m$};
		\node [style=none] (47) at (1.125, -1.075) {};
		\node [style=none] (48) at (1.125, -1.925) {};
		\node [style=none] (49) at (1.375, -1.5) {\scriptsize $p$};
		\node [style={gn_phase}, scale=1] (50) at (1.45, 0.2) {$-k\theta$};
		\node [style=rn] (51) at (0.75, 0.2) {};
		\node [style={gn_phase}, scale=1] (52) at (1.45, -0.55) {$-k\theta$};
		\node [style=rn] (53) at (0.75, -0.55) {};
		\node [style=none] (54) at (0.975, 1.25) {$\vdots$};
		\node [style=none] (55) at (2, 0.375) {};
		\node [style=none] (56) at (2, -0.725) {};
		\node [style=none] (57) at (2.575, -0.175) {\scriptsize $n-m$};
	\end{pgfonlayer}
	\begin{pgfonlayer}{edgelayer}
		\draw (10.center) to (9.center);
		\draw (10.center) to (0.center);
		\draw (0.center) to (2.center);
		\draw (2.center) to (9.center);
		\draw (13.center) to (15.center);
		\draw (12.center) to (14.center);
		\draw (18.center) to (20.center);
		\draw (21.center) to (23.center);
		\draw (3) to (8.center);
		\draw (4) to (1.center);
		\draw (27) to (25.center);
		\draw (26) to (30.center);
		\draw [style=braceedge] (39.center) to (38.center);
		\draw [style=braceedge] (42.center) to (41.center);
		\draw [style=braceedge] (44.center) to (45.center);
		\draw [style=braceedge] (47.center) to (48.center);
		\draw (50) to (51);
		\draw (52) to (53);
		\draw [style=braceedge] (55.center) to (56.center);
	\end{pgfonlayer}
\end{tikzpicture}
}%
	\endpgfgraphicnamed
 \]
	Then, the result immediately follows from Theorem \ref{thm:circ-int}.
\end{proof}

\subsection{Examples}

To conclude, we explicitly give the integration diagrams for the cases $n=1,2,3$ to illustrate and compare our integration approach with the work in  \cite{Zhao2021analyzingbarren}.

\begin{example}\label{1foldintglm}
  Let $k$ be a non-zero integer and $M$ a diagram with no occurrence of $\theta$. Then
  $$
  	\frac{1}{2\pi}\int_{-\pi}^{\pi} %
	\beginpgfgraphicnamed{1foldintgv2}
	\InputIfFileExists{1foldintgv2.tikz}{}{\input{./figures/1foldintgv2.tikz}}%
	\endpgfgraphicnamed
 ~d\alpha 
  	~\overset{\ref{thm:circ-int}}{=}~ %
	\beginpgfgraphicnamed{1foldintg2v2}
	\InputIfFileExists{1foldintg2v2.tikz}{}{\input{./figures/1foldintg2v2.tikz}}%
	\endpgfgraphicnamed

  $$ 
  This equation has been proved in \cite{Zhao2021analyzingbarren} for $k=1$.
  Here we obtain the result directly from Theorem \ref{thm:circ-int} using the fact that $W_1$ is the identity.
\end{example}

\begin{example}
	\label{2varibleintthm}
	Let $k$ be a non-zero integer and $M$ a diagram with no occurrence of $\theta$. Then
	$$
		\frac{1}{2\pi}\bigintsss_{-\pi}^{\pi}%
	\beginpgfgraphicnamed{anylinearmaplaphaintv2}
	\InputIfFileExists{anylinearmaplaphaintv2.tikz}{}{\input{./figures/anylinearmaplaphaintv2.tikz}}%
	\endpgfgraphicnamed
 d\theta
		~\overset{\ref{thm:circ-int}}{=} %
	\beginpgfgraphicnamed{2varibleintv4}
	\InputIfFileExists{2varibleintv4.tikz}{}{\input{./figures/2varibleintv4.tikz}}%
	\endpgfgraphicnamed
~
		~\overset{\ref{int2simp}}{=} %
	\beginpgfgraphicnamed{2varibleintv3}
	\InputIfFileExists{2varibleintv3.tikz}{}{\input{./figures/2varibleintv3.tikz}}%
	\endpgfgraphicnamed

	$$
	The corresponding result of this theorem is shown as Lemma 2 in  \cite{Zhao2021analyzingbarren} where there are three diagrammatic sum terms after integration, which results in their computation of variance of gradients becoming rather complicated. Here we only obtain a single diagram after integration. 
\end{example}

\begin{example} \label{3varibleintthm}
	Let $k$ be a non-zero integer and $M$ a diagram with no occurrence of $\theta$. Then
	$$
		\frac{1}{2\pi}\bigintsss_{-\pi}^{\pi}%
	\beginpgfgraphicnamed{anyzxrmapl3aphaint}
	\InputIfFileExists{anyzxrmapl3aphaint.tikz}{}{\input{./figures/anyzxrmapl3aphaint.tikz}}%
	\endpgfgraphicnamed
 d\theta 
		~\overset{\ref{thm:circ-int}}{=}~ %
	\beginpgfgraphicnamed{3varibleint}
	\InputIfFileExists{3varibleint.tikz}{}{\input{./figures/3varibleint.tikz}}%
	\endpgfgraphicnamed

	$$
	Zhao and Gao only considered diagrams with up to 2 occurrences of $\pm\theta$, so an equivalent result does not exist in \cite{Zhao2021analyzingbarren}.
\end{example}

\section{Example Application: Quantum Machine Learning}\label{sec:qml}

In the NISQ era of quantum computing \cite{Preskill2018quantumcomputingin}, many applications require the optimisation of parameterised quantum circuits: in quantum chemistry, variational quantum eigensolvers \cite{vqe2019} are optimised to find the ground state of a Hamiltonian; in quantum machine learning, a circuit ansatz is optimised against a cost function \cite{lewenstein1994quantum}, much alike how neural networks are optimised in classical machine learning. 

However, while the approach of using gradient-based methods to optimise deep neural networks has been consistently effective \cite{choromanska2015loss}, gradient-based optimisation of parameterised quantum circuits often suffer from barren plateaus: the training landscape of many circuit ans\"{a}tze have been shown to be exponentially flat with respect to circuit size, making gradient descent impossible~\cite{mcclean2018barren}. Therefore, it is crucial to develop techniques to detect and avoid barren plateaus.

So far, there has not been a fully diagrammatic analysis of barren plateaus using ZX-calculus. We believe the main obstacle to the analysis is the lack of techniques for manipulating sums of diagrams: an expectation of a Hamiltonian contains at least two occurrences of each circuit parameter, so the derivative of the expectation with respect to that parameter requires the product rule. Similarly, the rule for integrating diagrams in \cite{Zhao2021analyzingbarren} introduces three terms, after $n$ integrals there would be $3^n$ terms. Using these rules, the analysis of barren plateaus becomes exponentially costly, as the number of diagrams to be evaluated is exponential to the number of parameters in the circuit.

In this section, as a demonstration of the new differentiation and integration techniques of this paper,  we show that the analysis of barren plateaus by Zhao et al.~\cite{Zhao2021analyzingbarren} can be done entirely within the framework of ZX without introducing sums of diagrams.
Using the same setup, we consider an $n$-qubit parameterised quantum circuit $U(\bm{\theta})$ and a Hamiltonian $H$, obtaining the expectation value $\braket{H} =\bra{0}U^{\dagger}(\bm{\theta})HU(\bm{\theta})\ket{0}$. If $\mathbf{Var}\left(\frac{ \partial \braket{H}}{\partial\theta_j}\right) \approx 0$ for all $j$, then the quantum circuit is likely to start in a barren plateau where the circuit gradients $\frac{ \partial \braket{H}}{\partial\theta_j}$ are close to 0.

\subsection{Method}

Similar to Zhao et al.~\cite{Zhao2021analyzingbarren}, we make some assumptions on the circuit $U(\bm{\theta})$:
\begin{enumerate}
	\item The parameterised gates in $U$ are $R_X$, $R_Y$, or $R_Z$ and do not share parameters.
	\item The parameters $\bm\theta$ are independently and uniformly distributed on the interval $[\pi, -\pi]$.
\end{enumerate}
This allows us to represent $U(\bm{\theta})$ as a ZX diagram where each parameter $\theta_j$ only occurs once.
Thus, we can use the following notation for the expectation value $\braket{H} =\bra{0}U^{\dagger}(\bm{\theta})HU(\bm{\theta})\ket{0}$ where the parametrised spiders are fused out.

\begin{equation}\label{expectationgeneq}
	\braket{H}=%
	\beginpgfgraphicnamed{expectationgen}
	\InputIfFileExists{expectationgen.tikz}{}{\input{./figures/expectationgen.tikz}}%
	\endpgfgraphicnamed

\end{equation}

\begin{restatable}{lemma}{onefoldexpectintglm}\label{1foldexpectintglm}
\cite{Zhao2021analyzingbarren} Given $\braket{H}$ in the form of (\ref{expectationgeneq}), we have
$\mathbf{E}\left(\frac{ \partial \braket{H}}{\partial\theta_j}\right)=0$, for $j=1, \ldots, m$.
\end{restatable}
As a consequence, we have
\begin{align*}
\mathbf{Var}\left(\frac{ \partial \braket{H}}{\partial\theta_j}\right)
&= \mathbf{E}\left(\left(\frac{ \partial \braket{H}}{\partial\theta_j}\right)^2\right) \\
&= \frac{1}{(2\pi)^m}\int_{-\pi}^{\pi} \cdots\int_{-\pi}^{\pi}  \left(\frac{ \partial \braket{H}}{\partial\theta_j}\right)^2d\theta_1\cdots d\theta_m, j=1, \cdots, m.
\end{align*}
Note that we can use squares here since the expectation value $\langle H\rangle$ is a real variable.

\begin{restatable}{lemma}{onefoldexpectsquareintglm}
\label{1foldexpectsquareintglm}
\cite{Zhao2021analyzingbarren}
Given $\braket{H}$ in the form of (\ref{expectationgeneq}), we have
\[
\frac{1}{2\pi}\int_{-\pi}^{\pi}  \left(\frac{ \partial \braket{H}}{\partial\theta_j}\right)^2d\theta_j= %
	\beginpgfgraphicnamed{1foldexpectsquareintg}
	\InputIfFileExists{1foldexpectsquareintg.tikz}{}{\input{./figures/1foldexpectsquareintg.tikz}}%
	\endpgfgraphicnamed

\]
where the cycle connects the legs of $E_H$ that correspond to the positions of the $\pm\theta_j$ spiders in (\ref{expectationgeneq}).
\end{restatable}
\begin{remark}
An equivalent version of this lemma is given on page 12 in \cite{Zhao2021analyzingbarren}.
Here, we obtain the result using our general differentiation and integration method.
\end{remark}

\begin{theorem} \label{variancegenthm}
Given  $\braket{H}$ in the form of (\ref{expectationgeneq}), we have
\[
\mathbf{Var}\left(\frac{ \partial \braket{H}}{\partial\theta_j}\right)\quad=\quad %
	\beginpgfgraphicnamed{variancegen}
	\InputIfFileExists{variancegen.tikz}{}{\input{./figures/variancegen.tikz}}%
	\endpgfgraphicnamed

\]
\end{theorem}
\begin{proof}
We start from the diagram as shown in Lemma \ref{1foldexpectsquareintglm}, then drag out variables iteratively according to Example \ref{2varibleintthm}, and the result follows. 
\end{proof}

\begin{remark}
The variance computed in \cite{Zhao2021analyzingbarren} is based on a sum over $3^{m-1}$ terms ($m$ is the number of parameters in the considered circuit), so when $m$ is large it becomes infeasible to analyse the variance purely within ZX. Thus they have to resort to tensor networks which goes beyond the ZX method. In contrast, we avoid this exponential explosion by integrating without sums using algebraic ZX-calculus. 
\end{remark}

\subsection{Example} \label{sec:barren-example}

Finally, we demonstrate Theorem \ref{variancegenthm} by analysing an example ansatz for the existence of barren plateaus.
Concretely, we study the following $n$-qubit hardware efficient ansatz from~\cite{sim2019expressibility}, which to the best of our knowledge has not been analysed before.
\[ 
	U(\bm\theta) :=~ %
	\beginpgfgraphicnamed{sim9/ansatz}
	\InputIfFileExists{sim9/ansatz.tikz}{}{\input{./figures/sim9/ansatz.tikz}}%
	\endpgfgraphicnamed
 ~=~ %
	\beginpgfgraphicnamed{sim9/ansatz-zx}
	\InputIfFileExists{sim9/ansatz-zx.tikz}{}{\input{./figures/sim9/ansatz-zx.tikz}}%
	\endpgfgraphicnamed
 ~=~ %
	\beginpgfgraphicnamed{sim9/ansatz-zx2}
	\InputIfFileExists{sim9/ansatz-zx2.tikz}{}{\input{./figures/sim9/ansatz-zx2.tikz}}%
	\endpgfgraphicnamed

\]
Here, we used the standard ZX representation of quantum gates, i.e.
\[
	H = %
	\beginpgfgraphicnamed{Hadamard}
	\begin{tikzpicture}
	\begin{pgfonlayer}{nodelayer}
		\node [style=none] (0) at (-0.5, 0) {};
		\node [style=none] (1) at (0.5, 0) {};
		\node [style=H box] (2) at (0, 0) {};
	\end{pgfonlayer}
	\begin{pgfonlayer}{edgelayer}
		\draw (0.center) to (1.center);
	\end{pgfonlayer}
\end{tikzpicture}}%
	\endpgfgraphicnamed
, \qquad R_X(\theta)=%
	\beginpgfgraphicnamed{xrotate}
	\begin{tikzpicture}
	\begin{pgfonlayer}{nodelayer}
		\node [style=none] (0) at (0.825, 0) {};
		\node [style=none] (1) at (-0.8, 0) {};
		\node [style={gn_phase}] (2) at (0, 0) {$\theta$};
		\node [style=H box] (3) at (0.5, 0) {};
		\node [style=H box] (4) at (-0.5, 0) {};
	\end{pgfonlayer}
	\begin{pgfonlayer}{edgelayer}
		\draw (1.center) to (0.center);
	\end{pgfonlayer}
\end{tikzpicture}
}%
	\endpgfgraphicnamed
, \qquad CZ= %
	\beginpgfgraphicnamed{cz}
	\InputIfFileExists{cz.tikz}{}{\input{./figures/cz.tikz}}%
	\endpgfgraphicnamed

\]
and we used blue dashed lines %
	\beginpgfgraphicnamed{bluedashed}
	\begin{tikzpicture}
	\begin{pgfonlayer}{nodelayer}
		\node [style=none] (0) at (0, 0) {};
		\node [style=none] (1) at (1, 0) {};
	\end{pgfonlayer}
	\begin{pgfonlayer}{edgelayer}
		\draw [style=hadamard edge] (0.center) to (1.center);
	\end{pgfonlayer}
\end{tikzpicture}
}%
	\endpgfgraphicnamed
 to denote wires %
	\beginpgfgraphicnamed{had}
	\begin{tikzpicture}
	\begin{pgfonlayer}{nodelayer}
		\node [style=none] (0) at (0, 0) {};
		\node [style=none] (1) at (1, 0) {};
		\node [style=H box] (2) at (0.5, 0) {};
	\end{pgfonlayer}
	\begin{pgfonlayer}{edgelayer}
		\draw (0.center) to (1.center);
	\end{pgfonlayer}
\end{tikzpicture}
}%
	\endpgfgraphicnamed
 with a Hadamard gate on them.
For this example, we will consider the Hamiltonian $H = Z^{\otimes n}$ which corresponds to measuring all qubits in the computational basis.
This gives us
\[
	\braket{H} =~ %
	\beginpgfgraphicnamed{sim9/expval-1}
	\InputIfFileExists{sim9/expval-1.tikz}{}{\input{./figures/sim9/expval-1.tikz}}%
	\endpgfgraphicnamed
 ~\overset{X}{=}~ %
	\beginpgfgraphicnamed{sim9/expval-2}
	\InputIfFileExists{sim9/expval-2.tikz}{}{\input{./figures/sim9/expval-2.tikz}}%
	\endpgfgraphicnamed
 ~=~ %
	\beginpgfgraphicnamed{sim9/expval-3}
	\InputIfFileExists{sim9/expval-3.tikz}{}{\input{./figures/sim9/expval-3.tikz}}%
	\endpgfgraphicnamed

\]
Now, we can use Theorem \ref{variancegenthm} to get a single ZX diagram that represents the variance of $\braket{H}$`s gradient w.r.t. some parameter $\theta_j$.
\begin{equation}\label{eqn:sim9var}
	\mathbf{Var}\left(\frac{ \partial \braket{H}}{\partial\theta_j}\right) ~\overset{\ref{variancegenthm}}{=}~~ %
	\beginpgfgraphicnamed{sim9/var}
	\InputIfFileExists{sim9/var.tikz}{}{\input{./figures/sim9/var.tikz}}%
	\endpgfgraphicnamed

\end{equation}

\begin{restatable}{lemma}{cycle} \label{lem:cycle}
	The cycles in diagram (\ref{eqn:sim9var}) can be broken up as follows.
	\[ %
	\beginpgfgraphicnamed{sim9/cycle2}
	\InputIfFileExists{sim9/cycle2.tikz}{}{\input{./figures/sim9/cycle2.tikz}}%
	\endpgfgraphicnamed
 \qquad\qquad %
	\beginpgfgraphicnamed{sim9/cycle1}
	\InputIfFileExists{sim9/cycle1.tikz}{}{\input{./figures/sim9/cycle1.tikz}}%
	\endpgfgraphicnamed
 \]
\end{restatable}

\begin{proposition}
	The ansatz $U(\bm\theta)$ suffers from barren plateaus for $H = Z^{\otimes n}$.
\end{proposition}
\begin{proof}
	Using Lemma \ref{lem:cycle}, we simplify the variance diagram (\ref{eqn:sim9var}) to
	\[
		%
	\beginpgfgraphicnamed{sim9/var-2}
	\InputIfFileExists{sim9/var-2.tikz}{}{\input{./figures/sim9/var-2.tikz}}%
	\endpgfgraphicnamed

		~\overset{\ref{gridsimp}}{=}~ %
	\beginpgfgraphicnamed{sim9/var-3}
	\InputIfFileExists{sim9/var-3.tikz}{}{\input{./figures/sim9/var-3.tikz}}%
	\endpgfgraphicnamed

		~\overset{\ref{gridsimp}}{=}~ %
	\beginpgfgraphicnamed{sim9/var-4}
	\InputIfFileExists{sim9/var-4.tikz}{}{\input{./figures/sim9/var-4.tikz}}%
	\endpgfgraphicnamed

		~\overset{\ref{gridsimp}}{=}~ ...
	\]
	The initial diagram contains $5n$ spiders connected by Hadamard edges.
	We repeatedly apply Lemma \ref{gridsimp}, each time reducing the number of spiders by 10 and collecting a scalar of $\frac{1}{4}$.
	Depending on the parity of $n$, we will end up with 5 or 10 spiders scaled by $\frac{1}{2^n}$.
	Since the final diagram size is constant in $n$, we can conclude that the overall scaling is in $O\left(\frac{1}{2^n}\right)$.
	Thus, $U(\bm\theta)$ suffers from barren plateaus.
\end{proof}

Our diagrammatic differentiation and integration approach allowed us to carry out this analysis purely using simple reasoning and rewriting on polynomially-sized ZX diagrams.
Compared to Zhao et al.~\cite{Zhao2021analyzingbarren}, we did not need to resort to general tensor networks or consider diagrams with an exponential number of terms or exponential size.

For more advanced examples of barren plateau detection using Theorem \ref{variancegenthm} we refer to \cite{koch2022quantum} where more ans\"atze from \cite{sim2019expressibility} as well as IQP ans\"atze \cite{shepherd2009temporally} are analysed using our approach.

\section{Conclusion and further work}
We have elevated ZX-calculus from a graphical language for algebraic calculations to a new graphical tool for analytical reasoning.
For example, it can now be used for tackling quantum optimisation problems and reasoning about quantum mechanics.
We believe these techniques will extend the applicability of ZX-calculus to more problems related to quantum computing.
With more work, ZX-calculus can become a general tool for graphical differential calculus.

There are many directions for future work:
\begin{enumerate}
  \item \textbf{Generalisation of the results to qudit and qufinite cases \cite{qwangqufinite}:}
  The ideas in this paper can be extended beyond qubits, giving us more diagrammatic analytical tools.
  \item \textbf{Indefinite integration:}
  		This paper only gives the definite integral for ZX diagrams between $\pm\pi$. Indefinite integration of arbitrary ZX diagrams would in a sense complete the analytical ZX-calculus.
  \item \textbf{Parameter shift rules:}
  		As pointed out in Corollary \ref{diffthm2tmcr}, we recovered a graphical version of the parameter shift rule from \cite{schuld2019evaluating}.
  		Our diagrammatic approach might prove useful to discover more general shift rules or other gradient computation methods.
  		See \cite{koch2022quantum} for some initial investigations in this area.
  \item \textbf{Numerical barren plateau detection:}
  		Since the diagram in Theorem \ref{variancegenthm} is mostly Clifford, we can use stabiliser decomposition methods \cite{kissinger2022simulating, kissinger2022classical} to evaluate it.
  		As shown in \cite{koch2023speedy}, this allows us to numerically detect barren plateaus more efficiently than previous sampling-based methods.
\end{enumerate}

  \bibliographystyle{eptcs}
\bibliography{generic}

\appendix

\section{Proofs and Lemmas}
In this appendix, we include all the lemmas with their proofs which have been essentially existed (up to scalars) in previous papers. The lemmas are given in the order which they appear in this paper.

\piwtopicap*
\begin{proof}
\[  %
	\beginpgfgraphicnamed{piwtopicapprf}
	\InputIfFileExists{piwtopicapprf.tikz}{}{\input{./figures/piwtopicapprf.tikz}}%
	\endpgfgraphicnamed
 \]
\end{proof}

\dgngen*
\begin{proof}
\begin{align*}
  \frac{\partial}{\partial\theta}\left[%
	\beginpgfgraphicnamed{functiongn}
	\begin{tikzpicture}
	\begin{pgfonlayer}{nodelayer}
		\node [style=none] (0) at (0, -0.5) {};
		\node [style={gn_phase}] (1) at (0, 0.25) {${f(\theta)}$  };
	\end{pgfonlayer}
	\begin{pgfonlayer}{edgelayer}
		\draw (1) to (0.center);
	\end{pgfonlayer}
\end{tikzpicture}
}%
	\endpgfgraphicnamed
\right] &= 
  \frac{\partial}{\partial\theta}\left[\ket{0} + e^{if(\theta)}\ket{1}\right]
  =  if^{\prime}(\theta)e^{if(\theta)}\ket{1}\\
  &=\quad%
	\beginpgfgraphicnamed{functiongn1}
	\InputIfFileExists{functiongn1.tikz}{}{\input{./figures/functiongn1.tikz}}%
	\endpgfgraphicnamed
 \overset{B3}{=} %
	\beginpgfgraphicnamed{functiongn2}
	\InputIfFileExists{functiongn2.tikz}{}{\input{./figures/functiongn2.tikz}}%
	\endpgfgraphicnamed

\end{align*}
\end{proof}

\rgboxtrianglelm*
\begin{proof}
\[
	\beginpgfgraphicnamed{rgboxtriangleprf}
	\InputIfFileExists{rgboxtriangleprf.tikz}{}{\input{./figures/rgboxtriangleprf.tikz}}%
	\endpgfgraphicnamed

\]
\end{proof}

\dgboxsingle*
\begin{proof}
\begin{align*}
  \frac{\partial}{\partial \theta}\left[%
	\beginpgfgraphicnamed{functiongboxsingle}
	\begin{tikzpicture}
	\begin{pgfonlayer}{nodelayer}
		\node [style=gbox] (0) at (0, 0.25) {${f(\theta)}$  };
		\node [style=none] (1) at (0, -0.5) {};
	\end{pgfonlayer}
	\begin{pgfonlayer}{edgelayer}
		\draw (0) to (1.center);
	\end{pgfonlayer}
\end{tikzpicture}}%
	\endpgfgraphicnamed
\right] &= 
  \frac{\partial}{\partial \theta}\left[\ket{0} +f(\theta)\ket{1}\right]
  =  f^{\prime}(\theta)\ket{1}\\
  &=\quad%
	\beginpgfgraphicnamed{functiongbox1}
	\InputIfFileExists{functiongbox1.tikz}{}{\input{./figures/functiongbox1.tikz}}%
	\endpgfgraphicnamed
\end{align*}
\end{proof}

\begin{lemma}\label{zerobox}
	\[ %
	\beginpgfgraphicnamed{zerobox}
	\InputIfFileExists{zerobox.tikz}{}{\input{./figures/zerobox.tikz}}%
	\endpgfgraphicnamed
 \]
\end{lemma}
\begin{proof}
	\[ %
	\beginpgfgraphicnamed{zeroboxprf}
	\InputIfFileExists{zeroboxprf.tikz}{}{\input{./figures/zeroboxprf.tikz}}%
	\endpgfgraphicnamed
 \]
\end{proof}

\begin{lemma} \label{rdecomp}
	\[ %
	\beginpgfgraphicnamed{rdecomp}
	\InputIfFileExists{rdecomp.tikz}{}{\input{./figures/rdecomp.tikz}}%
	\endpgfgraphicnamed
 \]
\end{lemma}
\begin{proof}
	\[ %
	\beginpgfgraphicnamed{rdecompprf}
	\InputIfFileExists{rdecompprf.tikz}{}{\input{./figures/rdecompprf.tikz}}%
	\endpgfgraphicnamed
 \]
\end{proof}

\begin{lemma}\label{bas0t}
	\[ %
	\beginpgfgraphicnamed{bas0t}
	\InputIfFileExists{bas0t.tikz}{}{\input{./figures/bas0t.tikz}}%
	\endpgfgraphicnamed
 \]
\end{lemma}
\begin{proof}
	\[ %
	\beginpgfgraphicnamed{bas0tprf}
	\InputIfFileExists{bas0tprf.tikz}{}{\input{./figures/bas0tprf.tikz}}%
	\endpgfgraphicnamed
 \]
\end{proof}

\begin{lemma}\label{bas1t}
	\[ %
	\beginpgfgraphicnamed{bas1t}
	\InputIfFileExists{bas1t.tikz}{}{\input{./figures/bas1t.tikz}}%
	\endpgfgraphicnamed
 \]
\end{lemma}
\begin{proof}
	\[ %
	\beginpgfgraphicnamed{bas1tprf}
	\InputIfFileExists{bas1tprf.tikz}{}{\input{./figures/bas1tprf.tikz}}%
	\endpgfgraphicnamed
 \]
\end{proof}

\begin{lemma} \label{trihopf}
	\[ %
	\beginpgfgraphicnamed{trihopf}
	\InputIfFileExists{trihopf.tikz}{}{\input{./figures/trihopf.tikz}}%
	\endpgfgraphicnamed
 \]
\end{lemma}
\begin{proof}
	\[ %
	\beginpgfgraphicnamed{trihopfprf}
	\InputIfFileExists{trihopfprf.tikz}{}{\input{./figures/trihopfprf.tikz}}%
	\endpgfgraphicnamed
 \]
\end{proof}

\begin{lemma} \label{trihopf2}
	\[ %
	\beginpgfgraphicnamed{trihopf2}
	\InputIfFileExists{trihopf2.tikz}{}{\input{./figures/trihopf2.tikz}}%
	\endpgfgraphicnamed
 \]
\end{lemma}
\begin{proof}
	\[ %
	\beginpgfgraphicnamed{trihopf2prf}
	\InputIfFileExists{trihopf2prf.tikz}{}{\input{./figures/trihopf2prf.tikz}}%
	\endpgfgraphicnamed
 \]
\end{proof}

\begin{lemma} \label{triloop}
	\[ %
	\beginpgfgraphicnamed{triloop}
	\InputIfFileExists{triloop.tikz}{}{\input{./figures/triloop.tikz}}%
	\endpgfgraphicnamed
 \]
\end{lemma}
\begin{proof}
	\[ %
	\beginpgfgraphicnamed{triloopprf}
	\InputIfFileExists{triloopprf.tikz}{}{\input{./figures/triloopprf.tikz}}%
	\endpgfgraphicnamed
 \]
\end{proof}

\begin{lemma} \label{cnotstable}
	\[ %
	\beginpgfgraphicnamed{cnotstable}
	\InputIfFileExists{cnotstable.tikz}{}{\input{./figures/cnotstable.tikz}}%
	\endpgfgraphicnamed
 \]
\end{lemma}
\begin{proof}
	\[ %
	\beginpgfgraphicnamed{cnotstableprf}
	\InputIfFileExists{cnotstableprf.tikz}{}{\input{./figures/cnotstableprf.tikz}}%
	\endpgfgraphicnamed
 \]
\end{proof}

\begin{lemma} \label{triloop2}
	\[ %
	\beginpgfgraphicnamed{triloop2}
	\InputIfFileExists{triloop2.tikz}{}{\input{./figures/triloop2.tikz}}%
	\endpgfgraphicnamed
 \]
\end{lemma}
\begin{proof}
	\[ %
	\beginpgfgraphicnamed{triloop2prf}
	\InputIfFileExists{triloop2prf.tikz}{}{\input{./figures/triloop2prf.tikz}}%
	\endpgfgraphicnamed
 \]
\end{proof}

\begin{lemma} \label{triloop3}
	\[ %
	\beginpgfgraphicnamed{triloop3}
	\InputIfFileExists{triloop3.tikz}{}{\input{./figures/triloop3.tikz}}%
	\endpgfgraphicnamed
 \]
\end{lemma}
\begin{proof}
	\[ %
	\beginpgfgraphicnamed{triloop3prf}
	\InputIfFileExists{triloop3prf.tikz}{}{\input{./figures/triloop3prf.tikz}}%
	\endpgfgraphicnamed
 \]
\end{proof}

\begin{lemma} \label{wplug}
	\[ %
	\beginpgfgraphicnamed{wplug}
	\InputIfFileExists{wplug.tikz}{}{\input{./figures/wplug.tikz}}%
	\endpgfgraphicnamed
 \]
\end{lemma}
\begin{proof}
	\[ %
	\beginpgfgraphicnamed{wplugprf}
	\InputIfFileExists{wplugprf.tikz}{}{\input{./figures/wplugprf.tikz}}%
	\endpgfgraphicnamed
 \]
\end{proof}

\begin{lemma} \label{wplugrpi}
	\[ %
	\beginpgfgraphicnamed{wplugrpi}
	\InputIfFileExists{wplugrpi.tikz}{}{\input{./figures/wplugrpi.tikz}}%
	\endpgfgraphicnamed
 \]
\end{lemma}
\begin{proof}
	\[ %
	\beginpgfgraphicnamed{wplugrpiprf}
	\InputIfFileExists{wplugrpiprf.tikz}{}{\input{./figures/wplugrpiprf.tikz}}%
	\endpgfgraphicnamed
 \]
\end{proof}

\begin{lemma} \label{lem:W-join}
	\[ %
	\beginpgfgraphicnamed{integrate/W-join/stmt}
	\begin{tikzpicture}
	\begin{pgfonlayer}{nodelayer}
		\node [style=dbspider] (1) at (0, 0.375) {};
		\node [style=gn] (2) at (0, -0.375) {};
		\node [style=none] (3) at (0, 0.875) {};
		\node [style=none] (4) at (0, -0.75) {};
		\node [style=none] (5) at (0.75, 0) {$=$};
		\node [style=none] (6) at (1.5, 0.875) {};
		\node [style=none] (7) at (1.5, -0.75) {};
		\node [style=rn] (8) at (1.5, 0.5) {};
		\node [style=rn] (9) at (1.5, -0.375) {};
	\end{pgfonlayer}
	\begin{pgfonlayer}{edgelayer}
		\draw [bend left=45] (2) to (1);
		\draw [bend right=45] (2) to (1);
		\draw (3.center) to (1);
		\draw (4.center) to (2);
		\draw (9) to (7.center);
		\draw (8) to (6.center);
	\end{pgfonlayer}
\end{tikzpicture}
}%
	\endpgfgraphicnamed
 \]
\end{lemma}
\begin{proof}
	\[ %
	\beginpgfgraphicnamed{integrate/W-join/proof}
	\begin{tikzpicture}
	\begin{pgfonlayer}{nodelayer}
		\node [style=dbspider] (1) at (0.125, 0.375) {};
		\node [style=gn] (2) at (0.125, -0.375) {};
		\node [style=none] (3) at (0.125, 0.875) {};
		\node [style=none] (4) at (0.125, -0.75) {};
		\node [style=none] (5) at (0.875, 0) {$=$};
		\node [style=gn] (6) at (2, -0.75) {};
		\node [style=none] (7) at (2, -1.125) {};
		\node [style=gn] (8) at (2, -0.25) {};
		\node [style=rn] (9) at (1.5, 0.25) {};
		\node [style=triangle] (10) at (2, 0.25) {};
		\node [style=gn] (11) at (2, 0.75) {};
		\node [style=none] (12) at (2, 1.1) {};
		\node [style=none] (13) at (2.75, 0) {$=$};
		\node [style=none] (15) at (4, -0.625) {};
		\node [style=rn] (17) at (3.5, -0.25) {};
		\node [style=triangle] (18) at (4, -0.25) {};
		\node [style=gn] (19) at (4, 0.25) {};
		\node [style=none] (20) at (4, 0.6) {};
		\node [style=none] (21) at (4.625, 0) {$=$};
		\node [style=none] (22) at (5.25, -0.875) {};
		\node [style=rn] (23) at (5.25, 0) {};
		\node [style=triangle] (24) at (5.25, -0.5) {};
		\node [style=none] (26) at (5.25, 0.85) {};
		\node [style=rn] (27) at (5.25, 0.5) {};
		\node [style=none] (28) at (6, 0) {$=$};
		\node [style=none] (29) at (6.75, 0.875) {};
		\node [style=none] (30) at (6.75, -0.75) {};
		\node [style=rn] (31) at (6.75, 0.5) {};
		\node [style=rn] (32) at (6.75, -0.375) {};
		\node [style=none] (33) at (2.75, 0.25) {\scriptsize S1};
		\node [style=none] (34) at (2.75, 0.5) {\scriptsize Hopf};
		\node [style=none] (35) at (4.625, 0.25) {\scriptsize B1};
		\node [style=none] (36) at (6, 0.25) {\scriptsize Bas0};
	\end{pgfonlayer}
	\begin{pgfonlayer}{edgelayer}
		\draw [bend left=45] (2) to (1);
		\draw [bend right=45] (2) to (1);
		\draw (3.center) to (1);
		\draw (4.center) to (2);
		\draw (7.center) to (6);
		\draw (9) to (8);
		\draw (8) to (6);
		\draw (11) to (8);
		\draw (9) to (11);
		\draw (12.center) to (11);
		\draw (6) to (9);
		\draw (17) to (19);
		\draw (20.center) to (19);
		\draw (15.center) to (19);
		\draw (23) to (22.center);
		\draw (27) to (26.center);
		\draw (32) to (30.center);
		\draw (31) to (29.center);
	\end{pgfonlayer}
\end{tikzpicture}
}%
	\endpgfgraphicnamed
 \]
\end{proof}

\begin{lemma} \label{wsum}
	Let $a,b \in \mathbb C$. Then
	\[ %
	\beginpgfgraphicnamed{wsum}
	\InputIfFileExists{wsum.tikz}{}{\input{./figures/wsum.tikz}}%
	\endpgfgraphicnamed
 \]
\end{lemma}
\begin{proof}
	If $b = 0$, then
	\[ %
	\beginpgfgraphicnamed{wsumprf0}
	\InputIfFileExists{wsumprf0.tikz}{}{\input{./figures/wsumprf0.tikz}}%
	\endpgfgraphicnamed
 \]
	If $b \neq 0$, then
	\[ %
	\beginpgfgraphicnamed{wsumprf}
	\InputIfFileExists{wsumprf.tikz}{}{\input{./figures/wsumprf.tikz}}%
	\endpgfgraphicnamed
 \]
\end{proof}

\begin{lemma} \label{wtgloop}
	\[ %
	\beginpgfgraphicnamed{wtgloop}
	\InputIfFileExists{wtgloop.tikz}{}{\input{./figures/wtgloop.tikz}}%
	\endpgfgraphicnamed
 \]
\end{lemma}
\begin{proof}
	\[ %
	\beginpgfgraphicnamed{wtgloopprf}
	\InputIfFileExists{wtgloopprf.tikz}{}{\input{./figures/wtgloopprf.tikz}}%
	\endpgfgraphicnamed
 \]
	\end{proof}

\begin{lemma} \label{twtransform}
	Let $0 \neq x, y \in \mathbb C$. Then
	\[ %
	\beginpgfgraphicnamed{twtransform}
	\InputIfFileExists{twtransform.tikz}{}{\input{./figures/twtransform.tikz}}%
	\endpgfgraphicnamed
 \]
\end{lemma}
\begin{proof}
	\[ %
	\beginpgfgraphicnamed{twtransformprf}
	\InputIfFileExists{twtransformprf.tikz}{}{\input{./figures/twtransformprf.tikz}}%
	\endpgfgraphicnamed
 \]
	\end{proof}

\begin{lemma} \label{wwsum}
	Let $a,b, c, d \in \mathbb C$. Then
	\[ %
	\beginpgfgraphicnamed{wwsum}
	\InputIfFileExists{wwsum.tikz}{}{\input{./figures/wwsum.tikz}}%
	\endpgfgraphicnamed
 \]
\end{lemma}
\begin{proof}
If $b=0$, then
	\[ %
	\beginpgfgraphicnamed{wwsumprf1}
	\InputIfFileExists{wwsumprf1.tikz}{}{\input{./figures/wwsumprf1.tikz}}%
	\endpgfgraphicnamed
 \]
If $b \neq 0, c=d=0$, then
	\[ %
	\beginpgfgraphicnamed{wwsumprf2}
	\InputIfFileExists{wwsumprf2.tikz}{}{\input{./figures/wwsumprf2.tikz}}%
	\endpgfgraphicnamed
 \]
If $b \neq 0, c=0, d\neq 0$, then
	\[ %
	\beginpgfgraphicnamed{wwsumprf3}
	\InputIfFileExists{wwsumprf3.tikz}{}{\input{./figures/wwsumprf3.tikz}}%
	\endpgfgraphicnamed
 \]	
If $b \neq 0, c \neq 0, a=0$, then the proof is similar to the case when $b=0$.

If $b \neq 0, c \neq 0, a \neq 0, d=0$, then the proof is similar to the case when  $b \neq 0, c=0, d\neq 0$.

If $b \neq 0, c \neq 0, a \neq 0, d \neq 0$, then 
\[ %
	\beginpgfgraphicnamed{wwsumprf4}
	\InputIfFileExists{wwsumprf4.tikz}{}{\input{./figures/wwsumprf4.tikz}}%
	\endpgfgraphicnamed
 \]		
	\end{proof}

\begin{lemma}\label{w2pitritorpignslm}
  \[  %
	\beginpgfgraphicnamed{w2pitritorpigns}
	\InputIfFileExists{w2pitritorpigns.tikz}{}{\input{./figures/w2pitritorpigns.tikz}}%
	\endpgfgraphicnamed
\]
\end{lemma}
\begin{proof}
	\[ %
	\beginpgfgraphicnamed{w2pitritorpignsprf}
	\InputIfFileExists{w2pitritorpignsprf.tikz}{}{\input{./figures/w2pitritorpignsprf.tikz}}%
	\endpgfgraphicnamed
 \]
\end{proof}

\begin{lemma} \label{wloop}
	\[ %
	\beginpgfgraphicnamed{wloop}
	\InputIfFileExists{wloop.tikz}{}{\input{./figures/wloop.tikz}}%
	\endpgfgraphicnamed
 \]
\end{lemma}
\begin{proof}
	\[ %
	\beginpgfgraphicnamed{wloopprf}
	\InputIfFileExists{wloopprf.tikz}{}{\input{./figures/wloopprf.tikz}}%
	\endpgfgraphicnamed
 \]
\end{proof}

\begin{lemma} \label{tribrk}
	\[ %
	\beginpgfgraphicnamed{tribrk}
	\InputIfFileExists{tribrk.tikz}{}{\input{./figures/tribrk.tikz}}%
	\endpgfgraphicnamed
 \]
\end{lemma}
\begin{proof}
	\[ %
	\beginpgfgraphicnamed{tribrkprf}
	\InputIfFileExists{tribrkprf.tikz}{}{\input{./figures/tribrkprf.tikz}}%
	\endpgfgraphicnamed
 \]
\end{proof}

\begin{lemma}\label{diffexampleprf}
	\beginpgfgraphicnamed{diffexampletop-res}
	\InputIfFileExists{diffexampletop-res.tikz}{}{\input{./figures/diffexampletop-res.tikz}}%
	\endpgfgraphicnamed

\end{lemma}
\begin{proof}
	We use the triangle to do the change of basis from $\ket{+}$ and $\ket{1}$ to $\ket{0}$ and $\ket{1}$.
	\[ %
	\beginpgfgraphicnamed{diffexampletop}
	\InputIfFileExists{diffexampletop.tikz}{}{\input{./figures/diffexampletop.tikz}}%
	\endpgfgraphicnamed
 \]
	The rest of the proof is identical to the proof of Lemma \ref{w2pitritorpignslm}.
\end{proof}

\begin{lemma}\label{2gn1rpilm}
	\[ %
	\beginpgfgraphicnamed{2gn1rpi}
	\InputIfFileExists{2gn1rpi.tikz}{}{\input{./figures/2gn1rpi.tikz}}%
	\endpgfgraphicnamed
 \]
\end{lemma}
\begin{proof}
	\[ %
	\beginpgfgraphicnamed{2gn1rpiprf}
	\InputIfFileExists{2gn1rpiprf.tikz}{}{\input{./figures/2gn1rpiprf.tikz}}%
	\endpgfgraphicnamed
 \]
\end{proof}

\begin{lemma}\label{2gn1rpi1r0lm}
	\[ %
	\beginpgfgraphicnamed{2gn1rpi1r0}
	\InputIfFileExists{2gn1rpi1r0.tikz}{}{\input{./figures/2gn1rpi1r0.tikz}}%
	\endpgfgraphicnamed
 \]
\end{lemma}
\begin{proof}
	\[ %
	\beginpgfgraphicnamed{2gn1rpi1r0prf}
	\InputIfFileExists{2gn1rpi1r0prf.tikz}{}{\input{./figures/2gn1rpi1r0prf.tikz}}%
	\endpgfgraphicnamed
 \]
\end{proof}

\begin{lemma}\label{2gn2rpislm}
	\[ %
	\beginpgfgraphicnamed{2gn2rpis}
	\InputIfFileExists{2gn2rpis.tikz}{}{\input{./figures/2gn2rpis.tikz}}%
	\endpgfgraphicnamed
 \]
\end{lemma}
\begin{proof}
	\[ %
	\beginpgfgraphicnamed{2gn2rpisprf}
	\InputIfFileExists{2gn2rpisprf.tikz}{}{\input{./figures/2gn2rpisprf.tikz}}%
	\endpgfgraphicnamed
 \]
\end{proof}

\section{Integration proofs}
  
\circsum*
\begin{proof}
	We have
	\begin{equation} \label{eqn:theta-def}
	\beginpgfgraphicnamed{integrate/circ-sum/proof/def1}
	\begin{tikzpicture}
	\begin{pgfonlayer}{nodelayer}
		\node [style={gn_phase}] (0) at (0, 0) {$k\theta$};
		\node [style=none] (1) at (0.625, 0) {};
	\end{pgfonlayer}
	\begin{pgfonlayer}{edgelayer}
		\draw (1.center) to (0);
	\end{pgfonlayer}
\end{tikzpicture}
}%
	\endpgfgraphicnamed
 
		~\overset{(\text{\ref{gninterp}})}{=}~ %
	\beginpgfgraphicnamed{integrate/circ-sum/proof/def2}
	\begin{tikzpicture}
	\begin{pgfonlayer}{nodelayer}
		\node [style=rn] (0) at (0, 0) {};
		\node [style=none] (1) at (0.5, 0) {};
	\end{pgfonlayer}
	\begin{pgfonlayer}{edgelayer}
		\draw (1.center) to (0);
	\end{pgfonlayer}
\end{tikzpicture}
}%
	\endpgfgraphicnamed
 + e^{ik\theta}~ %
	\beginpgfgraphicnamed{integrate/circ-sum/proof/def3}
	\begin{tikzpicture}
	\begin{pgfonlayer}{nodelayer}
		\node [style={rn_phase}] (0) at (0, 0) {$\pi$};
		\node [style=none] (1) at (0.625, 0) {};
	\end{pgfonlayer}
	\begin{pgfonlayer}{edgelayer}
		\draw (1.center) to (0);
	\end{pgfonlayer}
\end{tikzpicture}
}%
	\endpgfgraphicnamed

		~= \sum_{x\in\{0, 1\}} e^{ik\theta x} ~ %
	\beginpgfgraphicnamed{integrate/circ-sum/proof/def4}
	\begin{tikzpicture}
	\begin{pgfonlayer}{nodelayer}
		\node [style={rn_phase}] (0) at (0, 0) {$x\pi$};
		\node [style=none] (1) at (0.625, 0) {};
	\end{pgfonlayer}
	\begin{pgfonlayer}{edgelayer}
		\draw (1.center) to (0);
	\end{pgfonlayer}
\end{tikzpicture}
}%
	\endpgfgraphicnamed

	\end{equation}
	Thus, we can decompose our diagram as follows:
	\begin{align*}
	\beginpgfgraphicnamed{integrate/circ-sum/proof/simp1}
	\begin{tikzpicture}
	\begin{pgfonlayer}{nodelayer}
		\node [style=none] (0) at (0.5, 1.25) {};
		\node [style=none] (1) at (-0.25, 1) {};
		\node [style=none] (2) at (0.5, -1.375) {};
		\node [style={gn_phase}, scale=1] (3) at (1.075, 1) {$-k\theta$};
		\node [style={gn_phase}, scale=1] (4) at (-0.75, 1) {$k\theta$};
		\node [style=none] (6) at (0.125, -0.05) {$M$};
		\node [style=none] (8) at (0.5, 1) {};
		\node [style=none] (9) at (-0.25, -1.375) {};
		\node [style=none] (10) at (-0.25, 1.25) {};
		\node [style=none] (12) at (-0.625, -1.125) {};
		\node [style=none] (13) at (-0.25, -0.375) {};
		\node [style=none] (14) at (-0.25, -1.125) {};
		\node [style=none] (15) at (-0.625, -0.375) {};
		\node [style=none] (16) at (-0.425, -0.625) {$\vdots$};
		\node [style=none] (18) at (0.875, -0.375) {};
		\node [style=none] (20) at (0.5, -0.375) {};
		\node [style=none] (21) at (0.5, -1.125) {};
		\node [style=none] (23) at (0.875, -1.125) {};
		\node [style=none] (24) at (0.675, -0.625) {$\vdots$};
		\node [style=none] (25) at (-0.25, 0.25) {};
		\node [style={gn_phase}, scale=1] (26) at (1.075, 0.25) {$-k\theta$};
		\node [style={gn_phase}, scale=1] (27) at (-0.75, 0.25) {$k\theta$};
		\node [style=none] (30) at (0.5, 0.25) {};
		\node [style=none] (31) at (-0.425, 0.75) {$\vdots$};
		\node [style=none] (32) at (0.675, 0.75) {$\vdots$};
	\end{pgfonlayer}
	\begin{pgfonlayer}{edgelayer}
		\draw (10.center) to (9.center);
		\draw (10.center) to (0.center);
		\draw (0.center) to (2.center);
		\draw (2.center) to (9.center);
		\draw (13.center) to (15.center);
		\draw (12.center) to (14.center);
		\draw (18.center) to (20.center);
		\draw (21.center) to (23.center);
		\draw (3) to (8.center);
		\draw (4) to (1.center);
		\draw (27) to (25.center);
		\draw (26) to (30.center);
	\end{pgfonlayer}
\end{tikzpicture}
}%
	\endpgfgraphicnamed

		~\overset{(\ref{eqn:theta-def})}{=}& \sum_{\vec x \in \{0,1\}^n} e^{ik\theta w(\vec x)} ~ %
	\beginpgfgraphicnamed{integrate/circ-sum/proof/simp2}
	\begin{tikzpicture}
	\begin{pgfonlayer}{nodelayer}
		\node [style=none] (0) at (0.5, 1.25) {};
		\node [style=none] (1) at (-0.25, 1) {};
		\node [style=none] (2) at (0.5, -1.375) {};
		\node [style={gn_phase}, scale=1] (3) at (1.075, 1) {$-k\theta$};
		\node [style={rn_phase}, scale=1] (4) at (-0.75, 1) {$x_1\pi$};
		\node [style=none] (6) at (0.125, -0.05) {$M$};
		\node [style=none] (8) at (0.5, 1) {};
		\node [style=none] (9) at (-0.25, -1.375) {};
		\node [style=none] (10) at (-0.25, 1.25) {};
		\node [style=none] (12) at (-0.625, -1.125) {};
		\node [style=none] (13) at (-0.25, -0.375) {};
		\node [style=none] (14) at (-0.25, -1.125) {};
		\node [style=none] (15) at (-0.625, -0.375) {};
		\node [style=none] (16) at (-0.425, -0.625) {$\vdots$};
		\node [style=none] (18) at (0.875, -0.375) {};
		\node [style=none] (20) at (0.5, -0.375) {};
		\node [style=none] (21) at (0.5, -1.125) {};
		\node [style=none] (23) at (0.875, -1.125) {};
		\node [style=none] (24) at (0.675, -0.625) {$\vdots$};
		\node [style=none] (25) at (-0.25, 0.25) {};
		\node [style={gn_phase}, scale=1] (26) at (1.075, 0.25) {$-k\theta$};
		\node [style={rn_phase}, scale=1] (27) at (-0.75, 0.25) {$x_n\pi$};
		\node [style=none] (30) at (0.5, 0.25) {};
		\node [style=none] (31) at (-0.425, 0.75) {$\vdots$};
		\node [style=none] (32) at (0.675, 0.75) {$\vdots$};
	\end{pgfonlayer}
	\begin{pgfonlayer}{edgelayer}
		\draw (10.center) to (9.center);
		\draw (10.center) to (0.center);
		\draw (0.center) to (2.center);
		\draw (2.center) to (9.center);
		\draw (13.center) to (15.center);
		\draw (12.center) to (14.center);
		\draw (18.center) to (20.center);
		\draw (21.center) to (23.center);
		\draw (3) to (8.center);
		\draw (4) to (1.center);
		\draw (27) to (25.center);
		\draw (26) to (30.center);
	\end{pgfonlayer}
\end{tikzpicture}
}%
	\endpgfgraphicnamed
 \\
		~\overset{(\ref{eqn:theta-def})}{=}& \sum_{\vec x, \vec y \in \{0,1\}^n} e^{ik\theta (w(\vec x) - w(\vec y))} ~ %
	\beginpgfgraphicnamed{integrate/circ-sum/proof/simp3}
	\begin{tikzpicture}
	\begin{pgfonlayer}{nodelayer}
		\node [style=none] (0) at (0.5, 1.25) {};
		\node [style=none] (1) at (-0.25, 1) {};
		\node [style=none] (2) at (0.5, -1.375) {};
		\node [style={rn_phase}, scale=1] (3) at (1.075, 1) {$y_1\pi$};
		\node [style={rn_phase}, scale=1] (4) at (-0.75, 1) {$x_1\pi$};
		\node [style=none] (6) at (0.125, -0.05) {$M$};
		\node [style=none] (8) at (0.5, 1) {};
		\node [style=none] (9) at (-0.25, -1.375) {};
		\node [style=none] (10) at (-0.25, 1.25) {};
		\node [style=none] (12) at (-0.625, -1.125) {};
		\node [style=none] (13) at (-0.25, -0.375) {};
		\node [style=none] (14) at (-0.25, -1.125) {};
		\node [style=none] (15) at (-0.625, -0.375) {};
		\node [style=none] (16) at (-0.425, -0.625) {$\vdots$};
		\node [style=none] (18) at (0.875, -0.375) {};
		\node [style=none] (20) at (0.5, -0.375) {};
		\node [style=none] (21) at (0.5, -1.125) {};
		\node [style=none] (23) at (0.875, -1.125) {};
		\node [style=none] (24) at (0.675, -0.625) {$\vdots$};
		\node [style=none] (25) at (-0.25, 0.25) {};
		\node [style={rn_phase}, scale=1] (26) at (1.075, 0.25) {$y_n\pi$};
		\node [style={rn_phase}, scale=1] (27) at (-0.75, 0.25) {$x_n\pi$};
		\node [style=none] (30) at (0.5, 0.25) {};
		\node [style=none] (31) at (-0.425, 0.75) {$\vdots$};
		\node [style=none] (32) at (0.675, 0.75) {$\vdots$};
	\end{pgfonlayer}
	\begin{pgfonlayer}{edgelayer}
		\draw (10.center) to (9.center);
		\draw (10.center) to (0.center);
		\draw (0.center) to (2.center);
		\draw (2.center) to (9.center);
		\draw (13.center) to (15.center);
		\draw (12.center) to (14.center);
		\draw (18.center) to (20.center);
		\draw (21.center) to (23.center);
		\draw (3) to (8.center);
		\draw (4) to (1.center);
		\draw (27) to (25.center);
		\draw (26) to (30.center);
	\end{pgfonlayer}
\end{tikzpicture}
}%
	\endpgfgraphicnamed
  \addtocounter{equation}{1}\tag{\theequation}\label{eqn:theta-simp}
	\end{align*}
	Furthermore, for every integer $a \in \mathbb Z$, we have
	\begin{equation} \label{eqn:int-simp}
		 \frac{1}{2\pi}\int_{-\pi}^{\pi}e^{ia\theta}d\theta 
		 = \begin{cases}
		 	1 & \text{if } a = 0 \\
		 	\frac{2\sin(a\pi)}{a} = 0 & \text{if } a \ne 0
		 \end{cases}
	\end{equation}
	Therefore, we can conclude
	\begin{align*}
		\frac{1}{2\pi} \int_{-\pi}^\pi %
	\beginpgfgraphicnamed{integrate/circ-sum/proof/simp1}
	\begin{tikzpicture}
	\begin{pgfonlayer}{nodelayer}
		\node [style=none] (0) at (0.5, 1.25) {};
		\node [style=none] (1) at (-0.25, 1) {};
		\node [style=none] (2) at (0.5, -1.375) {};
		\node [style={gn_phase}, scale=1] (3) at (1.075, 1) {$-k\theta$};
		\node [style={gn_phase}, scale=1] (4) at (-0.75, 1) {$k\theta$};
		\node [style=none] (6) at (0.125, -0.05) {$M$};
		\node [style=none] (8) at (0.5, 1) {};
		\node [style=none] (9) at (-0.25, -1.375) {};
		\node [style=none] (10) at (-0.25, 1.25) {};
		\node [style=none] (12) at (-0.625, -1.125) {};
		\node [style=none] (13) at (-0.25, -0.375) {};
		\node [style=none] (14) at (-0.25, -1.125) {};
		\node [style=none] (15) at (-0.625, -0.375) {};
		\node [style=none] (16) at (-0.425, -0.625) {$\vdots$};
		\node [style=none] (18) at (0.875, -0.375) {};
		\node [style=none] (20) at (0.5, -0.375) {};
		\node [style=none] (21) at (0.5, -1.125) {};
		\node [style=none] (23) at (0.875, -1.125) {};
		\node [style=none] (24) at (0.675, -0.625) {$\vdots$};
		\node [style=none] (25) at (-0.25, 0.25) {};
		\node [style={gn_phase}, scale=1] (26) at (1.075, 0.25) {$-k\theta$};
		\node [style={gn_phase}, scale=1] (27) at (-0.75, 0.25) {$k\theta$};
		\node [style=none] (30) at (0.5, 0.25) {};
		\node [style=none] (31) at (-0.425, 0.75) {$\vdots$};
		\node [style=none] (32) at (0.675, 0.75) {$\vdots$};
	\end{pgfonlayer}
	\begin{pgfonlayer}{edgelayer}
		\draw (10.center) to (9.center);
		\draw (10.center) to (0.center);
		\draw (0.center) to (2.center);
		\draw (2.center) to (9.center);
		\draw (13.center) to (15.center);
		\draw (12.center) to (14.center);
		\draw (18.center) to (20.center);
		\draw (21.center) to (23.center);
		\draw (3) to (8.center);
		\draw (4) to (1.center);
		\draw (27) to (25.center);
		\draw (26) to (30.center);
	\end{pgfonlayer}
\end{tikzpicture}
}%
	\endpgfgraphicnamed
 ~ d\theta
		&\overset{(\ref{eqn:theta-simp})}{=} \sum_{\vec x, \vec y \in \{0,1\}^n} \left( \frac{1}{2\pi} \int_{-\pi}^\pi e^{ik\theta (w(\vec x) - w(\vec y))} d\theta \right) ~ %
	\beginpgfgraphicnamed{integrate/circ-sum/proof/simp3}
	\begin{tikzpicture}
	\begin{pgfonlayer}{nodelayer}
		\node [style=none] (0) at (0.5, 1.25) {};
		\node [style=none] (1) at (-0.25, 1) {};
		\node [style=none] (2) at (0.5, -1.375) {};
		\node [style={rn_phase}, scale=1] (3) at (1.075, 1) {$y_1\pi$};
		\node [style={rn_phase}, scale=1] (4) at (-0.75, 1) {$x_1\pi$};
		\node [style=none] (6) at (0.125, -0.05) {$M$};
		\node [style=none] (8) at (0.5, 1) {};
		\node [style=none] (9) at (-0.25, -1.375) {};
		\node [style=none] (10) at (-0.25, 1.25) {};
		\node [style=none] (12) at (-0.625, -1.125) {};
		\node [style=none] (13) at (-0.25, -0.375) {};
		\node [style=none] (14) at (-0.25, -1.125) {};
		\node [style=none] (15) at (-0.625, -0.375) {};
		\node [style=none] (16) at (-0.425, -0.625) {$\vdots$};
		\node [style=none] (18) at (0.875, -0.375) {};
		\node [style=none] (20) at (0.5, -0.375) {};
		\node [style=none] (21) at (0.5, -1.125) {};
		\node [style=none] (23) at (0.875, -1.125) {};
		\node [style=none] (24) at (0.675, -0.625) {$\vdots$};
		\node [style=none] (25) at (-0.25, 0.25) {};
		\node [style={rn_phase}, scale=1] (26) at (1.075, 0.25) {$y_n\pi$};
		\node [style={rn_phase}, scale=1] (27) at (-0.75, 0.25) {$x_n\pi$};
		\node [style=none] (30) at (0.5, 0.25) {};
		\node [style=none] (31) at (-0.425, 0.75) {$\vdots$};
		\node [style=none] (32) at (0.675, 0.75) {$\vdots$};
	\end{pgfonlayer}
	\begin{pgfonlayer}{edgelayer}
		\draw (10.center) to (9.center);
		\draw (10.center) to (0.center);
		\draw (0.center) to (2.center);
		\draw (2.center) to (9.center);
		\draw (13.center) to (15.center);
		\draw (12.center) to (14.center);
		\draw (18.center) to (20.center);
		\draw (21.center) to (23.center);
		\draw (3) to (8.center);
		\draw (4) to (1.center);
		\draw (27) to (25.center);
		\draw (26) to (30.center);
	\end{pgfonlayer}
\end{tikzpicture}
}%
	\endpgfgraphicnamed
 \\
		&\overset{(\ref{eqn:int-simp})}{=} \sum_{\substack{\vec x,\vec y\in\{0, 1\}^n \\ w(\vec x) = w(\vec y)}} %
	\beginpgfgraphicnamed{integrate/circ-sum/proof/simp3}
	\begin{tikzpicture}
	\begin{pgfonlayer}{nodelayer}
		\node [style=none] (0) at (0.5, 1.25) {};
		\node [style=none] (1) at (-0.25, 1) {};
		\node [style=none] (2) at (0.5, -1.375) {};
		\node [style={rn_phase}, scale=1] (3) at (1.075, 1) {$y_1\pi$};
		\node [style={rn_phase}, scale=1] (4) at (-0.75, 1) {$x_1\pi$};
		\node [style=none] (6) at (0.125, -0.05) {$M$};
		\node [style=none] (8) at (0.5, 1) {};
		\node [style=none] (9) at (-0.25, -1.375) {};
		\node [style=none] (10) at (-0.25, 1.25) {};
		\node [style=none] (12) at (-0.625, -1.125) {};
		\node [style=none] (13) at (-0.25, -0.375) {};
		\node [style=none] (14) at (-0.25, -1.125) {};
		\node [style=none] (15) at (-0.625, -0.375) {};
		\node [style=none] (16) at (-0.425, -0.625) {$\vdots$};
		\node [style=none] (18) at (0.875, -0.375) {};
		\node [style=none] (20) at (0.5, -0.375) {};
		\node [style=none] (21) at (0.5, -1.125) {};
		\node [style=none] (23) at (0.875, -1.125) {};
		\node [style=none] (24) at (0.675, -0.625) {$\vdots$};
		\node [style=none] (25) at (-0.25, 0.25) {};
		\node [style={rn_phase}, scale=1] (26) at (1.075, 0.25) {$y_n\pi$};
		\node [style={rn_phase}, scale=1] (27) at (-0.75, 0.25) {$x_n\pi$};
		\node [style=none] (30) at (0.5, 0.25) {};
		\node [style=none] (31) at (-0.425, 0.75) {$\vdots$};
		\node [style=none] (32) at (0.675, 0.75) {$\vdots$};
	\end{pgfonlayer}
	\begin{pgfonlayer}{edgelayer}
		\draw (10.center) to (9.center);
		\draw (10.center) to (0.center);
		\draw (0.center) to (2.center);
		\draw (2.center) to (9.center);
		\draw (13.center) to (15.center);
		\draw (12.center) to (14.center);
		\draw (18.center) to (20.center);
		\draw (21.center) to (23.center);
		\draw (3) to (8.center);
		\draw (4) to (1.center);
		\draw (27) to (25.center);
		\draw (26) to (30.center);
	\end{pgfonlayer}
\end{tikzpicture}
}%
	\endpgfgraphicnamed

	\end{align*}
\end{proof}

\begin{lemma} \label{lem:adder-sym}
	$\Sigma_1$ is symmetric:
	\[%
	\beginpgfgraphicnamed{integrate/adder/sym}
	\begin{tikzpicture}
	\begin{pgfonlayer}{nodelayer}
		\node [style=none] (0) at (1.275, 0) {$\Sigma_1$};
		\node [style=none] (1) at (0.875, -0.875) {};
		\node [style=none] (2) at (0.875, 0.875) {};
		\node [style=none] (3) at (0.875, 0) {};
		\node [style=none] (4) at (1.625, -0.525) {};
		\node [style=none] (5) at (1.625, 0.525) {};
		\node [style=none] (6) at (0.875, 0.5) {};
		\node [style=none] (7) at (0.875, -0.5) {};
		\node [style=none] (8) at (0.625, -0.5) {};
		\node [style=none] (9) at (0.625, 0) {};
		\node [style=none] (10) at (0.625, 0.5) {};
		\node [style=none] (11) at (1.625, 0.375) {};
		\node [style=none] (12) at (1.625, -0.375) {};
		\node [style=none] (13) at (1.875, -0.375) {};
		\node [style=none] (14) at (1.875, 0.375) {};
		\node [style=none] (15) at (2.25, 0) {$=$};
		\node [style=none] (16) at (3.65, 0) {$\Sigma_1$};
		\node [style=none] (17) at (3.25, -0.875) {};
		\node [style=none] (18) at (3.25, 0.875) {};
		\node [style=none] (19) at (3.25, 0) {};
		\node [style=none] (20) at (4, -0.525) {};
		\node [style=none] (21) at (4, 0.525) {};
		\node [style=none] (22) at (3.25, 0.5) {};
		\node [style=none] (23) at (3.25, -0.5) {};
		\node [style=none] (24) at (2.625, -0.5) {};
		\node [style=none] (25) at (2.625, 0.5) {};
		\node [style=none] (26) at (2.625, 0) {};
		\node [style=none] (27) at (4, 0.375) {};
		\node [style=none] (28) at (4, -0.375) {};
		\node [style=none] (29) at (4.25, -0.375) {};
		\node [style=none] (30) at (4.25, 0.375) {};
		\node [style=none] (31) at (4.65, 0) {$=$};
		\node [style=none] (32) at (6.05, 0) {$\Sigma_1$};
		\node [style=none] (33) at (5.65, -0.875) {};
		\node [style=none] (34) at (5.65, 0.875) {};
		\node [style=none] (35) at (5.65, 0) {};
		\node [style=none] (36) at (6.4, -0.525) {};
		\node [style=none] (37) at (6.4, 0.525) {};
		\node [style=none] (38) at (5.65, -0.5) {};
		\node [style=none] (39) at (5.65, 0.5) {};
		\node [style=none] (40) at (5.025, 0.5) {};
		\node [style=none] (41) at (5.025, -0.5) {};
		\node [style=none] (42) at (5.025, 0) {};
		\node [style=none] (43) at (6.4, 0.375) {};
		\node [style=none] (44) at (6.4, -0.375) {};
		\node [style=none] (45) at (6.65, -0.375) {};
		\node [style=none] (46) at (6.65, 0.375) {};
		\node [style=none] (47) at (7.025, 0) {$=$};
		\node [style=none] (48) at (8.425, 0) {$\Sigma_1$};
		\node [style=none] (49) at (8.025, -0.875) {};
		\node [style=none] (50) at (8.025, 0.875) {};
		\node [style=none] (51) at (8.025, 0.5) {};
		\node [style=none] (52) at (8.775, -0.525) {};
		\node [style=none] (53) at (8.775, 0.525) {};
		\node [style=none] (54) at (8.025, -0.5) {};
		\node [style=none] (55) at (8.025, 0) {};
		\node [style=none] (56) at (7.4, 0) {};
		\node [style=none] (57) at (7.4, -0.5) {};
		\node [style=none] (58) at (7.4, 0.5) {};
		\node [style=none] (59) at (8.775, 0.375) {};
		\node [style=none] (60) at (8.775, -0.375) {};
		\node [style=none] (61) at (9.025, -0.375) {};
		\node [style=none] (62) at (9.025, 0.375) {};
	\end{pgfonlayer}
	\begin{pgfonlayer}{edgelayer}
		\draw (2.center) to (1.center);
		\draw (5.center) to (4.center);
		\draw (4.center) to (1.center);
		\draw (2.center) to (5.center);
		\draw (8.center) to (7.center);
		\draw (9.center) to (3.center);
		\draw (10.center) to (6.center);
		\draw (14.center) to (11.center);
		\draw (12.center) to (13.center);
		\draw (18.center) to (17.center);
		\draw (21.center) to (20.center);
		\draw (20.center) to (17.center);
		\draw (18.center) to (21.center);
		\draw (24.center) to (23.center);
		\draw [in=180, out=0, looseness=0.75] (25.center) to (19.center);
		\draw [in=180, out=0, looseness=0.75] (26.center) to (22.center);
		\draw (30.center) to (27.center);
		\draw (28.center) to (29.center);
		\draw (34.center) to (33.center);
		\draw (37.center) to (36.center);
		\draw (36.center) to (33.center);
		\draw (34.center) to (37.center);
		\draw (40.center) to (39.center);
		\draw [in=-180, out=0, looseness=0.75] (41.center) to (35.center);
		\draw [in=-180, out=0, looseness=0.75] (42.center) to (38.center);
		\draw (46.center) to (43.center);
		\draw (44.center) to (45.center);
		\draw (50.center) to (49.center);
		\draw (53.center) to (52.center);
		\draw (52.center) to (49.center);
		\draw (50.center) to (53.center);
		\draw (56.center) to (55.center);
		\draw [in=-180, out=0, looseness=0.75] (57.center) to (51.center);
		\draw [in=-180, out=0, looseness=0.75] (58.center) to (54.center);
		\draw (62.center) to (59.center);
		\draw (60.center) to (61.center);
	\end{pgfonlayer}
\end{tikzpicture}
}%
	\endpgfgraphicnamed
\]
\end{lemma}
\begin{proof}
	Immediately follows from the definition of $\Sigma_1$ and $(Sym)$.
\end{proof}

\sigmacorrect*
\begin{proof}
	The inductive construction follows the structure of a basic ripple-carry adder \cite{mano1972digital}.
	It suffices to verify that $\Sigma_1$ acts like a full-adder on computational basis states.
	
	For the all-zero state, we have
	\[ %
	\beginpgfgraphicnamed{integrate/adder/correct/base0}
	\begin{tikzpicture}
	\begin{pgfonlayer}{nodelayer}
		\node [style=none] (2) at (0.4, 0) {$\Sigma_1$};
		\node [style=none] (3) at (0, -0.875) {};
		\node [style=none] (4) at (0, 0.875) {};
		\node [style=none] (5) at (0, 0) {};
		\node [style=none] (6) at (0.75, -0.525) {};
		\node [style=none] (7) at (0.75, 0.525) {};
		\node [style=none] (14) at (0, 0.5) {};
		\node [style=none] (15) at (0, -0.5) {};
		\node [style=rn] (16) at (-0.375, -0.5) {};
		\node [style=rn] (17) at (-0.375, 0) {};
		\node [style=rn] (18) at (-0.375, 0.5) {};
		\node [style=none] (19) at (0.75, 0.375) {};
		\node [style=none] (20) at (0.75, -0.375) {};
		\node [style=none] (21) at (1, -0.375) {};
		\node [style=none] (22) at (1, 0.375) {};
		\node [style=none] (23) at (1.375, 0) {$=$};
		\node [style=rn] (27) at (2, 0.5) {};
		\node [style=rn] (28) at (2, 0) {};
		\node [style=rn] (29) at (2, -0.5) {};
		\node [style=bspider, rotate=90] (30) at (2.625, 0.5) {};
		\node [style=gn] (31) at (3.25, -0.5) {};
		\node [style=rn] (32) at (3.25, 0.5) {};
		\node [style=none] (33) at (3.625, 0.5) {};
		\node [style=none] (34) at (3.625, -0.5) {};
		\node [style=none] (35) at (4.125, 0) {$=$};
		\node [style=rn] (41) at (6, 0.375) {};
		\node [style=none] (42) at (6.375, 0.375) {};
		\node [style=none] (43) at (6.375, -0.375) {};
		\node [style=gn] (44) at (4.625, 0.375) {};
		\node [style=bspider, rotate=90] (46) at (5.375, 0.375) {};
		\node [style=gn] (48) at (6, -0.375) {};
		\node [style=none] (49) at (6.875, 0) {$=$};
		\node [style=rn] (59) at (9.425, 0.375) {};
		\node [style=none] (60) at (9.8, 0.375) {};
		\node [style=none] (61) at (9.8, -0.375) {};
		\node [style=gn] (62) at (7.5, 0.375) {};
		\node [style=bspider, rotate=90] (63) at (8.8, 0.375) {};
		\node [style=gn] (65) at (9.425, -0.375) {};
		\node [style=rn] (66) at (7.95, 0.625) {};
		\node [style=rn] (67) at (8.375, 0.625) {};
		\node [style=none] (68) at (10.3, 0) {$=$};
		\node [style=none] (70) at (11.875, 0.375) {};
		\node [style=none] (71) at (11.875, -0.375) {};
		\node [style=bspider, rotate=90] (73) at (11.375, 0.375) {};
		\node [style=rn] (74) at (10.9, 0.125) {};
		\node [style=rn] (75) at (11.5, -0.375) {};
		\node [style=rn] (77) at (10.9, 0.625) {};
		\node [style=none] (78) at (12.375, 0) {$=$};
		\node [style=rn] (79) at (12.95, 0.375) {};
		\node [style=none] (80) at (13.325, 0.375) {};
		\node [style=none] (81) at (13.325, -0.375) {};
		\node [style=rn] (84) at (12.95, -0.375) {};
		\node [style=none] (85) at (1.375, 0.25) {\scriptsize \ref{def:sum}};
		\node [style=none] (86) at (4.125, 0.25) {\scriptsize S2};
		\node [style=none] (87) at (6.875, 0.25) {\scriptsize \ref{lem:W-join}};
		\node [style=none] (88) at (6.875, 0.5) {\scriptsize Aso};
		\node [style=none] (89) at (10.3, 0.25) {\scriptsize B1};
		\node [style=none] (90) at (12.375, 0.25) {\scriptsize \ref{wplug}};
	\end{pgfonlayer}
	\begin{pgfonlayer}{edgelayer}
		\draw (4.center) to (3.center);
		\draw (7.center) to (6.center);
		\draw (6.center) to (3.center);
		\draw (4.center) to (7.center);
		\draw (16) to (15.center);
		\draw (17) to (5.center);
		\draw (18) to (14.center);
		\draw (22.center) to (19.center);
		\draw (20.center) to (21.center);
		\draw (27) to (30);
		\draw (28) to (30);
		\draw (30) to (29);
		\draw (28) to (31);
		\draw (29) to (31);
		\draw (31) to (27);
		\draw (30) to (32);
		\draw (32) to (33.center);
		\draw (32) to (31);
		\draw (31) to (34.center);
		\draw (41) to (42.center);
		\draw (44) to (46);
		\draw [bend left] (44) to (46);
		\draw (41) to (46);
		\draw (48) to (41);
		\draw (48) to (43.center);
		\draw [in=-90, out=180, looseness=0.75] (48) to (44);
		\draw (59) to (60.center);
		\draw (59) to (63);
		\draw (65) to (59);
		\draw (65) to (61.center);
		\draw [in=-90, out=180, looseness=0.75] (65) to (62);
		\draw [bend left=15, looseness=0.75] (67) to (63);
		\draw [bend right=15] (66) to (62);
		\draw (75) to (71.center);
		\draw (79) to (80.center);
		\draw (84) to (81.center);
		\draw (73) to (70.center);
		\draw (74) to (73);
		\draw (73) to (77);
		\draw [bend right] (44) to (46);
		\draw [bend right] (62) to (63);
	\end{pgfonlayer}
\end{tikzpicture}
}%
	\endpgfgraphicnamed
 \]
	For states with one bit turned on, we have
	\[ %
	\beginpgfgraphicnamed{integrate/adder/correct/base1}
	\begin{tikzpicture}
	\begin{pgfonlayer}{nodelayer}
		\node [style=none] (2) at (0.4, 0) {$\Sigma_1$};
		\node [style=none] (3) at (0, -0.875) {};
		\node [style=none] (4) at (0, 0.875) {};
		\node [style=none] (5) at (0, 0) {};
		\node [style=none] (6) at (0.75, -0.525) {};
		\node [style=none] (7) at (0.75, 0.525) {};
		\node [style=none] (14) at (0, 0.5) {};
		\node [style=none] (15) at (0, -0.5) {};
		\node [style={rn_phase}] (16) at (-0.375, -0.5) {$\pi$};
		\node [style=rn] (17) at (-0.375, 0) {};
		\node [style=rn] (18) at (-0.375, 0.5) {};
		\node [style=none] (19) at (0.75, 0.375) {};
		\node [style=none] (20) at (0.75, -0.375) {};
		\node [style=none] (21) at (1, -0.375) {};
		\node [style=none] (22) at (1, 0.375) {};
		\node [style=none] (23) at (1.375, 0) {$=$};
		\node [style=rn] (27) at (2, 0.5) {};
		\node [style=rn] (28) at (2, 0) {};
		\node [style={rn_phase}] (29) at (2, -0.5) {$\pi$};
		\node [style=bspider, rotate=90] (30) at (2.625, 0.5) {};
		\node [style=gn] (31) at (3.25, -0.5) {};
		\node [style=rn] (32) at (3.25, 0.5) {};
		\node [style=none] (33) at (3.625, 0.5) {};
		\node [style=none] (34) at (3.625, -0.5) {};
		\node [style=none] (35) at (4.125, 0) {$=$};
		\node [style=rn] (41) at (6.5, 0.375) {};
		\node [style=none] (42) at (6.875, 0.375) {};
		\node [style=none] (43) at (6.875, -0.375) {};
		\node [style=gn] (44) at (4.625, 0.375) {};
		\node [style=bspider, rotate=90] (46) at (5.875, 0.375) {};
		\node [style={rn_phase}] (47) at (5.25, 0.125) {$\pi$};
		\node [style=gn] (48) at (6.5, -0.375) {};
		\node [style=none] (49) at (-3.125, -2) {$=$};
		\node [style=rn] (59) at (-0.575, -1.625) {};
		\node [style=none] (60) at (-0.2, -1.625) {};
		\node [style=none] (61) at (-0.2, -2.375) {};
		\node [style=gn] (62) at (-2.5, -1.625) {};
		\node [style=bspider, rotate=90] (63) at (-1.2, -1.625) {};
		\node [style={rn_phase}] (64) at (-1.825, -1.875) {$\pi$};
		\node [style=gn] (65) at (-0.575, -2.375) {};
		\node [style=rn] (66) at (-2.05, -1.375) {};
		\node [style=rn] (67) at (-1.625, -1.375) {};
		\node [style=none] (68) at (0.3, -2) {$=$};
		\node [style=none] (70) at (1.875, -1.625) {};
		\node [style=none] (71) at (1.875, -2.375) {};
		\node [style=bspider, rotate=90] (73) at (1.375, -1.625) {};
		\node [style={rn_phase}] (74) at (0.9, -1.875) {$\pi$};
		\node [style=rn] (75) at (1.375, -2.375) {};
		\node [style=rn] (77) at (0.9, -1.375) {};
		\node [style=none] (78) at (2.375, -2) {$=$};
		\node [style={rn_phase}] (79) at (2.95, -1.625) {$\pi$};
		\node [style=none] (80) at (3.325, -1.625) {};
		\node [style=none] (81) at (3.325, -2.375) {};
		\node [style=rn] (84) at (2.95, -2.375) {};
		\node [style=none] (85) at (-1.9, 0) {$\Sigma_1$};
		\node [style=none] (86) at (-2.3, -0.875) {};
		\node [style=none] (87) at (-2.3, 0.875) {};
		\node [style=none] (88) at (-2.3, 0) {};
		\node [style=none] (89) at (-1.55, -0.525) {};
		\node [style=none] (90) at (-1.55, 0.525) {};
		\node [style=none] (91) at (-2.3, 0.5) {};
		\node [style=none] (92) at (-2.3, -0.5) {};
		\node [style=rn] (93) at (-2.675, -0.5) {};
		\node [style={rn_phase}] (94) at (-2.675, 0) {$\pi$};
		\node [style=rn] (95) at (-2.675, 0.5) {};
		\node [style=none] (96) at (-1.55, 0.375) {};
		\node [style=none] (97) at (-1.55, -0.375) {};
		\node [style=none] (98) at (-1.3, -0.375) {};
		\node [style=none] (99) at (-1.3, 0.375) {};
		\node [style=none] (100) at (-0.925, 0) {$=$};
		\node [style=none] (101) at (-4.15, 0) {$\Sigma_1$};
		\node [style=none] (102) at (-4.55, -0.875) {};
		\node [style=none] (103) at (-4.55, 0.875) {};
		\node [style=none] (104) at (-4.55, 0) {};
		\node [style=none] (105) at (-3.8, -0.525) {};
		\node [style=none] (106) at (-3.8, 0.525) {};
		\node [style=none] (107) at (-4.55, 0.5) {};
		\node [style=none] (108) at (-4.55, -0.5) {};
		\node [style=rn] (109) at (-4.925, -0.5) {};
		\node [style=rn] (110) at (-4.925, 0) {};
		\node [style={rn_phase}] (111) at (-4.925, 0.5) {$\pi$};
		\node [style=none] (112) at (-3.8, 0.375) {};
		\node [style=none] (113) at (-3.8, -0.375) {};
		\node [style=none] (114) at (-3.55, -0.375) {};
		\node [style=none] (115) at (-3.55, 0.375) {};
		\node [style=none] (116) at (-3.175, 0) {$=$};
		\node [style=none] (117) at (1.375, 0.25) {\scriptsize \ref{def:sum}};
		\node [style=none] (118) at (4.125, 0.25) {\scriptsize S2};
		\node [style=none] (119) at (-3.125, -1.75) {\scriptsize \ref{lem:W-join}};
		\node [style=none] (120) at (-3.125, -1.5) {\scriptsize Aso};
		\node [style=none] (121) at (0.3, -1.75) {\scriptsize B1};
		\node [style=none] (122) at (2.375, -1.75) {\scriptsize \ref{wplug}};
		\node [style=none] (123) at (-3.175, 0.25) {\scriptsize \ref{lem:adder-sym}};
		\node [style=none] (124) at (-0.925, 0.25) {\scriptsize \ref{lem:adder-sym}};
	\end{pgfonlayer}
	\begin{pgfonlayer}{edgelayer}
		\draw (4.center) to (3.center);
		\draw (7.center) to (6.center);
		\draw (6.center) to (3.center);
		\draw (4.center) to (7.center);
		\draw (16) to (15.center);
		\draw (17) to (5.center);
		\draw (18) to (14.center);
		\draw (22.center) to (19.center);
		\draw (20.center) to (21.center);
		\draw (27) to (30);
		\draw (28) to (30);
		\draw (30) to (29);
		\draw (28) to (31);
		\draw (29) to (31);
		\draw (31) to (27);
		\draw (30) to (32);
		\draw (32) to (33.center);
		\draw (32) to (31);
		\draw (31) to (34.center);
		\draw (41) to (42.center);
		\draw [bend right=15] (47) to (46);
		\draw [bend left=15] (47) to (44);
		\draw [bend left=15] (44) to (46);
		\draw [bend left=45, looseness=0.75] (44) to (46);
		\draw (41) to (46);
		\draw (48) to (41);
		\draw (48) to (43.center);
		\draw [in=-90, out=180, looseness=0.75] (48) to (44);
		\draw (59) to (60.center);
		\draw [bend right=15] (64) to (63);
		\draw [bend left=15] (64) to (62);
		\draw (59) to (63);
		\draw (65) to (59);
		\draw (65) to (61.center);
		\draw [in=-90, out=180, looseness=0.75] (65) to (62);
		\draw [bend left=15, looseness=0.75] (67) to (63);
		\draw [bend right=15] (66) to (62);
		\draw (75) to (71.center);
		\draw (79) to (80.center);
		\draw (84) to (81.center);
		\draw (73) to (70.center);
		\draw (74) to (73);
		\draw (73) to (77);
		\draw (87.center) to (86.center);
		\draw (90.center) to (89.center);
		\draw (89.center) to (86.center);
		\draw (87.center) to (90.center);
		\draw (93) to (92.center);
		\draw (94) to (88.center);
		\draw (95) to (91.center);
		\draw (99.center) to (96.center);
		\draw (97.center) to (98.center);
		\draw (103.center) to (102.center);
		\draw (106.center) to (105.center);
		\draw (105.center) to (102.center);
		\draw (103.center) to (106.center);
		\draw (109) to (108.center);
		\draw (110) to (104.center);
		\draw (111) to (107.center);
		\draw (115.center) to (112.center);
		\draw (113.center) to (114.center);
	\end{pgfonlayer}
\end{tikzpicture}
}%
	\endpgfgraphicnamed
 \]
	For states with two bits turned on, we have
	\[ %
	\beginpgfgraphicnamed{integrate/adder/correct/base2}
	\begin{tikzpicture}
	\begin{pgfonlayer}{nodelayer}
		\node [style=none] (2) at (0.4, 0) {$\Sigma_1$};
		\node [style=none] (3) at (0, -0.875) {};
		\node [style=none] (4) at (0, 0.875) {};
		\node [style=none] (5) at (0, 0) {};
		\node [style=none] (6) at (0.75, -0.525) {};
		\node [style=none] (7) at (0.75, 0.525) {};
		\node [style=none] (14) at (0, 0.5) {};
		\node [style=none] (15) at (0, -0.5) {};
		\node [style={rn_phase}] (16) at (-0.375, -0.5) {$\pi$};
		\node [style={rn_phase}] (17) at (-0.375, 0) {$\pi$};
		\node [style=rn] (18) at (-0.375, 0.5) {};
		\node [style=none] (19) at (0.75, 0.375) {};
		\node [style=none] (20) at (0.75, -0.375) {};
		\node [style=none] (21) at (1, -0.375) {};
		\node [style=none] (22) at (1, 0.375) {};
		\node [style=none] (23) at (1.375, 0) {$=$};
		\node [style=rn] (27) at (2, 0.525) {};
		\node [style={rn_phase}] (28) at (2, 0.025) {$\pi$};
		\node [style={rn_phase}] (29) at (2, -0.475) {$\pi$};
		\node [style=bspider, rotate=90] (30) at (2.625, 0.525) {};
		\node [style=gn] (31) at (3.25, -0.475) {};
		\node [style=rn] (32) at (3.25, 0.525) {};
		\node [style=none] (33) at (3.625, 0.525) {};
		\node [style=none] (34) at (3.625, -0.475) {};
		\node [style=none] (35) at (-3.125, -1.975) {$=$};
		\node [style={rn_phase}] (41) at (-0.75, -1.6) {$\pi$};
		\node [style=none] (42) at (0.125, -1.6) {};
		\node [style=none] (43) at (0.125, -2.35) {};
		\node [style=gn] (44) at (-2.625, -1.6) {};
		\node [style=bspider, rotate=90] (46) at (-1.375, -1.6) {};
		\node [style={rn_phase}] (47) at (-2, -1.35) {$\pi$};
		\node [style=gn] (48) at (-0.75, -2.35) {};
		\node [style=none] (49) at (0.625, -1.975) {$=$};
		\node [style={rn_phase}] (59) at (3.175, -1.6) {$\pi$};
		\node [style=none] (60) at (3.925, -1.6) {};
		\node [style=none] (61) at (3.925, -2.35) {};
		\node [style=gn] (62) at (1.25, -1.6) {};
		\node [style=bspider, rotate=90] (63) at (2.55, -1.6) {};
		\node [style={rn_phase}] (64) at (1.925, -1.35) {$\pi$};
		\node [style=gn] (65) at (3.175, -2.35) {};
		\node [style=rn] (66) at (1.7, -1.85) {};
		\node [style=rn] (67) at (2.125, -1.85) {};
		\node [style=none] (68) at (4.425, -1.975) {$=$};
		\node [style=none] (70) at (6.5, -1.6) {};
		\node [style=none] (71) at (6.5, -2.35) {};
		\node [style=bspider, rotate=90] (73) at (5.5, -1.6) {};
		\node [style={rn_phase}] (74) at (5.025, -1.35) {$\pi$};
		\node [style={rn_phase}] (75) at (6.075, -2.35) {$\pi$};
		\node [style=rn] (77) at (5.025, -1.85) {};
		\node [style=none] (78) at (7, -1.975) {$=$};
		\node [style={rn_phase}] (79) at (7.575, -2.35) {$\pi$};
		\node [style=none] (80) at (7.95, -2.35) {};
		\node [style=none] (81) at (7.95, -1.6) {};
		\node [style=rn] (84) at (7.575, -1.6) {};
		\node [style=none] (85) at (-1.9, 0) {$\Sigma_1$};
		\node [style=none] (86) at (-2.3, -0.875) {};
		\node [style=none] (87) at (-2.3, 0.875) {};
		\node [style=none] (88) at (-2.3, 0) {};
		\node [style=none] (89) at (-1.55, -0.525) {};
		\node [style=none] (90) at (-1.55, 0.525) {};
		\node [style=none] (91) at (-2.3, 0.5) {};
		\node [style=none] (92) at (-2.3, -0.5) {};
		\node [style={rn_phase}] (93) at (-2.675, -0.5) {$\pi$};
		\node [style=rn] (94) at (-2.675, 0) {};
		\node [style={rn_phase}] (95) at (-2.675, 0.5) {$\pi$};
		\node [style=none] (96) at (-1.55, 0.375) {};
		\node [style=none] (97) at (-1.55, -0.375) {};
		\node [style=none] (98) at (-1.3, -0.375) {};
		\node [style=none] (99) at (-1.3, 0.375) {};
		\node [style=none] (100) at (-0.925, 0) {$=$};
		\node [style=none] (101) at (-4.15, 0) {$\Sigma_1$};
		\node [style=none] (102) at (-4.55, -0.875) {};
		\node [style=none] (103) at (-4.55, 0.875) {};
		\node [style=none] (104) at (-4.55, 0) {};
		\node [style=none] (105) at (-3.8, -0.525) {};
		\node [style=none] (106) at (-3.8, 0.525) {};
		\node [style=none] (107) at (-4.55, 0.5) {};
		\node [style=none] (108) at (-4.55, -0.5) {};
		\node [style=rn] (109) at (-4.925, -0.5) {};
		\node [style={rn_phase}] (110) at (-4.925, 0) {$\pi$};
		\node [style={rn_phase}] (111) at (-4.925, 0.5) {$\pi$};
		\node [style=none] (112) at (-3.8, 0.375) {};
		\node [style=none] (113) at (-3.8, -0.375) {};
		\node [style=none] (114) at (-3.55, -0.375) {};
		\node [style=none] (115) at (-3.55, 0.375) {};
		\node [style=none] (116) at (-3.175, 0) {$=$};
		\node [style=none] (117) at (4.325, 0.025) {$=$};
		\node [style=rn] (118) at (6.7, 0.4) {};
		\node [style=none] (119) at (7.075, 0.4) {};
		\node [style=none] (120) at (7.075, -0.35) {};
		\node [style=gn] (121) at (4.825, 0.4) {};
		\node [style=bspider, rotate=90] (122) at (6.075, 0.4) {};
		\node [style={rn_phase}] (123) at (5.45, 0.075) {$\pi$};
		\node [style=gn] (124) at (6.7, -0.35) {};
		\node [style={rn_phase}] (125) at (5.45, 0.575) {$\pi$};
		\node [style={rn_phase}] (126) at (-0.3, -2.35) {$\pi$};
		\node [style={rn_phase}] (127) at (3.575, -2.35) {$\pi$};
		\node [style={rn_phase}] (128) at (6.075, -1.6) {$\pi$};
		\node [style=none] (129) at (1.375, 0.25) {\scriptsize \ref{def:sum}};
		\node [style=none] (130) at (4.325, 0.25) {\scriptsize S2};
		\node [style=none] (131) at (0.625, -1.75) {\scriptsize \ref{lem:W-join}};
		\node [style=none] (132) at (0.625, -1.5) {\scriptsize Aso};
		\node [style=none] (133) at (4.425, -1.75) {\scriptsize B1};
		\node [style=none] (134) at (7, -1.75) {\scriptsize \ref{wplug}};
		\node [style=none] (135) at (-3.125, -1.75) {\scriptsize B3};
		\node [style=none] (136) at (-3.175, 0.25) {\scriptsize \ref{lem:adder-sym}};
		\node [style=none] (137) at (-0.925, 0.25) {\scriptsize \ref{lem:adder-sym}};
	\end{pgfonlayer}
	\begin{pgfonlayer}{edgelayer}
		\draw (4.center) to (3.center);
		\draw (7.center) to (6.center);
		\draw (6.center) to (3.center);
		\draw (4.center) to (7.center);
		\draw (16) to (15.center);
		\draw (17) to (5.center);
		\draw (18) to (14.center);
		\draw (22.center) to (19.center);
		\draw (20.center) to (21.center);
		\draw (27) to (30);
		\draw (28) to (30);
		\draw (30) to (29);
		\draw (28) to (31);
		\draw (29) to (31);
		\draw (31) to (27);
		\draw (30) to (32);
		\draw (32) to (33.center);
		\draw (32) to (31);
		\draw (31) to (34.center);
		\draw (41) to (42.center);
		\draw [bend left=15] (47) to (46);
		\draw [bend right=15] (47) to (44);
		\draw [bend right=15] (44) to (46);
		\draw [bend right=45, looseness=0.75] (44) to (46);
		\draw (41) to (46);
		\draw (48) to (41);
		\draw (48) to (43.center);
		\draw [in=-90, out=180, looseness=0.75] (48) to (44);
		\draw (59) to (60.center);
		\draw [bend left=15] (64) to (63);
		\draw [bend right=15] (64) to (62);
		\draw (59) to (63);
		\draw (65) to (59);
		\draw (65) to (61.center);
		\draw [in=-90, out=180, looseness=0.75] (65) to (62);
		\draw [bend right=15, looseness=0.75] (67) to (63);
		\draw [bend left=15] (66) to (62);
		\draw (75) to (71.center);
		\draw (79) to (80.center);
		\draw (84) to (81.center);
		\draw (73) to (70.center);
		\draw (74) to (73);
		\draw (73) to (77);
		\draw (87.center) to (86.center);
		\draw (90.center) to (89.center);
		\draw (89.center) to (86.center);
		\draw (87.center) to (90.center);
		\draw (93) to (92.center);
		\draw (94) to (88.center);
		\draw (95) to (91.center);
		\draw (99.center) to (96.center);
		\draw (97.center) to (98.center);
		\draw (103.center) to (102.center);
		\draw (106.center) to (105.center);
		\draw (105.center) to (102.center);
		\draw (103.center) to (106.center);
		\draw (109) to (108.center);
		\draw (110) to (104.center);
		\draw (111) to (107.center);
		\draw (115.center) to (112.center);
		\draw (113.center) to (114.center);
		\draw (118) to (119.center);
		\draw [bend right=15] (123) to (122);
		\draw [bend left=15] (123) to (121);
		\draw [bend left=60, looseness=1.25] (121) to (122);
		\draw (118) to (122);
		\draw (124) to (118);
		\draw (124) to (120.center);
		\draw [in=-90, out=180] (124) to (121);
		\draw [bend left=15] (121) to (125);
		\draw [bend left=15] (125) to (122);
	\end{pgfonlayer}
\end{tikzpicture}
}%
	\endpgfgraphicnamed
 \]
	Finally, for the all-one state, we have
	\[ %
	\beginpgfgraphicnamed{integrate/adder/correct/base3}
	\begin{tikzpicture}
	\begin{pgfonlayer}{nodelayer}
		\node [style=none] (2) at (0.4, 0) {$\Sigma_1$};
		\node [style=none] (3) at (0, -0.875) {};
		\node [style=none] (4) at (0, 0.875) {};
		\node [style=none] (5) at (0, 0) {};
		\node [style=none] (6) at (0.75, -0.525) {};
		\node [style=none] (7) at (0.75, 0.525) {};
		\node [style=none] (14) at (0, 0.5) {};
		\node [style=none] (15) at (0, -0.5) {};
		\node [style={rn_phase}] (16) at (-0.375, -0.5) {$\pi$};
		\node [style={rn_phase}] (17) at (-0.375, 0) {$\pi$};
		\node [style={rn_phase}] (18) at (-0.375, 0.5) {$\pi$};
		\node [style=none] (19) at (0.75, 0.375) {};
		\node [style=none] (20) at (0.75, -0.375) {};
		\node [style=none] (21) at (1, -0.375) {};
		\node [style=none] (22) at (1, 0.375) {};
		\node [style=none] (23) at (1.375, 0) {$=$};
		\node [style={rn_phase}] (27) at (2, 0.5) {$\pi$};
		\node [style={rn_phase}] (28) at (2, 0) {$\pi$};
		\node [style={rn_phase}] (29) at (2, -0.5) {$\pi$};
		\node [style=bspider, rotate=90] (30) at (2.625, 0.5) {};
		\node [style=gn] (31) at (3.25, -0.5) {};
		\node [style=rn] (32) at (3.25, 0.5) {};
		\node [style=none] (33) at (3.625, 0.5) {};
		\node [style=none] (34) at (3.625, -0.5) {};
		\node [style=none] (35) at (4.125, 0) {$=$};
		\node [style={rn_phase}] (41) at (9.5, 0.375) {$\pi$};
		\node [style=none] (42) at (10.375, 0.375) {};
		\node [style=none] (43) at (10.375, -0.375) {};
		\node [style=gn] (44) at (7.875, 0.375) {};
		\node [style=bspider, rotate=90] (46) at (8.875, 0.375) {};
		\node [style=gn] (48) at (9.5, -0.375) {};
		\node [style=none] (49) at (1.375, -2) {$=$};
		\node [style={rn_phase}] (59) at (3.925, -1.625) {$\pi$};
		\node [style=none] (60) at (4.8, -1.625) {};
		\node [style=none] (61) at (4.8, -2.375) {};
		\node [style=gn] (62) at (2, -1.625) {};
		\node [style=bspider, rotate=90] (63) at (3.3, -1.625) {};
		\node [style=gn] (65) at (3.925, -2.375) {};
		\node [style=rn] (66) at (2.45, -1.375) {};
		\node [style=rn] (67) at (2.875, -1.375) {};
		\node [style=none] (68) at (5.3, -2) {$=$};
		\node [style=none] (70) at (7.375, -1.625) {};
		\node [style=none] (71) at (7.375, -2.375) {};
		\node [style=bspider, rotate=90] (73) at (6.375, -1.625) {};
		\node [style=rn] (74) at (5.9, -1.875) {};
		\node [style={rn_phase}] (75) at (6.95, -2.375) {$\pi$};
		\node [style=rn] (77) at (5.9, -1.375) {};
		\node [style=none] (78) at (7.875, -2) {$=$};
		\node [style={rn_phase}] (79) at (8.45, -1.625) {$\pi$};
		\node [style=none] (80) at (8.825, -1.625) {};
		\node [style=none] (81) at (8.825, -2.375) {};
		\node [style={rn_phase}] (84) at (8.45, -2.375) {$\pi$};
		\node [style=rn] (85) at (6.45, 0.375) {};
		\node [style=none] (86) at (6.825, 0.375) {};
		\node [style=none] (87) at (6.825, -0.625) {};
		\node [style=gn] (88) at (4.575, 0.375) {};
		\node [style=bspider, rotate=90] (89) at (5.825, 0.375) {};
		\node [style=gn] (90) at (6.45, -0.625) {};
		\node [style=none] (91) at (7.325, 0) {$=$};
		\node [style={rn_phase}] (92) at (5.2, 0.375) {$\pi$};
		\node [style={rn_phase}] (93) at (5.2, 0.875) {$\pi$};
		\node [style={rn_phase}] (94) at (5.2, -0.125) {$\pi$};
		\node [style={rn_phase}] (95) at (10, -0.375) {$\pi$};
		\node [style={rn_phase}] (96) at (4.4, -2.375) {$\pi$};
		\node [style={rn_phase}] (97) at (6.95, -1.625) {$\pi$};
		\node [style=none] (98) at (1.375, 0.25) {\scriptsize \ref{def:sum}};
		\node [style=none] (99) at (4.125, 0.25) {\scriptsize S2};
		\node [style=none] (100) at (1.375, -1.75) {\scriptsize \ref{lem:W-join}};
		\node [style=none] (101) at (1.375, -1.5) {\scriptsize Aso};
		\node [style=none] (102) at (5.3, -1.75) {\scriptsize B1};
		\node [style=none] (103) at (7.85, -1.75) {\scriptsize \ref{wplug}};
		\node [style=none] (104) at (7.325, 0.25) {\scriptsize B3};
	\end{pgfonlayer}
	\begin{pgfonlayer}{edgelayer}
		\draw (4.center) to (3.center);
		\draw (7.center) to (6.center);
		\draw (6.center) to (3.center);
		\draw (4.center) to (7.center);
		\draw (16) to (15.center);
		\draw (17) to (5.center);
		\draw (18) to (14.center);
		\draw (22.center) to (19.center);
		\draw (20.center) to (21.center);
		\draw (27) to (30);
		\draw (28) to (30);
		\draw (30) to (29);
		\draw (28) to (31);
		\draw (29) to (31);
		\draw (31) to (27);
		\draw (30) to (32);
		\draw (32) to (33.center);
		\draw (32) to (31);
		\draw (31) to (34.center);
		\draw (41) to (42.center);
		\draw (44) to (46);
		\draw [bend left=45] (44) to (46);
		\draw (41) to (46);
		\draw (48) to (41);
		\draw (48) to (43.center);
		\draw [in=-90, out=180, looseness=0.75] (48) to (44);
		\draw (59) to (60.center);
		\draw (59) to (63);
		\draw (65) to (59);
		\draw (65) to (61.center);
		\draw [in=-90, out=180, looseness=0.75] (65) to (62);
		\draw [bend left=15, looseness=0.75] (67) to (63);
		\draw [bend right=15] (66) to (62);
		\draw (75) to (71.center);
		\draw (79) to (80.center);
		\draw (84) to (81.center);
		\draw (73) to (70.center);
		\draw (74) to (73);
		\draw (73) to (77);
		\draw [bend right=45] (44) to (46);
		\draw [bend right] (62) to (63);
		\draw (85) to (86.center);
		\draw (85) to (89);
		\draw (90) to (85);
		\draw (90) to (87.center);
		\draw [in=-90, out=180] (90) to (88);
		\draw [bend left] (88) to (93);
		\draw (88) to (92);
		\draw [bend left] (94) to (88);
		\draw (92) to (89);
		\draw [bend right] (89) to (93);
		\draw [bend right] (94) to (89);
	\end{pgfonlayer}
\end{tikzpicture}
}%
	\endpgfgraphicnamed
 \]
\end{proof}

\wcorrect*
\begin{proof}
	By induction on $n$.
	$W_1$ is trivially correct since the Hamming weight of a single bit is equal to its value.
	Now suppose that the lemma holds for all $l < n$.
	If $n$ is even, i.e. $n = 2t$ for some $t < n$, then
	\[ %
	\beginpgfgraphicnamed{integrate/W/even-correct1}
	\InputIfFileExists{integrate/W/even-correct1.tikz}{}{\input{./figures/integrate/W/even-correct1.tikz}}%
	\endpgfgraphicnamed
 \]
	where $k = \lfloor \log(t) \rfloor + 1$, $[\vec a] = w(x_1, ..., x_t)$, and $[\vec b] = w(x_{t+1}, ..., x_{2t})$. Then
	\[ %
	\beginpgfgraphicnamed{integrate/W/even-correct2}
	\InputIfFileExists{integrate/W/even-correct2.tikz}{}{\input{./figures/integrate/W/even-correct2.tikz}}%
	\endpgfgraphicnamed
 \]
	where $[\vec z] = [\vec a] + \vec[\vec b] + 0 =  w(x_1, ..., x_t) + w(x_{t+1}, ..., x_{2t}) = w(\vec x)$ as required.
	
	The case where $n$ is odd follows similarly.
\end{proof}

\wsize*
\begin{proof}
	We denote the size of a ZX diagram $D$ as $S(D)$.
	In the following, we only consider the size in big-O notation.
	
	Clearly, we have $S(\Sigma_n) \in O(n)$.
	Following the definition of $W_n$, we thus have
	\begin{align*}
		S(W_1) = O(1), \qquad\qquad
		S(W_n) &= 2 S(W_{n/2}) + S(\Sigma_{\lfloor\log n\rfloor + 1})  && \text{for } n > 1 \\
		&= 2 S(W_{n/2}) + O(\lfloor\log n\rfloor + 1)
	\end{align*}
	The Master theorem \cite{bentley1980general} implies that this recurrence relation satisfies $S(W_n) \in O(n)$.
\end{proof}

\begin{lemma} \label{int2simp}
	\[ %
	\beginpgfgraphicnamed{int2simp}
	\InputIfFileExists{int2simp.tikz}{}{\input{./figures/int2simp.tikz}}%
	\endpgfgraphicnamed
 \]
\end{lemma}
\begin{proof}
	\[ %
	\beginpgfgraphicnamed{int2simpprf}
	\InputIfFileExists{int2simpprf.tikz}{}{\input{./figures/int2simpprf.tikz}}%
	\endpgfgraphicnamed
 \]
\end{proof}

\section{Barren Plateau Analysis}
\onefoldexpectintglm*
  \begin{proof} 
  By integrating over the uniform distribution, we have
  \[
    \mathbf{E}\left(\frac{ \partial \braket{H}}{\partial\theta_j}\right)=\frac{1}{(2\pi)^m}\int_{-\pi}^{\pi} \cdots\int_{-\pi}^{\pi}  \frac{ \partial \braket{H}}{\partial\theta_j}d\theta_1\ldots d\theta_m.
  \] 
  We prove the theorem by showing that
  $\frac{1}{2\pi}\int_{-\pi}^{\pi} \frac{ \partial \braket{H}}{\partial\theta_j}d\theta_j= 0$.
  \begin{align*}
  \frac{ \partial \braket{H}}{\partial\theta_j} \quad&\overset{\ref{diffthm2tmcr}}{=}\quad %
	\beginpgfgraphicnamed{1foldexpectintgprf1}
	\InputIfFileExists{1foldexpectintgprf1.tikz}{}{\input{./figures/1foldexpectintgprf1.tikz}}%
	\endpgfgraphicnamed
 \\
  \frac{1}{2\pi}\int_{-\pi}^{\pi} \frac{ \partial \braket{H}}{\partial\theta_j}d\theta_j\quad
  &\overset{\ref{1foldintglm}}{=}\quad%
	\beginpgfgraphicnamed{1foldexpectintgprf1-5}
	\InputIfFileExists{1foldexpectintgprf1-5.tikz}{}{\input{./figures/1foldexpectintgprf1-5.tikz}}%
	\endpgfgraphicnamed
 \\
  &\overset{\ref{2gn1rpilm}}{=}%
	\beginpgfgraphicnamed{1foldexpectintgprf2}
	\InputIfFileExists{1foldexpectintgprf2.tikz}{}{\input{./figures/1foldexpectintgprf2.tikz}}%
	\endpgfgraphicnamed
 
  \end{align*}
  Since the unconnected pink $\pi$ spider is equal to the zero scalar, the entire diagram evaluates to zero.
  \end{proof}

\onefoldexpectsquareintglm*
\begin{proof}
	By Example \ref{2varibleintthm},
	
	\[
	\frac{1}{2\pi}\int_{-\pi}^{\pi}  \left(\frac{ \partial \braket{H}}{\partial\theta_j}\right)^2d\theta_j=
	\]
	\[
	\beginpgfgraphicnamed{1foldexpectsquareintgprf}
	\InputIfFileExists{1foldexpectsquareintgprf.tikz}{}{\input{./figures/1foldexpectsquareintgprf.tikz}}%
	\endpgfgraphicnamed
 ~ %
	\beginpgfgraphicnamed{1foldexpectsquareintgprf2}
	\InputIfFileExists{1foldexpectsquareintgprf2.tikz}{}{\input{./figures/1foldexpectsquareintgprf2.tikz}}%
	\endpgfgraphicnamed

	\] 
	\[
	\beginpgfgraphicnamed{1foldexpectsquareintgprf3}
	\InputIfFileExists{1foldexpectsquareintgprf3.tikz}{}{\input{./figures/1foldexpectsquareintgprf3.tikz}}%
	\endpgfgraphicnamed
 ~ %
	\beginpgfgraphicnamed{1foldexpectsquareintgprf4}
	\InputIfFileExists{1foldexpectsquareintgprf4.tikz}{}{\input{./figures/1foldexpectsquareintgprf4.tikz}}%
	\endpgfgraphicnamed

	\] 
	\[
	\beginpgfgraphicnamed{1foldexpectsquareintgprf5}
	\InputIfFileExists{1foldexpectsquareintgprf5.tikz}{}{\input{./figures/1foldexpectsquareintgprf5.tikz}}%
	\endpgfgraphicnamed

	\quad\overset{\ref{2gn2rpislm}}{=}\quad %
	\beginpgfgraphicnamed{1foldexpectsquareintgprf6}
	\InputIfFileExists{1foldexpectsquareintgprf6.tikz}{}{\input{./figures/1foldexpectsquareintgprf6.tikz}}%
	\endpgfgraphicnamed

	\]
\end{proof}

\cycle*
\begin{proof}
	\[ %
	\beginpgfgraphicnamed{sim9/cycle1prf}
	\InputIfFileExists{sim9/cycle1prf.tikz}{}{\input{./figures/sim9/cycle1prf.tikz}}%
	\endpgfgraphicnamed
 \]
	\[ %
	\beginpgfgraphicnamed{sim9/cycle2prf}
	\InputIfFileExists{sim9/cycle2prf.tikz}{}{\input{./figures/sim9/cycle2prf.tikz}}%
	\endpgfgraphicnamed
 \]
\end{proof}

\begin{lemma} \label{gridsimp}
	For all $a \in \{0,1\}$, we have
	\[ %
	\beginpgfgraphicnamed{sim9/gridsimp}
	\InputIfFileExists{sim9/gridsimp.tikz}{}{\input{./figures/sim9/gridsimp.tikz}}%
	\endpgfgraphicnamed
 \]
\end{lemma}
\begin{proof}
	\[ %
	\beginpgfgraphicnamed{sim9/gridsimpprf}
	\InputIfFileExists{sim9/gridsimpprf.tikz}{}{\input{./figures/sim9/gridsimpprf.tikz}}%
	\endpgfgraphicnamed
 \]
\end{proof}

\section{Comparison with Jeandel et al.}
\label{zx-compare}

Jeandel, Perdrix, and Veshchezerova present an alternate method to represent derivatives with the ZX-calculus \cite{jeandel2022addition}.
We want to show that the two papers arrive at similar results through very different techniques, and that
some results in \cite{jeandel2022addition} can be more compactly represented by explicitly using the W spider.
The nodes used in \cite{jeandel2022addition} are similar to ours: the circle green nodes, the Hadamard node and the red nodes are exactly the same, but we additionally have pink nodes which are different to the red nodes up to a variational scalar, and have green box nodes which have the circle green nodes as special cases while can be turned into circle green nodes with the help of the yellow triangle node \cite{amarngwanglics}.
Also note that their black triangle corresponds to our yellow triangle and our black triangle corresponds to the W spider.

First, we show that the triangular-shaped diagram that features in their differentiation result can be represented compactly using the W spider:
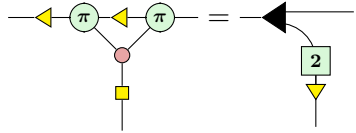
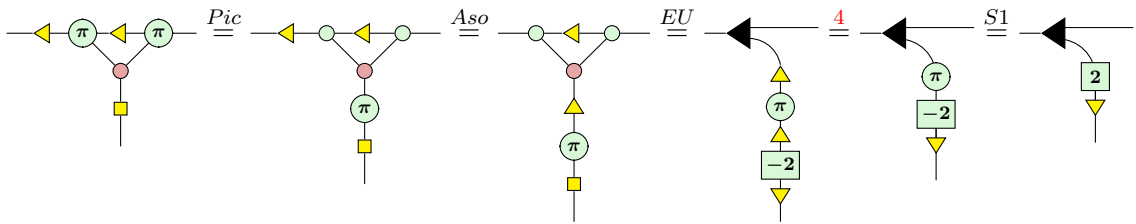
\begin{lemma}\label{bigtrianlgetow}
$$%
	\beginpgfgraphicnamed{compare/l1}
	\InputIfFileExists{compare/l1.tikz}{}{\input{./figures/compare/l1.tikz}}%
	\endpgfgraphicnamed
 = %
	\beginpgfgraphicnamed{compare/l6}
	\InputIfFileExists{compare/l6.tikz}{}{\input{./figures/compare/l6.tikz}}%
	\endpgfgraphicnamed
$$
\end{lemma}
\begin{proof}
$$
	\beginpgfgraphicnamed{compare/l1}
	\InputIfFileExists{compare/l1.tikz}{}{\input{./figures/compare/l1.tikz}}%
	\endpgfgraphicnamed
  \overset{Pic}{=}
	\beginpgfgraphicnamed{compare/l2}
	\InputIfFileExists{compare/l2.tikz}{}{\input{./figures/compare/l2.tikz}}%
	\endpgfgraphicnamed
 \overset{Aso}{=}
	\beginpgfgraphicnamed{compare/l3}
	\InputIfFileExists{compare/l3.tikz}{}{\input{./figures/compare/l3.tikz}}%
	\endpgfgraphicnamed

\overset{EU}{=}
	\beginpgfgraphicnamed{compare/l4}
	\InputIfFileExists{compare/l4.tikz}{}{\input{./figures/compare/l4.tikz}}%
	\endpgfgraphicnamed
 \overset{ \ref{traingleinverse}}{=}
	\beginpgfgraphicnamed{compare/l5}
	\InputIfFileExists{compare/l5.tikz}{}{\input{./figures/compare/l5.tikz}}%
	\endpgfgraphicnamed
 \overset{S1}{=}
	\beginpgfgraphicnamed{compare/l6}
	\InputIfFileExists{compare/l6.tikz}{}{\input{./figures/compare/l6.tikz}}%
	\endpgfgraphicnamed

$$
\end{proof}

\begin{remark}
The differentiation of ``linear'' ZX diagrams result obtained by
\cite{jeandel2022addition} is equivalent to our lemma \ref{diffthm2tm}.
\ctikzfig{compare/c1}
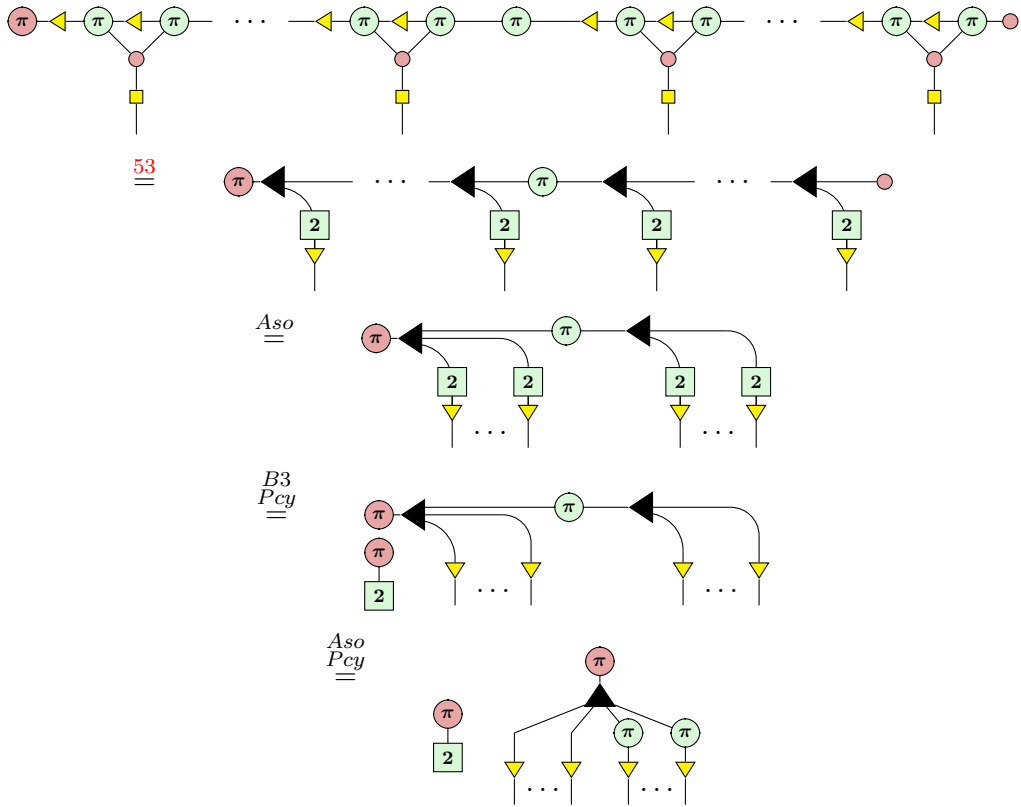
$$\overset{ \ref{bigtrianlgetow}}{=}\qquad%
	\beginpgfgraphicnamed{compare/c2}
	\InputIfFileExists{compare/c2.tikz}{}{\input{./figures/compare/c2.tikz}}%
	\endpgfgraphicnamed
$$
$$\overset{Aso}{=}\qquad%
	\beginpgfgraphicnamed{compare/c3}
	\InputIfFileExists{compare/c3.tikz}{}{\input{./figures/compare/c3.tikz}}%
	\endpgfgraphicnamed
$$
$$\overset{\substack{B3\\  Pcy}}{=}\qquad%
	\beginpgfgraphicnamed{compare/c4}
	\InputIfFileExists{compare/c4.tikz}{}{\input{./figures/compare/c4.tikz}}%
	\endpgfgraphicnamed
$$
$$\overset{\substack{Aso\\  Pcy}}{=}\qquad%
	\beginpgfgraphicnamed{compare/c5}
	\InputIfFileExists{compare/c5.tikz}{}{\input{./figures/compare/c5.tikz}}%
	\endpgfgraphicnamed
$$
Although the end results are equivalent,
we emphasise that their result is obtained through their theory of summing
controlled diagrams, whilst our result is obtained through our arbitrary
differentiation result.
\end{remark}

\begin{remark}
The general differentiation result in \cite{jeandel2022addition} requires an
inductive conversion to controlled diagrams, and is obtained through combining
diagrammatic addition and the Leibniz product rule. Because of this, the
resulting diagram will not resemble the original diagram.

In comparison, our Theorem \ref{diffthmgeneralgbox}
does not affect the topology of the original diagram and can be calculated
almost instantly.
\end{remark}

\begin{remark}
Both the general results on addition of diagrams and on differentiation of diagrams shown in  \cite{jeandel2022addition} are based on induction on generators. If such induction methods are allowed, then there is an alternative way to generally add two ZX diagrams or differentiate a ZX diagram: first rewrite inductively the diagrams into the compressed normal form  shown in \cite{qwangnormalformbit} as follows
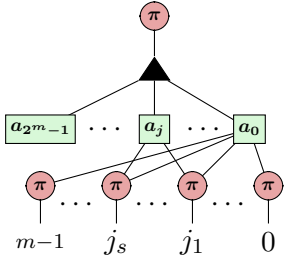
\[ %
	\beginpgfgraphicnamed{normalformcompress}
	\InputIfFileExists{normalformcompress.tikz}{}{\input{./figures/normalformcompress.tikz}}%
	\endpgfgraphicnamed
 \]
 which corresponding to the vector
\[
  \begin{pmatrix}
        a_0  \\  a_1\\
        \vdots \\ a_{2^m-2}\\
        a_{2^m-1} \end{pmatrix},
\]
then the sum of two diagrams can be obtained by  adding up the corresponding parameters in the two diagrams:
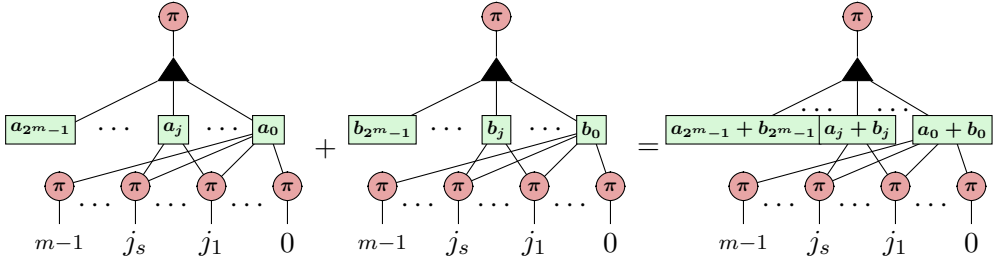
\[ %
	\beginpgfgraphicnamed{normalformsum}
	\InputIfFileExists{normalformsum.tikz}{}{\input{./figures/normalformsum.tikz}}%
	\endpgfgraphicnamed
 \]
and the differentiation can be obtained element-wisely:
\[\frac{\partial}{\partial t}\left[%
	\beginpgfgraphicnamed{normalformfunctions}
	\InputIfFileExists{normalformfunctions.tikz}{}{\input{./figures/normalformfunctions.tikz}}%
	\endpgfgraphicnamed
\right]\quad=%
	\beginpgfgraphicnamed{normalformfunctionsdiff}
	\InputIfFileExists{normalformfunctionsdiff.tikz}{}{\input{./figures/normalformfunctionsdiff.tikz}}%
	\endpgfgraphicnamed
\]

\end{remark}

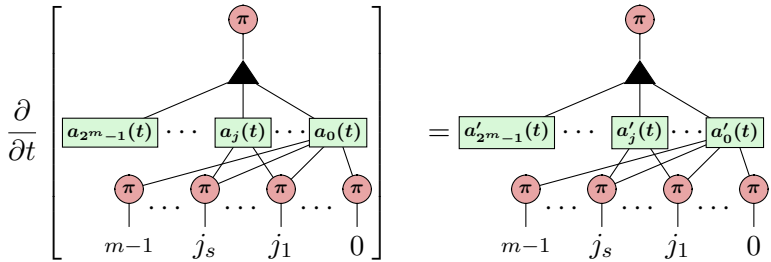
\end{document}